\documentclass[11pt]{article}

\usepackage{amssymb,amsmath,amsthm,amscd}
\usepackage{epsf,epsfig}%,amsfonts..epsf,
\usepackage{texdraw}%pstricks,

\newtheorem{Def}{Definition}[section]

\newtheorem{Pro}[Def]{Proposition}

\newtheorem{Rem}[Def]{Remark}

\newcommand{\br}{\begin{Rem}}
\newcommand{\er}{\end{Rem}}
\newtheorem{ex}[Def]{Example}
\newcommand{\bex}{\begin{ex}}
\newcommand{\eex}{\end{ex}}
\newcommand{\bd}{\begin{Def}}
\newcommand{\ed}{\end{Def}}

\newcommand{\be}{\begin{equation}}
\newcommand{\ee}{\end{equation}}
\newcommand{\bea}{\begin{eqnarray}}
\newcommand{\eea}{\end{eqnarray}}
\newcommand{\nn}{\nonumber}
\newcommand{\pa}{\partial}

\title{${\mathbb{Z}}_N$ graded discrete Lax pairs and discrete integrable systems}
\author{Allan P. Fordy and Pavlos Xenitidis \\ School of Mathematics, University of Leeds, Leeds LS2 9JT, UK}
\date{\today}

\begin{document}

\maketitle

\begin{abstract}
We introduce a class of ${\mathbb{Z}}_N$ graded discrete Lax pairs, with $N\times N$ matrices, linear in the spectral parameter.  We give a classification scheme for such Lax pairs and the associated discrete integrable systems.  We present two potential forms and completely classify the generic case.  Many well known examples belong to our scheme for $N=2$, so many of our systems may be regarded as generalisations of these. Even at $N=3$, several new integrable systems arise.  Many of our equations are mutually compatible, so can be used together to form ``coloured'' lattices.

We also present continuous isospectral deformations of our Lax pairs, giving compatible differential-difference systems, which play the role of continuous symmetries of our discrete systems.  We present master symmetries and a recursive formulae for their respective hierarchies, for the generic case.

We present two nonlocal symmetries of our discrete systems, which have a natural representation in terms of the potential forms.  These give rise to the two-dimensional Toda lattice, with our nonlinear symmetries being the B\"acklund transformations and our discrete system being the nonlinear superposition formula (for the generic case).
\end{abstract}

\emph{Keywords}: Discrete integrable system, Lax pair, symmetry, B\"acklund transformation, $2D$ Toda lattice.

%\tableofcontents

\section{Introduction}

The classification of discrete integrable systems is still rather primitive when compared with that of integrable PDEs.  The most famous is the ABS classification of integrable equations on quad-graphs \cite{ABS}, but this is only for scalar equations, associated with $2\times 2$ matrix Lax pairs.  Multi-component generalisations and scalar equations with $3\times 3$ Lax pairs are rather sporadic \cite{A,MX,NPCQ,NPbog,SHL}.

In this paper we introduce an $N\times N$ spectral problem, linear in the spectral parameter and with ${\mathbb{Z}}_N$ grading, which gives rise to a large class of discrete integrable systems.  In the $2\times 2$ specialisation, these include some well known examples (discrete potential, modified and Schwarzian KdV equations, Hirota's KdV equation and the discrete sine-Gordon equation).  The $3\times 3$ case includes the discrete versions of the Boussinesq and modified Boussinesq equations \cite{NPCQ}. The $N\times N$ case includes multi-component generalisations of them all.  Many completely new examples arise for $N\geq 3$.  Our {\em generic} system has two natural descriptions in terms of different {\em potential functions}, which, in the case of $N=2$, just reflect the discretisations of the ``KdV family''.  For $N=3$ we retrieve the ``Boussinesq family'', but also derive several completely new examples.  In particular the system
\begin{eqnarray}
\phi^{(0)}_{m+1,n+1} &=& \left(
\frac{\alpha\,\phi_{m+1,n}^{(1)}-\beta \phi^{(1)}_{m,n+1}}{\alpha \,\phi_{m+1,n}^{(0)}\phi^{(1)}_{m,n+1}-\beta \,\phi_{m,n+1}^{(0)}\phi^{(1)}_{m+1,n}} \right) \,\frac{1}{\phi^{(0)}_{m,n}}\,,   \nn\\
\phi^{(1)}_{m+1,n+1} &=&  \left(
\frac{\alpha\,\phi_{m,n+1}^{(0)}-\beta \phi^{(0)}_{m+1,n}}{\alpha \,\phi_{m+1,n}^{(0)}\phi^{(1)}_{m,n+1}-\beta \,\phi_{m,n+1}^{(0)}\phi^{(1)}_{m+1,n}} \right)\,\frac{1}{\phi^{(1)}_{m,n}}\,.  \nn
\end{eqnarray}
arises in the context of {\em quotient potentials} and can be {\em decoupled} to a nine-point scalar equation for either of the functions involved in it (see (\ref{eq:L3-U-mBSQ})) and can be {\em reduced} to a simple scalar equation on a quadrilateral (see (\ref{eq:MX2})).  On the other hand, the system
\bea
 \chi_{m+1,n+1}^{(0)} &=& \frac{(\chi^{(0)}_{m+1,n}-\chi^{(1)}_{m,n}) \chi^{(1)}_{m+1,n} - (\chi^{(0)}_{m,n+1}-\chi^{(1)}_{m,n}) \chi^{(1)}_{m,n+1}}{\chi^{(0)}_{m+1,n} - \chi^{(0)}_{m,n+1}},  \nn\\
 \chi^{(1)}_{m+1,n+1} &=& \chi^{(0)}_{m,n}\,+\,\frac{1}{\chi_{m+1,n}^{(1)} - \chi_{m,n+1}^{(1)}}\,\left(\frac{\alpha^3}{\chi_{m+1,n}^{(0)}-\chi^{(1)}_{m,n} }\,
                                         -\,\frac{\beta^3}{\chi_{m,n+1}^{(0)}-\chi^{(1)}_{m,n}}\right),  \nn
\eea
arises in the context of {\em additive potentials} and can be decoupled for either of the variables to the nine point scalar discrete Boussinesq equation.  We also consider a degenerate form of our Lax pair, which includes multicomponent generalisations of Hirota's KdV equation (see (\ref{eq:gen-dis-eq})).

Many of our systems are pairwise compatible, so can be used to build two and three dimensional consistent lattices, some of which have unusual features, giving rise to non-standard {\em initial value problems}.

We also consider {\em continuous deformations} of our discrete Lax matrices, giving rise to differential difference equations, which can be interpreted as continuous (local) symmetries of our fully discrete systems (see Section \ref{continuous-defs}).  Again, the ${\mathbb{Z}}_N$ grading gives us a way of systematically calculating general formulae for our generic systems.  We give the general form of the ``first'' generalised symmetry, together with a ``master symmetry'', with which it is possible to construct higher symmetries.

We present two {\em nonlocal symmetries} (see Section \ref{2dtoda}) for our general discrete system.  These are associated with forms of the $2D$ Toda lattice \cite{M1,FG3}), when we use the potential forms.  In particular, in the quotient potential form, the nonlocal symmetries act as B\"acklund transformations for this Toda lattice.  In the generic (non-reduced) case, our fully discrete system is just the corresponding nonlinear superposition principle.  This connection with the Toda lattice was to be expected, since this nonlinear superposition formula contains, as special cases, several well known examples of discrete integrable system, including the modified KdV and modified Boussinesq equations (see \cite{NC}).  The connection of the $2D$ Toda lattice to these ``modified'' (PDE) hierarchies was given in \cite{FG3} in the context of the factorisation of scalar Lax operators \cite{FG1,FG2} and the whole hierarchy shares the same Bianchi superposition formula.

In the next section we present the basic algebraic framework used throughout the paper.  One of the most important results is to provide a classification scheme which enables us to place all our examples in a coherent framework.  Most of the paper is concerned with systematically analysing each class of discrete system, together with the corresponding differential-difference symmetries.

A number of open problems are discussed in the conclusions.

\section{${\mathbb{Z}}_N$-Graded Lax Pairs} \label{sec:ZN-LP}

In this section we introduce the general framework for what follows in the paper.  We first introduce the idea of ${\mathbb{Z}}_N$-grading and {\em level structure}. We then introduce the general Lax pair and the corresponding discrete system.  There is a considerable amount of redundancy, so we introduce an {\em equivalence relation}, which enables us to classify all systems belonging to our framework.  As a consequence of the {\em level structure}, our systems fall naturally into two categories: {\em coprime} and {\em non-coprime} cases.  The {\em coprime} case further separates into the {\em generic} and the {\em degenerate} subcases.

In what follows we use the following convention for upper indices in parentheses: the notation $a \in {\mathbb{Z}}_N$ implies that $a$ is an integer such that $0 \le a \le N-1$ and ({\em by definition}) summation in ${\mathbb{Z}}_N$ is taken $\bmod{N}$.  The notation $u^{(i)}_{m,n}$ denotes that $u^{(i)}$ is located at lattice point $(m,n)$.  The symbol $\delta_{i,j}$ will denote the usual Kroneker delta.  The symbols ${\cal{S}}_m$ and ${\cal{S}}_n$ will respectively denote the shifts in the $m$ and $n$ directions: ${\cal{S}}_m u^{(i)}_{m,n}=u^{(i)}_{m+1,n}$ and ${\cal{S}}_n u^{(i)}_{m,n}=u^{(i)}_{m,n+1}$, with $\Delta_m = {\cal{S}}_m-1$ and $\Delta_n = {\cal{S}}_n-1$ the corresponding {\em differences}.

\subsection{${\mathbb{Z}}_N$-Grading}

To introduce ${\mathbb{Z}}_N$-grading, we need:
\begin{Def}[Matrix $\Omega$]
The $N \times N$ matrix $\Omega$ is defined by
$$
 (\Omega)_{i,j} \,=\, \delta_{j-i,1} +  \delta_{i-j,N-1},
$$
which will be said to have {\em level $1$}.
\end{Def}
This is a cyclic matrix, satisfying $\Omega^N={\rm{I}}_N$ (the $N\times N$ identity matrix), so defines a grading, with
$$
(\Omega^k)_{i,j} \,=\, \delta_{j-i,k} +  \delta_{i-j,N-k},\quad 0 \leq k \leq N-1,
$$
having {\em level $k$}. $\Omega^N$ has {\em level $0$}.  It also follows that $(\Omega^k)^{-1}=\Omega^{N-k}$.

\begin{Def}[A level $k$ matrix]\label{def:level-k-mat}
An $N \times N$ matrix $A$ of the form
$$
A \,=\, {\rm{diag}}\left(a^{(0)},\dots,a^{(N-1)} \right)\Omega^{k}
$$
will be said to have level $k$, written $lev(A) = k$.
\end{Def}

\noindent
It can be seen that if $(N,k)=1$, then $A^N=\left(\prod_{i=0}^{N-1} a^{(i)}\right)\, {\rm{I}}_N$.

If $B$ is another $N\times N$ matrix of a certain level, then
$$
lev(AB)\, = \, lev(BA)\,  = \, lev(A) + lev(B)\; (\bmod{N}).
$$

Let us denote by ${\rm{R}}$ the ring  of the $N \times N$ matrices, and by ${\rm{R}}_k$ the set of all $N \times N$ matrices of level $k$. Then, it follows that ${\rm{R}}$ is a ${\mathbb{Z}}_N$-graded ring as it is the direct sum decomposition
$$
{\rm{R}} \,=\, \bigoplus_{k \in {\mathbb{Z}}_N} {\rm{R}}_k\,,\qquad {\rm{R}}_k {\rm{R}}_\ell \subseteq {\rm{R}}_{k+\ell}.
$$

\subsection{The General Lax Pair}

We employ the above ${\mathbb{Z}}_N$ grading of ${\rm{R}}$ to build some particular classes of discrete Lax pairs, whose matrices belong to the polynomial ring ${\rm{R}}[\lambda]$.

Specifically, we consider a pair of matrix equations of the form
\begin{subequations}\label{eq:dLP-gen}
\begin{eqnarray}
&& \Psi_{m+1,n} \,=\, L_{m,n}\, \Psi_{m,n} \,\equiv\, \left( U_{m,n} \,+\, \lambda \,\Omega^{\ell_1}\right)\,\Psi_{m,n},\quad lev\left(U_{m,n}\right) = k_1 \ne \ell_1\,,\label{eq:dLP-gen-L} \\
&& \Psi_{m,n+1} \,=\, M_{m,n}\, \Psi_{m,n} \,\equiv\, \left( V_{m,n} \,+\, \lambda \,\Omega^{\ell_2}\right)\,\Psi_{m,n},\quad lev\left(V_{m,n}\right) = k_2 \ne \ell_2\,,\label{eq:dLP-gen-M}
\end{eqnarray}
\end{subequations}
which is characterised by the quadruple $\left(k_1,\ell_1;k_2,\ell_2 \right)$. We refer to it as {\emph{the level structure}} of system (\ref{eq:dLP-gen}) and derive necessary conditions for the system (\ref{eq:dLP-gen}) to be compatible.

Since matrices $U$, $V$ and $\Omega$ are independent of $\lambda$, the compatibility condition of (\ref{eq:dLP-gen}),
\begin{equation} \label{eq:dLP-gen-cc}
L_{m,n+1} M_{m,n} \,=\,M_{m+1,n} L_{m,n},
\end{equation}
splits into the system
\begin{subequations} \label{eq:dLP-gen-scc}
\begin{eqnarray}
U_{m,n+1} V_{m,n} &=& V_{m+1,n} U_{m,n}\,, \label{eq:dLP-gen-scc-1}\\
U_{m,n+1} \Omega^{\ell_2} - \Omega^{\ell_2} U_{m,n}  &=& V_{m+1,n} \Omega^{\ell_1} - \Omega^{\ell_1} V_{m,n} \,.\label{eq:dLP-gen-scc-2}
\end{eqnarray}
\end{subequations}
It is obvious that both sides of equation (\ref{eq:dLP-gen-scc-1}) have the same level $k_1 + k_2$. On the other hand, the left and right hand sides of equation (\ref{eq:dLP-gen-scc-2}) have respective levels $k_1 + \ell_2$ and $k_2 + \ell_1$. Hence, compatibility condition (\ref{eq:dLP-gen-scc-2}) yields nontrivial equations for the entries of matrices $U$ and $V$ if and only if
\begin{equation}\label{eq:dLP-nec-rel}
k_1 + \ell_2 \,\equiv\,k_2 + \ell_1 \; (\bmod N).
\end{equation}

\begin{Def}[The quadruple ${\cal{Q}}_N$]
We denote the set of quadruples $(k_1,\ell_1 ; k_2,\ell_2)$ which satisfy condition (\ref{eq:dLP-nec-rel}) as
$$
{\cal{Q}}_N\,=\,\left\{
(k_1,\ell_1 ; k_2,\ell_2) \in {\mathbb{Z}}_N^4 \, \big| \, k_1 \ne \ell_1,\,\, k_2 \ne \ell_2 ,\,\, k_1+\ell_2 \equiv k_2 + \ell_1\; (\bmod{N})
\right\},
$$
\end{Def}

\br
This condition implies that $\ell_2-k_2\equiv  \ell_1-k_1 \, (\bmod N)$, so the greatest common divisors $(N,\ell_1-k_1)$ and $(N,\ell_2-k_2)$ are equal.
\er

\bd[Coprime Case]
We define the Lax pair (\ref{eq:dLP-gen}) to be coprime if $(N,\ell_i-k_i)=1$.  Otherwise, we refer to it as non-coprime.
\ed

It is easily seen that in the coprime case, we have
\be\label{coprimeDet}
\det (L_{m,n}) = \prod_{i=0}^{N-1} u_{m,n}^{(i)} - (-\lambda)^N, \quad\mbox{and}\quad
                                \det (M_{m,n}) = \prod_{i=0}^{N-1} v_{m,n}^{(i)} - (-\lambda)^N.
\ee

\subsubsection{The General Discrete System}

Supposing that condition (\ref{eq:dLP-nec-rel}) is satisfied, let
\begin{equation} \label{eq:A-B-entries}
U_{m,n}\,=\, {\rm{diag}}\left(u^{(0)}_{m,n},\cdots,u^{(N-1)}_{m,n}\right) \Omega^{k_1}\,,\quad V_{m,n}\,=\, {\rm{diag}}\left(v^{(0)}_{m,n},\cdots,v^{(N-1)}_{m,n}\right) \Omega^{k_2}.
\end{equation}
In view of (\ref{eq:A-B-entries}), equations (\ref{eq:dLP-gen-scc}) can be written explicitly as
\begin{subequations}  \label{eq:dLP-ex-cc}
\begin{eqnarray}
u^{(i)}_{m,n+1} v_{m,n}^{(i+k_1)} &=& v^{(i)}_{m+1,n} u^{(i+k_2)}_{m,n}\,,\quad i \in {\mathbb{Z}}_N ,  \label{eq:dLP-ex-cc-1}\\
u^{(i)}_{m,n+1} - u_{m,n}^{(i+\ell_2)} &=& v^{(i)}_{m+1,n} - v^{(i+\ell_1)}_{m,n}\,,\quad i \in {\mathbb{Z}}_N ,   \label{eq:dLP-ex-cc-2}
\end{eqnarray}
\end{subequations}
or, in a solved form, as
\begin{equation}  \label{eq:dLP-ex-cc-s}
u^{(i)}_{m,n+1} \,=\, \frac{u^{(i+\ell_2)}_{m,n} - v^{(i+\ell_1)}_{m,n}}{u^{(i+k_2)}_{m,n} - v^{(i+k_1)}_{m,n}}\, u^{(i+k_2)}_{m,n}\,, \quad
v^{(i)}_{m+1,n}\,=\, \frac{u^{(i+\ell_2)}_{m,n} - v^{(i+\ell_1)}_{m,n}}{u^{(i+k_2)}_{m,n} - v^{(i+k_1)}_{m,n}}\, v^{(i+k_1)}_{m,n},
\end{equation}
assuming that $u^{(i)}_{m,n} \ne v^{(j)}_{m,n}$ for all $i,j \in {\mathbb{Z}}_N$.

If we add equations (\ref{eq:dLP-ex-cc-2}) we obtain the {\em discrete local conservation law}
\be\label{lcl}
\Delta_n\, \left(\sum_{i=0}^{N-1} u^{(i)}_{m,n}\right) = \Delta_m\, \left(\sum_{i=0}^{N-1} v^{(i)}_{m,n}\right).
\ee
This could be obtained directly from the matrix equation (\ref{eq:dLP-gen-cc}) by first multiplying by $\Omega^{-k_1-\ell_2}$  and then taking the trace.  The matrix equation (\ref{eq:dLP-gen-cc}) also implies the following equation for the determinants:
\begin{equation} \label{eq:dLP-gen-cc-det}
\det(L_{m,n+1}) \,  \det(M_{m,n}) \,=\,\det(M_{m+1,n})\,  \det(L_{m,n}),
\end{equation}
which is polynomial in $\lambda$, so implies a separate condition for each coefficient.  The precise form of these determinants depends upon the choice of quadruple $(k_1,\ell_1 ; k_2,\ell_2)$.  In the {\em coprime} case we have
$$
\Delta_n(a)=\Delta_m(b), \quad\mbox{where}\quad
          a=\prod_{i=0}^{N-1} u_{m,n}^{(i)},\;\;\; b = \prod_{i=0}^{N-1} v_{m,n}^{(i)}\, ,
$$
and ${\cal{S}}_n(a) b = {\cal{S}}_m(b) a$, which, together, imply
\be \label{eq:det=c}
\Delta_n(a)=0 \quad\mbox{and}\quad \Delta_m(b)=0.
\ee
This can also be seen from looking directly at the form of equations (\ref{eq:dLP-ex-cc-s}).  The quantities $a$ and $b$, defined above, play an important part in this paper.  The non-coprime case is discussed in Section \ref{gcd}.

\subsubsection{Equivalent Lax Pairs}

We consider two transformations which give rise to equivalent Lax pairs:

\begin{enumerate}
\item
The interchange of lattice variables $(m,n) \mapsto (n,m)$ is a point transformation which corresponds to the interchange of matrices $L_{m,n}$ and $M_{m,n}$. Algebraically, this point transformation corresponds to the interchange of pairs $(k_1,\ell_1)$ and $(k_2,\ell_2)$ which does not affect condition (\ref{eq:dLP-nec-rel}).
\item
Consider the gauge transformation of Lax pair (\ref{eq:dLP-gen}) with the constant matrix $G$ defined as
\begin{equation} \label{eq:def-mat-G}
(G)_{i,j}= \delta_{i+j,2}+\delta_{i+j,N+2},\quad\mbox{for}\;\;    1\leq i,j \leq N
\end{equation}
which satisfies $G^{-1} = G$.

Since $G \Omega G^{-1} = \Omega^{-1}=\Omega^{N-1}$, this switches level  $k$ matrices with level $N-k$ matrices ($\bmod{N}$, so level $0$ stay as level $0$).

Applying this gauge transformation to Lax pair (\ref{eq:dLP-gen}), we will derive a system with level structure $(N-k_1,N-\ell_1;N-k_2,N-\ell_2)$. For the discrete system (\ref{eq:dLP-ex-cc}), this gauge transformation corresponds to a permutation of the dependent variables,
$$
\left( u^{(i)}_{m,n}\,,\, v^{(i)}_{m,n} \right) \,\mapsto \,  \left( u^{(N-i)}_{m,n}\,,\, v^{(N-i)}_{m,n} \right).
$$
Again, condition (\ref{eq:dLP-nec-rel}) is satisfied for the quadruple $(N-k_1,N-\ell_1;N-k_2,N-\ell_2)$ provided that it holds for $(k_1,\ell_1;k_2,\ell_2)$.
\end{enumerate}

Using the above transformations we define the following equivalence relation among Lax pairs (\ref{eq:dLP-gen}).
\begin{Def}[Equivalence relation] \label{def:equiv-LP}
Two discrete Lax pairs with level structures $\left(k_1,\ell_1;k_2,\ell_2 \right) \in {\cal{Q}}_N$ and $\left(k_1^\prime,\ell_1^\prime;k_2^\prime,\ell_2^\prime \right) \in {\cal{Q}}_N$ are equivalent and we write
$$\left(k_1^\prime,\ell_1^\prime;k_2^\prime,\ell_2^\prime \right) \,  \sim \,   \left(k_1,\ell_1;k_2,\ell_2 \right) \,,$$
if one quadruple can be mapped to the other by applying any of the following transformations.
\begin{equation} \label{eq:equiv-P}
\begin{array}{lcl}
{\cal{T}}_1 &:& \left(a,b;c,d \right)  \mapsto \left(c,d;a,b \right) \\
{\cal{T}}_2 &:& \left(a,b;c,d \right)  \mapsto \left(N-a,N-b;N-c,N-d \right).
\end{array}
\end{equation}
Otherwise, they will be called inequivalent.
\end{Def}
We can now reduce the problem of classification of Lax pairs to the classification of inequivalent classes of quadruples $\left(k_1,\ell_1;k_2,\ell_2 \right)$ in the quotient space ${\cal{Q}}_N/\sim$.

\subsection{The Greatest Common Divisor $(N,\ell-k)=p$}\label{gcd}

For this section, we define ${\bf u} = (u^{(0)},u^{(1)},\dots ,u^{(N-1)})$ and ${\cal D}_{N}^{\bf u}={\rm{diag}}(u^{(0)},u^{(1)},\dots ,u^{(N-1)})$ and denote the $N\times N$ matrix $\Omega$ by $\Omega_N$.  Our Lax matrix $L$ is of the form
\be\label{L-Dok}  %
L= {\cal D}_N^u\, \Omega_N^k+\lambda \, \Omega_N^\ell = \left({\cal D}_N^u +\lambda \, \Omega_N^{\ell-k}\right)\, \Omega_N^k.
\ee  %
For this section we are just writing $(k,\ell)$ instead of $(k_1,\ell_1)$ and we are suppressing the dependence on $(m,n)$.
We now consider the consequence of $(N,\ell-k)=p\neq 1$.  We have integers $q,r$, such that $N=pq,\, \ell-k=pr$, with $(q,r)=1$.

\bd[Permutation matrix]  %
Let the permutation matrix $P$ be such that
$$
P_{ij}=\left\{\begin{array}{ll}
                1 & \mbox{when}\; (i,j)\in \{(n+(m-1)q,m+(n-1) p),\; 1\leq m \leq p, 1\leq n \leq q\}, \\
                0 & \mbox{otherwise}.
              \end{array}  \right.
$$
\ed  %
Defining ${\bf u}_i = (u^{(i)},u^{(i+p)},\dots ,u^{(i+p(q-1))}),\; i=0, \dots ,p-1$ and
${\bf u}^b=({\bf u}_0,\dots , {\bf u}_{p-1})$, we have
$$
\left({\bf u}^b\right)^T=P\, {\bf u}^T \quad\mbox{and therefore}\quad {\bf u}^b= {\bf u} P^T={\bf u} P^{-1}.
$$
Thus, for any matrix $A$, the matrix $P A P^{-1}$ has a $p\times p$ block structure, with each block a $q\times q$ matrix.  The components of the $(i,j)$ block are $\{A_{i+(l-1)p,j+(m-1)p}\}_{l,m=1}^{q}$.

Let ${\cal D}_{q}^{{\bf u}_i}={\rm{diag}}\left(u^{(i)},u^{(i+p)},\dots ,u^{(i+p(q-1))} \right)$, a $q\times q$ diagonal matrix.  Then
$$
P {\cal D}_N^u P^{-1} = {\rm{diag}}\left({\cal D}_{q}^{{\bf u}_0},\dots,{\cal D}_{q}^{{\bf u}_{p-1}} \right),
$$
and $\omega_p=P \Omega_N P^{-1}$ has a $p\times p$ block structure, with $(\omega_p)_{i,i+1}=I_q,\, i=1,\dots ,p-1$ and $(\omega_p)_{p,1}=\Omega_q$. We then have
$$
\omega_p^p={\rm{diag}}\left(\Omega_q,\dots,\Omega_q \right),\quad\mbox{so}\;\; \omega_p^{\ell-k}=\omega_p^{pr}={\rm{diag}}\left(\Omega_q^r,\dots,\Omega_q^r \right).
$$

We can piece these formulae together in

\begin{Pro}[Block Structure when $(N,\ell-k)=p\neq1$] \label{prop:prop-block}
Let the permutation matrix $P$ be defined as above and $L$ be given by (\ref{L-Dok}).  Then
\be\label{plp-1}
P L P^{-1} = {\rm{diag}}\left(L^{(0)},\dots,L^{(p-1)} \right) \, \omega_p^k,
\ee
where
$$
L^{(i)} = {\cal D}_q^{{\bf u}_i} + \lambda \Omega_q^r \quad\mbox{and}\quad N=pq,\;\; \ell-k = pr.
$$
We then have
$$
\det\left(L\right)= (-1)^{(N-1)k} \prod_{i=0}^{p-1}
    \left( a_i- (-\lambda)^{q}\right),\quad\mbox{where}\;\; a_i=\prod_{j=0}^{q-1} u^{(i+j p)}.
$$
\end{Pro}
Note that this determinant follows from that of each (coprime) block, with
$$
\det(L^{(i)})=a_i-(-\lambda)^q \quad\mbox{and}\quad \det(\omega_p)=\det(\Omega_N)=(-1)^{N-1}.
$$

\subsubsection{The Lax Pair when $(N,\ell_i-k_i)=p$}\label{sect:pnot=1}

For the Lax pair (\ref{eq:dLP-gen}), we have seen that $k_i,\ell_i$ must satisfy (\ref{eq:dLP-nec-rel}).  As a consequence, $(N,\ell_1-k_1)=(N,\ell_2-k_2)=p$.  Suppose $p\neq 1$.  Then the {\em same} permutation matrix $P$ transforms {\em both} $L$ and $M$ to the form described in Proposition \ref{prop:prop-block}, with the {\em same} $p, q, r$, but $k_1$ and $k_2$ may be distinct.  Therefore the determinants have the {\em same} structure:
\begin{subequations}   \label{eq:detLM}
\begin{eqnarray}
&& \det\left(L_{m,n}\right)= (-1)^{(N-1)k_1} \prod_{i=0}^{p-1} \left( a_i -  (-\lambda)^{q}\right),
                          \quad\mbox{where} \;\; a_i=\prod_{j=0}^{q-1} u^{(i+j p)}_{m,n}, \label{eq:detL}\\
&& \nonumber \\
&&  \det\left(M_{m,n}\right) = (-1)^{(N-1)k_2} \prod_{i=0}^{p-1} \left( b_i- (-\lambda)^{q}\right)
   \quad\mbox{where}  \;\; b_i = \prod_{j=0}^{q-1} v^{(i+j p)}_{m,n}.  \label{eq:detM}
\end{eqnarray}
\end{subequations}
where
$$
b_i- (-\lambda)^{q} = |M^{(i)}| = |{\cal D}_q^{{\bf v}_i} + \lambda \Omega_q^r|.
$$
The formula (\ref{eq:det=c}) then implies that {\em symmetric functions} of $a_i$ and of $b_i$ are constants of the motion.

In fact, if we use $\hat a_i$ to denote $a_i$ when evaluated at $(m,n+1)$ and $\tilde b_i$ to denote $b_i$ when evaluated at $(m+1,n)$, then we have
$$
\hat a_i = a_{i+k_2} \quad\mbox{and}\quad   \tilde b_i = b_{i+k_1},
$$
where $i+k_1$ and $i+k_2$ are taken $\bmod{p}$.  These are simple permutations on finite sets.  Symmetric polynomials of their orbits are first integrals.

\begin{Rem}
When $p=1$ we just have that
$$
a_0=\prod_{j=0}^{N-1} u^{(j)} \quad\mbox{and}\quad b_0=\prod_{j=0}^{N-1} v^{(j)}
$$
are constants in $n$ and $m$ respectively.  This is easy to see by looking at the form of equations (\ref{eq:dLP-ex-cc-s}), since the product
$$
\prod_{i=0}^{N-1} \left(\frac{u^{(i+\ell_2)}_{m,n} - v^{(i+\ell_1)}_{m,n}}{u^{(i+k_2)}_{m,n} - v^{(i+k_1)}_{m,n}}\right) = 1.
$$
This is a consequence of equation (\ref{eq:dLP-nec-rel}), which implies that $\ell_2-\ell_1 = k_2-k_1$, so the product in the denominator is just a re-ordering of that in the numerator.
\end{Rem}

\subsection{Classification Problem}\label{sect:class-prob}

Summarising the above analysis, we can formulate the classification problem of Lax pairs as follows.

For every dimension $N$, find all the equivalence classes in the quotient space ${\cal{Q}}_N/\sim$. For the representatives of  the classified equivalence classes, consider the corresponding Lax pairs (\ref{eq:dLP-gen}) and analyse the resulting systems (\ref{eq:dLP-ex-cc}) depending on whether or not $N$ and $\ell_i-k_i$, $i=1,2$, are coprime.
\begin{enumerate}
\item  {\it{The coprime case:}} $ \big(N,\ell_1-k_1\big) =  \big(N,\ell_2-k_2\big) =1$. \\
This involves Lax pairs which satisfy
\begin{equation} \label{eq:dLP-coprime-case}
\prod_{j=0}^{N-1} u^{(j)}_{m,n} \,=\,a,\quad
      \prod_{j=0}^{N-1} v^{(j)}_{m,n} \,=\,b,\quad a, \,b \in {\mathbb{C}}.
\end{equation}
The above relations allow us to express one function from each set in terms of the remaining ones. The coprime case is further subdivided into:
\begin{itemize}
\item {\it{The generic subcase :}} $a b \ne 0$,
\item {\it{The degenerate subcase :}} $a \ne 0$, $ b = 0$.
\end{itemize}
Lax pairs with $a =0$, $b \ne 0$ are equivalent to the above degenerate case by a change of independent variables. Finally, the fully degenerate case $a = b  =0$ is empty. \\

\item {\it{The non-coprime case:}} $ \big(N,\ell_1-k_1\big) =  \big(N,\ell_2-k_2\big) = p > 1$ \\
This case corresponds to the Lax pairs discussed in section (\ref{sect:pnot=1}), which allows us to reduce the number of functions in each set by $p$.

Depending on the values of $k_1$ and $k_2$, we may obtain completely decoupled systems of lower dimension.  Otherwise, we obtain a {\em coupled system} of coprime cases of lower dimension.
\end{enumerate}

\section{The Coprime Case}\label{sect:coprime}
\setcounter{equation}{0}

We analyse system (\ref{eq:dLP-ex-cc}) for the coprime case with $\big(N,\ell_i-k_i\big) =1$.

For the {\em generic case}, with $a b \ne 0$, relations (\ref{eq:dLP-coprime-case}) imply that none of the functions $u$ and $v$ can be identically zero.  This allows us to express functions $u$ and $v$ in terms of some potentials so that either equations (\ref{eq:dLP-ex-cc-1}) or equations (\ref{eq:dLP-ex-cc-2}) hold identically.  Sections \ref{sect:coprime-quotient} and \ref{sect:coprime-additive} analyse these two basic cases.  We give a list of all inequivalent systems in $2$ and $3$ dimensions and extend some of these to $N$ dimensions.  For $N=2$ we recover many of the standard scalar systems, including the discrete modified KdV and potential KdV equations, the discrete Schwarzian KdV (see \cite{NC} and references therein) and the Hirota's discrete sine-Gordon equation (see \cite{DJM} and references therein).  In Section \ref{BTpotentials} we discuss the B\"acklund relation between the two potential cases.

We finish this section by discussing the {\em degenerate case}, for which $a\ne 0, b=0$.  This means that at least one of the functions $v^{(i)}_{m,n}$ must be identically zero.  The degenerate case is generally much more complex, so we cannot yet give a full classification.  After some general calculations, we restrict ourselves to some examples, the simplest of which is the Hirota KdV equation.  We generalise this to $N$ dimensions and give a scalar reduction in each case.

\subsection{Quotient Potentials}\label{sect:coprime-quotient}

Equations (\ref{eq:dLP-ex-cc-1}) hold identically if we set
\begin{equation} \label{eq:dLP-gen-ch-1}
u^{(i)}_{m,n}\,= \, \alpha\,\frac{\phi^{(i)}_{m+1,n}}{\phi^{(i+k_1)}_{m,n}}\,,\quad v^{(i)}_{m,n}\,= \, \beta\,\frac{\phi^{(i)}_{m,n+1}}{\phi^{(i+k_2)}_{m,n}}\,,\quad i \in {\mathbb{Z}}_N,
\end{equation}
where, from (\ref{eq:dLP-coprime-case}), $a=\alpha^N,\, b = \beta^N$.  Equations (\ref{eq:dLP-ex-cc-2}) then take the form
\begin{equation} \label{eq:dLP-gen-sys-1}
\alpha \left(\frac{\phi^{(i)}_{m+1,n+1}}{\phi^{(i+k_1)}_{m,n+1}}\,-\,\frac{\phi^{(i+\ell_2)}_{m+1,n}}{\phi^{(i+\ell_2+k_1)}_{m,n}} \right) \,=\,
  \beta \left(\frac{\phi^{(i)}_{m+1,n+1}}{\phi^{(i+k_2)}_{m+1,n}}\,-\,\frac{\phi^{(i+\ell_1)}_{m,n+1}}{\phi^{(i+\ell_1+k_2)}_{m,n}} \right)\,,
      \quad  i \in {\mathbb{Z}}_N,
\end{equation}
and their solved form (\ref{eq:dLP-ex-cc-s}) is written as
\begin{equation} \label{eq:dLP-gen-sys-1-a}
\phi^{(i)}_{m+1,n+1}\,=\, \frac{\phi_{m,n+1}^{(i+k_1)} \phi_{m+1,n}^{(i+k_2)}}{\phi_{m,n}^{(i+k_1+\ell_2)}}\,\left(
\frac{\alpha\,\phi_{m+1,n}^{(i+\ell_2)}- \beta \,\phi_{m,n+1}^{(i+\ell_1)}}{\alpha\,\phi_{m+1,n}^{(i+k_2)}- \beta \,\phi_{m,n+1}^{(i+k_1)}}
\right)\,,\quad  i  \in {\mathbb{Z}}_N.
\end{equation}
In this potential form, the Lax pair (\ref{eq:dLP-gen}) can be written
\begin{subequations} \label{eq:LP-ir-g-rat}
\begin{equation} \label{eq:LP-ir-g-rat-1}
\begin{array}{l} \Psi_{m+1,n} \,=\, \left( \alpha {\boldsymbol{\phi}}_{m+1,n} \Omega^{k_1} {\boldsymbol{\phi}}_{m,n}^{-1} + \lambda \Omega^{\ell_1}\right) \Psi_{m,n},\\[3mm]
 \Psi_{m,n+1} \,=\, \left( \beta {\boldsymbol{\phi}}_{m,n+1} \Omega^{k_2} {\boldsymbol{\phi}}_{m,n}^{-1} + \lambda \Omega^{\ell_2}\right) \Psi_{m,n}, \end{array}
\end{equation}
where
\begin{equation} \label{eq:LP-ir-g-rat-2}
{\boldsymbol{\phi}}_{m,n} \,:=\, {\rm{diag}}\left(\phi^{(0)}_{m,n},\cdots,\phi^{(N-1)}_{m,n}\right) \quad {\mbox{and}} \quad \det\left({\boldsymbol{\phi}}_{m,n}\right) = \prod_{i=0}^{N-1}\phi^{(i)}_{m,n}=1.
\end{equation}
\end{subequations}
We can then show that the Lax pair (\ref{eq:LP-ir-g-rat}) is compatible if and only if the system (\ref{eq:dLP-gen-sys-1}) holds.

Using the equivalence relation $\sim$ we can determine the inequivalent systems in every dimension $N$. However, in the quotient potential case we have a further equivalence relation at our disposal:
\begin{Pro} \label{prop:rat-pot-trans}
We denote system (\ref{eq:dLP-gen-sys-1}) by ${\cal{R}}(\phi^{(i)};k_1,\ell_1,\alpha;k_2,\ell_2,\beta)$. Consider also the system which follows from (\ref{eq:dLP-gen-sys-1}) by interchanging indices $k_i$ and $\ell_i$ and replacing $(\phi^{(i)},\alpha,\beta)$ with $(\widetilde{\phi}^{(i)},\tilde\alpha,\tilde\beta)$, i.e. system ${\cal{R}}(\widetilde{\phi}^{(i)};\ell_1,k_1,\tilde\alpha;\ell_2,k_2,\tilde\beta)$. Then,
\begin{enumerate}
\item
Solutions of systems ${\cal{R}}(\phi^{(i)};k_1,\ell_1,\alpha;k_2,\ell_2,\beta)$ and ${\cal{R}}(\widetilde{\phi}^{(i)};\ell_1,k_1,\tilde\alpha;\ell_2,k_2,\tilde\beta)$ are related by the point transformation
$$
{\cal{I}}\,:\, \left\{ \phi^{(i)}_{m,n} \widetilde{\phi}^{(i)}_{m,n}\,=\,1\,,\quad
         \alpha\,\tilde\alpha\,=\,1,\quad \beta\,\tilde\beta\,=\,1\right\}.
$$
\item
The Lax pairs (\ref{eq:LP-ir-g-rat}) for systems ${\cal{R}}(\phi^{(i)};k_1,\ell_1,\alpha;k_2,\ell_2,\beta)$ and  ${\cal{R}}(\widetilde{\phi}^{(i)};\ell_1,k_1,\tilde\alpha;\ell_2,k_2,\tilde\beta)$ are related by the gauge transformation $\widetilde\Psi_{m,n} = \alpha^{-m} \beta^{-n} \lambda^{-m-n} {\boldsymbol{\phi}}_{m,n}^{-1} \Psi_{m,n}$, along with the above point transformation and the inversion $\lambda \mapsto \lambda^{-1}$.
\end{enumerate}
\end{Pro}

Thus, two Lax pairs of the form (\ref{eq:LP-ir-g-rat}) with level structures $(k_1,\ell_1;k_2,\ell_2)$ and $(\ell_1,k_1;\ell_2,k_2)$ can be considered equivalent as they are related by a point transformation. Taking into consideration this observation, we present the inequivalent systems in two and three dimensions. In particular, in the two-dimensional case there exist only two classes, and we find only four inequivalent systems when $N=3$.

\subsubsection{Integrable Systems in Two Dimensions}

We start with the two-dimensional case where there exist only two inequivalent classes of Lax pairs (\ref{eq:LP-ir-g-rat}) and corresponding equations.\\

\noindent {\emph{Equivalence class $\left[(0,1;0,1)\right]$}}
\begin{equation} \label{eq:2D-0101}
\alpha\,\left(\phi_{m,n}\phi_{m,n+1}-\phi_{m+1,n}\phi_{m+1,n+1}\right)\,-\, \beta\,\left(\phi_{m,n}\phi_{m+1,n}-\phi_{m,n+1}\phi_{m+1,n+1}\right)\,=\,0,
\end{equation}
where $\phi_{m,n} = \phi^{(0)}_{m,n} = 1/\phi^{(1)}_{m,n}$.  This equation is known as the discrete modified KdV equation, or H3 with $\delta=0$ (see \cite{ABS,NC}).\\

\noindent {\emph{Equivalence class $\left[(0,1;1,0)\right]$}}
\begin{equation} \label{eq:2D-0110}
\alpha\,\left(\phi_{m,n}\phi_{m+1,n+1}-\phi_{m+1,n}\phi_{m,n+1}\right)\,-\, \beta\,\left(\phi_{m,n}\phi_{m+1,n} \phi_{m,n+1}\phi_{m+1,n+1}-1\right)\,=\,0,
\end{equation}
where $\phi_{m,n} = \phi^{(0)}_{m,n} = 1/\phi^{(1)}_{m,n}$. This is Hirota's discrete sine-Gordon equation, see \cite{DJM} and references therein. \\

\subsubsection{Integrable Systems in Three Dimensions}

\begin{Rem} \label{rem:rat-pot-1}
The systems in the following list derive from the corresponding systems (\ref{eq:dLP-gen-sys-1}), with the following substitution
\begin{equation} \label{eq:rat-pot-transf}
\left( \phi^{(0)}_{m,n}\,,\, \phi^{(1)}_{m,n}\,,\, \phi^{(2)}_{m,n} \right) \,\mapsto\,
     \left( \frac{1}{\phi^{(0)}_{m,n}}\,,\, \phi^{(1)}_{m,n}\,,\, \frac{\phi^{(0)}_{m,n}}{\phi^{(1)}_{m,n}} \right),
\end{equation}
which is a convenient choice (and incorporates the constraint $\phi^{(0)}_{m,n} \phi^{(1)}_{m,n} \phi^{(2)}_{m,n}= 1$).
\end{Rem}

\noindent {\emph{Equivalence class $\left[(0,1;0,1)\right]$}}
\begin{subequations} \label{eq:3D-0101}
\begin{eqnarray}
\phi^{(0)}_{m+1,n+1} &=& \left(
\frac{\alpha\,\phi_{m,n+1}^{(0)} - \beta \,\phi^{(0)}_{m+1,n}}{\alpha\,\phi_{m+1,n}^{(1)} - \beta \,\phi^{(1)}_{m,n+1}}
    \right)\,\phi^{(1)}_{m,n}\,,\\[3mm]
\phi^{(1)}_{m+1,n+1} &=& \left(
\frac{\alpha\,\phi^{(0)}_{m+1,n}\phi_{m,n+1}^{(1)} - \beta \,\phi^{(0)}_{m,n+1}\phi^{(1)}_{m+1,n}}
             {\alpha\,\phi_{m+1,n}^{(1)} - \beta \,\phi^{(1)}_{m,n+1}}\right)
             \,\frac{\phi^{(1)}_{m,n}}{\phi^{(0)}_{m,n}}\,.
\end{eqnarray}
\end{subequations}
This is the three dimensional analog of equation (\ref{eq:2D-0101}) and it is a well known example of a two-component discrete integrable system.  It can be decoupled for either of the functions involved in it to a nine-point scalar equation known as the modified Boussinesq equation \cite{NPCQ}. The hierarchies of its symmetries and conservation laws were studied in \cite{XN}. \\

\noindent {\emph{Equivalence class $\left[(0,1;1,2)\right]$}}
\begin{subequations} \label{eq:3D-0112}
\begin{eqnarray}
\phi^{(0)}_{m+1,n+1} &=& \left(
\frac{\alpha\,\phi_{m,n+1}^{(0)}\phi^{(1)}_{m+1,n}-\beta}{\alpha\,\phi_{m+1,n}^{(0)}-\beta \,\phi^{(1)}_{m+1,n} \phi^{(1)}_{m,n+1}}\right) \,\frac{\phi^{(0)}_{m,n}}{\phi^{(1)}_{m,n}}\,,\\
\phi^{(1)}_{m+1,n+1} &=& \left(
\frac{\alpha\,\phi^{(1)}_{m,n+1}-\beta \,\phi_{m+1,n}^{(0)}\phi^{(0)}_{m,n+1}}{\alpha\,\phi_{m+1,n}^{(0)}-\beta \,\phi^{(1)}_{m+1,n} \phi^{(1)}_{m,n+1}}\right) \,\phi^{(0)}_{m,n}\,.
\end{eqnarray}
\end{subequations}
This is a new integrable system (but see also \cite{MX1}) which cannot be decoupled to a (local) scalar equation on a bigger stencil for any of the variables. \\

\noindent {\emph{Equivalence class $\left[(0,1;2,0)\right]$}}
\begin{subequations} \label{eq:3D-0120}
\begin{eqnarray}
\phi^{(0)}_{m+1,n+1} &=& \left(
\frac{\alpha\,\phi_{m+1,n}^{(0)}\phi^{(0)}_{m,n+1} - \beta \phi^{(1)}_{m+1,n}}{\alpha- \beta \,\phi_{m+1,n}^{(0)}\phi^{(1)}_{m,n+1}} \right) \,\frac{1}{\phi^{(0)}_{m,n}}\,,\\
\phi^{(1)}_{m+1,n+1} &=& \left(
\frac{\alpha\,\phi_{m+1,n}^{(1)}\phi^{(1)}_{m,n+1} - \beta \phi^{(0)}_{m,n+1}}{\alpha- \beta \,\phi_{m+1,n}^{(0)}\phi^{(1)}_{m,n+1}} \right) \,\frac{1}{\phi^{(1)}_{m,n}}\,.
\end{eqnarray}
\end{subequations}
This is another new integrable system (but see also \cite{MX1}) which occurs only in three dimensions. Again, this system cannot be decoupled to a scalar equation on a bigger stencil for either $\phi^{(0)}$ or $\phi^{(1)}$.\\

\noindent {\emph{Equivalence class $\left[(1,2;1,2)\right]$}}
\begin{subequations} \label{eq:3D-1212}
\begin{eqnarray}
\phi^{(0)}_{m+1,n+1} &=& \left(
\frac{\alpha\,\phi_{m+1,n}^{(1)}-\beta \phi^{(1)}_{m,n+1}}{\alpha \,\phi_{m+1,n}^{(0)}\phi^{(1)}_{m,n+1}-\beta \,\phi_{m,n+1}^{(0)}\phi^{(1)}_{m+1,n}} \right) \,\frac{1}{\phi^{(0)}_{m,n}}\,,\\
\phi^{(1)}_{m+1,n+1} &=&  \left(
\frac{\alpha\,\phi_{m,n+1}^{(0)}-\beta \phi^{(0)}_{m+1,n}}{\alpha \,\phi_{m+1,n}^{(0)}\phi^{(1)}_{m,n+1}-\beta \,\phi_{m,n+1}^{(0)}\phi^{(1)}_{m+1,n}} \right)\,\frac{1}{\phi^{(1)}_{m,n}}\,.
\end{eqnarray}
\end{subequations}
This is a new integrable system which can be decoupled to a nine-point scalar equation for either of the functions involved in it. This scalar equation for $\phi^{(0)}$ can be written as
\begin{subequations} \label{eq:L3-U-mBSQ}
\begin{equation}
\left(\frac{\alpha \phi^{(0)}_{m+1,n+1} - \beta \phi^{(0)}_{m+2,n}}{\alpha \phi^{(0)}_{m,n+2} - \beta \phi^{(0)}_{m+1,n+1}}\right) \, \left(\frac{\alpha^2 \phi^{(0)}_{m+1,n+1}-\beta^2 \phi^{(0)}_{m,n+2}}{\alpha^2 \phi^{(0)}_{m+2,n}-\beta^2 \phi^{(0)}_{m+1,n+1}}\right)\,=\, \frac{{\cal{M}}_{m,n} {\cal{M}}_{m,n+1} {\cal{M}}_{m+1,n+1}}{{\cal{N}}_{m,n} {\cal{N}}_{m+1,n} {\cal{N}}_{m+1,n+1}}\,,
\end{equation}
where
\begin{equation}
{\cal{M}}_{m,n} := \alpha \phi^{(0)}_{m,n} \phi^{(0)}_{m+1,n} \phi^{(0)}_{m+1,n+1} +\beta,\quad {\cal{N}}_{m,n} := \beta \phi^{(0)}_{m,n}
\phi^{(0)}_{m,n+1} \phi^{(0)}_{m+1,n+1} + \alpha\,,
\end{equation}
\end{subequations}
while the corresponding equation for $\phi^{(1)}$ follows from (\ref{eq:L3-U-mBSQ}) by replacing $\phi^{(0)}_{m,n}$ with $1/\phi^{(1)}_{m,n}$.

There is an interesting reduction of system (\ref{eq:3D-1212}) to the integrable equation \cite{MX}
\begin{equation} \label{eq:MX2}
u_{m,n} u_{m+1,n+1} \left( u_{m+1,n} + u_{m,n+1} \right) + 1 =0.
\end{equation}
Specifically, system (\ref{eq:3D-1212}) reduces to equation (\ref{eq:MX2}) by setting
\begin{equation} \label{eq:red-3D1212-MX2}
\phi^{(0)}_{m,n} \,= \, \phi^{(1)}_{m,n} \, = \, \frac{-1}{2^{1/3} u_{m,n}}\,,\quad \beta \,=\,- \alpha.
\end{equation}
In particular, in view of the above reduction and setting for convenience $\alpha = 2^{1/3}$, the Lax pair for system (\ref{eq:3D-1212}) becomes
\begin{eqnarray}
\Psi_{m+1,n} &=&  \left(\begin{array}{ccc}
                   0 & \frac{1}{u_{m+1,n}} & \lambda \\
                   \lambda & 0 & \frac{1}{u_{m,n}} \\
                   -2 u_{m,n} u_{m+1,n} & \lambda & 0
                   \end{array} \right)\,\Psi_{m,n}, \nonumber \\[3mm]
\Psi_{m,n+1} &=& \left(\begin{array}{ccc}
                       0 & \frac{-1}{u_{m,n+1}} & \lambda \\
                       \lambda & 0 & \frac{-1}{u_{m,n}} \\
                       2 u_{m,n} u_{m,n+1} & \lambda & 0
                       \end{array} \right)\,\Psi_{m,n},   \nonumber
\end{eqnarray}
providing us with another Lax pair for equation (\ref{eq:MX2}), compared with the one in \cite{MX}.

\subsubsection{Integrable Systems in $N$ Dimensions}

\noindent {\emph{Equivalence class $\left[(0,1;0,1)\right]$}}\\

In this case, (\ref{eq:dLP-gen-sys-1}) can be written as
\begin{subequations} \label{eq:h-0101-r}
\begin{eqnarray}
\phi_{m+1,n+1}^{(i)} &=& \frac{\phi_{m,n+1}^{(i)} \phi_{m+1,n}^{(i)}}{\phi_{m,n}^{(i+1)}}\,
\left(\frac{\alpha \phi_{m,n+1}^{(i+1)} - \beta \phi_{m+1,n}^{(i+1)}}{\alpha \phi_{m,n+1}^{(i)} - \beta \phi_{m+1,n}^{(i)}}\right)\,,
\quad i =0, \dots,N-3,\\
&& \nonumber \\
\phi_{m+1,n+1}^{(N-2)} &=& \left(
\frac{\phi_{m,n+1}^{(N-2)} \phi_{m+1,n}^{(N-2)}}{\alpha \phi_{m,n+1}^{(N-2)} - \beta \phi_{m+1,n}^{(N-2)}}\right)
\, H_{m,n}\, \left( \frac{\alpha}{H_{m,n+1}}\,-\, \frac{\beta}{H_{m+1,n}}\right)\,,
\end{eqnarray}
\end{subequations}
where $H_{m,n}:= \prod_{j=0}^{N-2}\phi_{m,n}^{(j)}$, by using constraint (\ref{eq:LP-ir-g-rat-2}) to replace $\phi^{(N-1)}_{m,n}$ in terms of the remaining potentials.

This is equivalent to the $N-$component nonlinear superposition formula for the two dimensional Toda lattice, given in \cite{FG3} (discussed in Section \ref{2dtoda} below) (see also \cite{ALN}).

\subsection{Additive Potentials}\label{sect:coprime-additive}

Equations (\ref{eq:dLP-ex-cc-2}) hold identically if we set
\begin{equation}\label{eq:dLP-gen-ch-2}
u^{(i)}_{m,n} = \chi_{m+1,n}^{(i)}-\chi_{m,n}^{(i+\ell_1)},\quad
v^{(i)}_{m,n} = \chi_{m,n+1}^{(i)}-\chi_{m,n}^{(i+\ell_2)}, \quad i \in {\mathbb{Z}}_N.
\end{equation}
Equations (\ref{eq:dLP-ex-cc-1}) then take the form
\begin{equation}\label{eq:dLP-gen-sys-2}
\frac{\left( \chi^{(i)}_{m+1,n+1} - \chi^{(i+\ell_1)}_{m,n+1}\right)}{\left( \chi^{(i+k_2)}_{m+1,n} - \chi^{(i+k_2+ \ell_1)}_{m,n}\right)}=
    \frac{\left( \chi^{(i)}_{m+1,n+1} - \chi^{(i+\ell_2)}_{m+1,n}\right)}{\left( \chi^{(i+k_1)}_{m,n+1} - \chi^{(i+k_1+ \ell_2)}_{m,n}\right)} ,
\end{equation}
for $i \in {\mathbb{Z}}_N$, and their solved form (\ref{eq:dLP-ex-cc-s}) is written as
\begin{equation}\label{eq:dLP-gen-sys-2-a}
\chi^{(i)}_{m+1,n+1}\,=\,\frac{\chi_{m,n+1}^{(i+k_1)} \chi_{m,n+1}^{(i+\ell_1)} - \chi_{m+1,n}^{(i+k_2)} \chi_{m+1,n}^{(i+\ell_2)} - \chi_{m,n}^{(i+k_1+\ell_2)} (\chi_{m,n+1}^{(i+\ell_1)} - \chi_{m+1,n}^{(i+\ell_2)}) }{\chi_{m,n+1}^{(i+k_1)} - \chi_{m+1,n}^{(i+k_2)}}\,.
\end{equation}
In this potential form, the Lax pair (\ref{eq:dLP-gen}) can be written
\begin{equation} \label{eq:LP-ir-g-add}
\begin{array}{l}
\Psi_{m+1,n} \,=\,
\Big(\left({\boldsymbol{\chi}}_{m+1,n}-\Omega^{\ell_1}{\boldsymbol{\chi}}_{m,n} \Omega^{-\ell_1}\right)  \Omega^{k_1}
            + \lambda \Omega^{\ell_1}\Big) \Psi_{m,n},\\[3mm]
\Psi_{m,n+1} \,=\,
\Big(\left({\boldsymbol{\chi}}_{m,n+1}-\Omega^{\ell_2}{\boldsymbol{\chi}}_{m,n} \Omega^{-\ell_2}\right)  \Omega^{k_2}
               + \lambda \Omega^{\ell_2}\Big) \Psi_{m,n},
\end{array}
\end{equation}
where
$$
{\boldsymbol{\chi}}_{m,n} \,:=\, {\rm{diag}}\left(\chi^{(0)}_{m,n},\cdots,\chi^{(N-1)}_{m,n}\right).
$$
We can then show that the Lax pair (\ref{eq:LP-ir-g-add}) is compatible if and only if the system (\ref{eq:dLP-gen-sys-2}) holds.

In this case, conditions (\ref{eq:dLP-coprime-case}) become the first integrals
\begin{equation}\label{eq:dLP-gen-ch-2-cd}
\prod_{i=0}^{N-1} \left(\chi^{(i)}_{m+1,n} - \chi^{(i+\ell_1)}_{m,n} \right)\,=\,\alpha^N,\quad
\prod_{i=0}^{N-1} \left(\chi^{(i)}_{m,n+1} - \chi^{(i+\ell_2)}_{m,n} \right)\,=\,\beta^N,
\end{equation}
where we have set $a=\alpha^N,\, b=\beta^N$.
Hence it is not always possible to reduce the number of potentials $\chi$ (in {\em local} terms) by employing these.

 \br
 There do exist cases where we are able to reduce the number of potentials in system (\ref{eq:dLP-gen-sys-2}) but not in Lax pair (\ref{eq:LP-ir-g-add}). In those cases, one may derive another local Lax pair for the reduced system which will not belong to the class we consider in this paper.
 \er

\subsubsection{Integrable Systems in Two Dimensions}

We present the inequivalent integrable systems which can be derived in two dimensions. \\

\noindent {\emph{Equivalence class $\left[(0,1;0,1)\right]$}}\\

Using constraints (\ref{eq:dLP-gen-ch-2-cd}), we can replace either $\chi^{(0)}$ or $\chi^{(1)}$, to obtain
\begin{equation} \label{eq:2D-0101-a}
\left(\chi_{m+1,n+1}-\chi_{m,n}\right) \left(\chi_{m+1,n}-\chi_{m,n+1}\right)\,=\,\alpha^2- \beta^2,
\end{equation}
which is the discrete potential KdV or H1 equation \cite{ABS,NC}.\\

\noindent {\emph{Equivalence class $\left[(0,1;1,0)\right]$}}\\

In this case, the corresponding system cannot be decoupled. It is omitted here as it follows from (\ref{eq:dLP-gen-sys-2}) by setting $k_1 = \ell_2 = 0$ and $\ell_1=k_2=1$. In the same way, its first integrals
$$
\left(\chi_{m+1,n}^{(0)}-\chi_{m,n}^{(1)}\right) \left(\chi_{m+1,n}^{(1)}-\chi_{m,n}^{(0)}\right) = \alpha^2,\quad \left(\chi_{m,n+1}^{(0)}-\chi_{m,n}^{(0)}\right) \left(\chi_{m,n+1}^{(1)}-\chi_{m,n}^{(1)}\right) = \beta^2,
$$
follow from (\ref{eq:dLP-gen-ch-2-cd}).\\

\noindent {\emph{Equivalence class $\left[(1,0;1,0)\right]$}}\\
This is another case where the first integrals can be used to eliminate one of the two variables. This leads to the Schwarzian KdV equation \cite{NC} (ie Q1 of \cite{ABS}, with $\delta=0$):
\begin{equation} \label{eq:2D-1010-a}
\alpha^2 \left(\chi_{m,n}- \chi_{m,n+1}\right) \left(\chi_{m+1,n}- \chi_{m+1,n+1}\right) - \beta^2 \left(\chi_{m,n}- \chi_{m+1,n}\right) \left(\chi_{m,n+1}- \chi_{m+1,n+1}\right) = 0.
\end{equation}

\subsubsection{Integrable Systems in Three Dimensions}

In the three-dimensional case $N=3$ there exist six inequivalent classes which can be divided into two categories.
\begin{enumerate}
\item
The first category contains two equivalence classes, namely $[(0,1;0,1)]$ and $[(1,0;1,0)]$,  for which we can reduce the number of potentials by employing the corresponding first integrals. More precisely, using the first integrals (\ref{eq:dLP-gen-ch-2-cd}), one can choose to eliminate potential $\chi^{(0)}_{m,n}$, and derive a two-component system for the remaining two potentials, denoted here by $P(\chi^{(1)}_{m,n},\chi^{(2)}_{m,n})$. In fact, working in the same way, the elimination of $\chi^{(1)}_{m,n}$ yields system $P(\chi^{(2)}_{m,n},\chi^{(0)}_{m,n})$, whereas the elimination of $\chi^{(2)}_{m,n}$ results to system $P(\chi^{(0)}_{m,n},\chi^{(1)}_{m,n})$. In this sense, the first integrals may be regarded as a periodic map of the corresponding two-component systems $P(\chi^{(j)}_{m,n},\chi^{(j+1)}_{m,n})$, $j \in {\mathbb{Z}}_3$. Moreover, for both classes, the corresponding two-component systems can be decoupled further to nine-point scalar equations, and, we may interpret the two-component systems as maps of the corresponding scalar equations.
\item
The second category contains classes $[(0,1;1,2)]$, $[(0,1;2,0)]$, $[(1,2;1,2)]$ and $[(1,2;2,0)]$. The common characteristic of these classes is that the corresponding systems involve three potentials, the number of which cannot be reduced using the first integrals (\ref{eq:dLP-gen-ch-2-cd}). To the best of our knowledge, these systems are new. Also, the equivalence classes $[(0,1;2,0)]$ and $[(1,2;2,0)]$, as well as the corresponding discrete systems, exist only in three dimensions, as a consequence of (\ref{eq:dLP-nec-rel}).
\end{enumerate}

\noindent {\emph{Equivalence class $\left[(0,1;0,1)\right]$}}\\

Consider the discrete system (\ref{eq:dLP-gen-sys-2}) and first integrals (\ref{eq:dLP-gen-ch-2-cd}) with $\left(k_1, \ell_1;k_2, \ell_2 \right) = (0,1;0,1)$. We can employ the first integrals in order to eliminate one of the potentials from the discrete system and derive a two-component system for the remaining potentials.  Following the above discussion, we eliminate $\chi^{(2)}$ to obtain a system for $(\chi^{(0)}_{m,n},\chi_{m,n}^{(1)})$:
\bea
 \chi_{m+1,n+1}^{(0)} &=& \frac{(\chi^{(0)}_{m+1,n}-\chi^{(1)}_{m,n}) \chi^{(1)}_{m+1,n} - (\chi^{(0)}_{m,n+1}-\chi^{(1)}_{m,n}) \chi^{(1)}_{m,n+1}}{\chi^{(0)}_{m+1,n} - \chi^{(0)}_{m,n+1}},  \nn\\[-2mm]
 \label{eq:3D-0101-a} \\[-2mm]
 \chi^{(1)}_{m+1,n+1} &=& \chi^{(0)}_{m,n}\,+\,\frac{1}{\chi_{m+1,n}^{(1)} - \chi_{m,n+1}^{(1)}}\,\left(\frac{\alpha^3}{\chi_{m+1,n}^{(0)}-\chi^{(1)}_{m,n} }\,-\,\frac{\beta^3}{\chi_{m,n+1}^{(0)}-\chi^{(1)}_{m,n}}\right),  \nn
\eea
which is a new integrable system.

On the other hand, using the first integrals we are not able to eliminate the potential $\chi^{(2)}_{m,n}$ from Lax pair (\ref{eq:LP-ir-g-add}). However, employing the three-dimensional consistency of system (\ref{eq:3D-0101-a}), we are able to construct the following Lax pair for system (\ref{eq:3D-0101-a}), which does not belong  in the class of Lax pairs considered in this paper:
\begin{eqnarray}
\Psi_{m+1,n} &=&   \left(\begin{array}{ccc}
                             \chi_{m+1,n}^{(1)}-\chi_{m,n}^{(0)} & 0 & -1 \\
                              -1 & \chi_{m+1,n}^{(0)}-\chi_{m,n}^{(1)} & 0 \\
                              F_{m,n} & \lambda &  \chi_{m+1,n}^{(0)}-\chi_{m,n}^{(0)}
                              \end{array} \right) \Psi_{m,n}\,,  \nonumber  \\
\Psi_{m,n+1} &=& \left(\begin{array}{ccc}
                   \chi_{m,n+1}^{(1)}-\chi_{m,n}^{(0)} & 0 & -1 \\
                   -1 & \chi_{m,n+1}^{(0)}-\chi_{m,n}^{(1)} & 0 \\
                   G_{m,n} & \lambda &  \chi_{m,n+1}^{(0)}-\chi_{m,n}^{(0)}
                   \end{array} \right) \Psi_{m,n},  \nonumber
\end{eqnarray}
where
\begin{eqnarray*}
&& F_{m,n} := \frac{\alpha^3}{\chi^{(0)}_{m+1,n}-\chi_{m,n}^{(1)}}- (\chi_{m+1,n}^{(0)}-\chi_{m,n}^{(0)}) (\chi_{m+1,n}^{(1)}-\chi_{m,n}^{(0)}),\\
&& G_{m,n} :=  \frac{\beta^3}{\chi^{(0)}_{m,n+1}-\chi_{m,n}^{(1)}}- (\chi_{m,n+1}^{(0)}-\chi_{m,n}^{(0)}) (\chi_{m,n+1}^{(1)}-\chi_{m,n}^{(0)}).
\end{eqnarray*}

\br[Reduction to the discrete Boussinesq equation]
This system can be decoupled for either of the variables to the nine point scalar equation known as discrete Boussinesq equation \cite{NPCQ}.
\er

\noindent {\emph{Equivalence class $\left[(1,0;1,0)\right]$}}\\

Working as with the previous equivalence class, we can derive a two-component system for any pair of potentials corresponding to the discrete system (\ref{eq:dLP-gen-sys-2}) and its first integrals (\ref{eq:dLP-gen-ch-2-cd}) with $N=3$ and $\left(k_1, \ell_1;k_2, \ell_2 \right) = (1,0;1,0)$. The system can be written in the following form.
\begin{eqnarray}
&& \chi_{m+1,n+1}^{(0)} = \frac{\chi_{m+1,n}^{(0)} \Delta_m(\chi^{(1)}_{m,n})  -
   \chi_{m,n+1}^{(0)} \Delta_n(\chi^{(1)}_{m,n})}{\chi_{m+1,n}^{(1)}-\chi_{m,n+1}^{(1)}}\,, \nonumber \\[-2mm]
&& \label{eq:3D-1010-a} \\[-2mm]
&& \chi_{m+1,n+1}^{(1)} = \frac{\alpha^3 \chi^{(1)}_{m+1,n}\, \Delta_n(\chi^{(0)}_{m,n})\, \Delta_n(\chi^{(1)}_{m,n}) - \beta^3 \chi^{(1)}_{m,n+1}\, \Delta_m(\chi^{(0)}_{m,n})\, \Delta_m(\chi^{(1)}_{m,n})}{\alpha^3\, \Delta_n(\chi^{(0)}_{m,n})\, \Delta_n(\chi^{(1)}_{m,n})
                       - \beta^3\, \Delta_m(\chi^{(0)}_{m,n})\, \Delta_m(\chi^{(1)}_{m,n})}\, . \nn
\end{eqnarray}
A Lax pair for system (\ref{eq:3D-1010-a}) is of the form (\ref{eq:LP-ir-g-add}), with $\chi^{(2)}$ replaced as discussed above.  The hierarchies of its symmetries and conservation laws were studied in \cite{XN1}.

\subsubsection{Integrable Systems in $N$ Dimensions}

The two equivalence classes we discussed above can be defined in any dimension $N$ and we can always eliminate one of the potentials from the discrete systems using the first integrals.  \\

\noindent {\emph{Equivalence class $\left[(0,1;0,1)\right]$}}\\

The following  $(N-1)$-component system is related to this equivalence class.
\begin{eqnarray}
 \chi_{m+1,n+1}^{(i)} &=& \frac{(\chi^{(i)}_{m,n+1}-\chi^{(i+1)}_{m,n}) \chi^{(i+1)}_{m,n+1} - (\chi^{(i)}_{m+1,n}-\chi^{(i+1)}_{m,n}) \chi^{(i+1)}_{m+1,n}}{\chi^{(i)}_{m,n+1} - \chi^{(i)}_{m+1,n}}\,,\quad i = 0,\dots,N-3,\nonumber  \\
&& \label{eq:h-0101-a} \\
 \chi^{(N-2)}_{m+1,n+1} &=& \chi^{(0)}_{m,n}\,+\,\frac{1}{\chi_{m+1,n}^{(N-2)} - \chi_{m,n+1}^{(N-2)}}\,\left(\frac{\alpha^N}{X}\,
                        -\,\frac{\beta^N}{Y} \right).\nonumber
\end{eqnarray}
where $X=\prod_{j=0}^{N-3}(\chi_{m+1,n}^{(j)}-\chi^{(j+1)}_{m,n})$ and $Y=\prod_{j=0}^{N-3}(\chi_{m,n+1}^{(j)}-\chi^{(j+1)}_{m,n})$.\\

It can be derived from the compatibility condition of the Lax pair
$$
\begin{array}{l}
\Psi_{m+1,n} \,=\, \left({\boldsymbol{\alpha}}_{m,n} - \Omega^{N-1} + J_{m,n} \right)\Psi_{m,n}\,,\\ \\ \Psi_{m,n+1} \,=\, \left({\boldsymbol{\beta}}_{m,n} - \Omega^{N-1} + K_{m,n} \right)\Psi_{m,n}\,,
\end{array}
$$
where ${\boldsymbol{\alpha}}_{m,n}$ and ${\boldsymbol{\beta}}_{m,n}$ are $N \times N$ diagonal matrices with entries
$$
\begin{array}{l}
\left({\boldsymbol{\alpha}}_{m,n}\right)_{i,i} \,=\, \left(1-\delta_{i,N}\right) \left(\chi_{m+1,n}^{(N-i-1)} - \chi_{m,n}^{(N-i)}\right) + \delta_{i,N}  \left(\chi^{(0)}_{m+1,n}-\chi^{(0)}_{m,n}\right), \\
\\
\left({\boldsymbol{\beta}}_{m,n}\right)_{i,i} \,=\, \left(1-\delta_{i,N}\right) \left(\chi_{m,n+1}^{(N-i-1)} - \chi_{m,n}^{(N-i)}\right) + \delta_{i,N} \left(\chi^{(0)}_{m,n+1}-\chi^{(0)}_{m,n}\right),
\end{array}
$$
with all upper indices being considered $\bmod{(N-1)}$, and
$$
\begin{array}{l}
(J_{m,n})_{i,j} = \delta_{i,N} \delta_{j,1} X_{m,n} + \delta_{i,N} \delta_{j,N-1} (\lambda+1),\\
\\
(K_{m,n})_{i,j} = \delta_{i,N} \delta_{j,1} Y_{m,n} + \delta_{i,N} \delta_{j,N-1} (\lambda+1),
\end{array}
$$
where $X_{m,n}$ and $Y_{m,n}$ are determined by the requirement
$$
\det\left( {\boldsymbol{\alpha}}_{m,n} - \Omega^{N-1} + J_{m,n}  \right) = \lambda + \alpha^N,\qquad
\det\left({\boldsymbol{\beta}}_{m,n} - \Omega^{N-1} + K_{m,n} \right) = \lambda + \beta^N.
$$

\noindent {\emph{Equivalence class $\left[(1,0;1,0)\right]$}}\\

With $(k_1,\ell_1;k_2,\ell_2) = (1,0;1,0)$, one may eliminate potential $\chi^{(N-1)}_{m,n}$ from system (\ref{eq:dLP-gen-sys-2}), using first integrals (\ref{eq:dLP-gen-ch-2-cd}), to derive
\begin{eqnarray}
\chi_{m+1,n+1}^{(i)} &=&
\frac{\chi_{m+1,n}^{(i)} (\chi_{m+1,n}^{(i+1)}-\chi_{m,n}^{(i+1)})  - \chi_{m,n+1}^{(i)} (\chi_{m,n+1}^{(i+1)}-\chi_{m,n}^{(i+1)})}{\chi_{m+1,n}^{(i+1)}-\chi_{m,n+1}^{(i+1)}}\,,\quad i =0, \dots,N-3,\nonumber \\[3mm]
\chi_{m+1,n+1}^{(N-2)} &=&
\frac{\alpha^N \chi_{m+1,n}^{(N-2)} A_{m,n} - \beta^N \chi_{m,n+1}^{(N-2)} B_{m,n}}{\alpha^N A_{m,n} - \beta^N B_{m,n}}\,,\nonumber
\end{eqnarray}
where
$$
A_{m,n} := \prod_{j=0}^{N-2}(\chi_{m,n+1}^{(j)}-\chi_{m,n}^{(j)}), \quad
     B_{m,n} := \prod_{j=0}^{N-2}(\chi_{m+1,n}^{(j)}-\chi_{m,n}^{(j)}).
$$
A Lax pair follows from (\ref{eq:LP-ir-g-add}) using (\ref{eq:dLP-gen-ch-2-cd}).

\subsection{B{\"a}cklund Transformations between Potential Forms}\label{BTpotentials}

The introduction of two different sets of potentials allows us to derive a B{\"a}cklund transformation between systems (\ref{eq:dLP-gen-sys-1}) and (\ref{eq:dLP-gen-sys-2}), which follows from the combination of relations (\ref{eq:dLP-gen-ch-1}) and (\ref{eq:dLP-gen-ch-2}). If we denote systems (\ref{eq:dLP-gen-sys-1}) and (\ref{eq:dLP-gen-sys-2}) respectively by ${\cal{R}}(\phi^{(i)};k_1,\ell_1,\alpha;k_2,\ell_2,\beta)$ and ${\cal{A}}(\chi^{(i)};k_1,\ell_1;k_2,\ell_2)$, then we have:
\begin{Pro} \label{prop:bt-rat-add}
The system of equations
$$
\alpha\,\frac{\phi^{(i)}_{m+1,n}}{\phi^{(i+k_1)}_{m,n}}\,=\,\chi^{(i)}_{m+1,n} - \chi^{(i+\ell_1)}_{m,n} \,,\qquad  \beta\,\frac{\phi^{(i)}_{m,n+1}}{\phi^{(i+k_2)}_{m,n}}\,=\, \chi^{(i)}_{m,n+1} - \chi^{(i+\ell_2)}_{m,n},
$$
which we call ${\mathbb{B}}_{RA}(\phi,\chi;k_1,\ell_1,\alpha ; k_2,\ell_2,\beta)$,
defines a B{\"a}cklund transformation between systems ${\cal{R}}(\phi^{(i)};k_1,\ell_1,\alpha;k_2,\ell_2,\beta)$ and ${\cal{A}}(\chi^{(i)};k_1,\ell_1;k_2,\ell_2)$.
\end{Pro}

Finally, if we combine Propositions \ref{prop:rat-pot-trans} and \ref{prop:bt-rat-add}, we have
\begin{Pro} \label{prop:bt-add-add}
Consider systems  ${\cal{A}}(\chi^{(i)};k_1,\ell_1;k_2,\ell_2)$ and
${\cal{A}}(\widetilde{\chi}^{(i)};\ell_1,k_1;\ell_2,k_2)$ with $k_i+\ell_i \ne N$.
Solutions of one system are mapped to solutions of the other through the B{\"a}cklund transformation
$$
{\mathbb{B}}_{AA}(\chi,\widetilde{\chi}) = \left\{ \begin{array}{l} \left(\chi_{m+1,n}^{(i)}-\chi_{m,n}^{(i+\ell_1)} \right) \left(\widetilde{\chi}_{m+1,n}^{(i)}-\widetilde{\chi}_{m,n}^{(i+k_1)} \right)\,=\,\frac{\phi_{m,n}^{(i+\ell_1)}}{\phi_{m,n}^{(i+k_1)}}\,=\,\frac{\widetilde{\phi}_{m,n}^{(i+k_1)}}{\widetilde{\phi}_{m,n}^{(i+\ell_1)}}\,,\\ \\  \left(\chi_{m,n+1}^{(i)}-\chi_{m,n}^{(i+\ell_2)} \right) \left(\widetilde{\chi}_{m,n+1}^{(i)}-\widetilde{\chi}_{m,n}^{(i+k_2)} \right)\,=\,\frac{\phi_{m,n}^{(i+\ell_2)}}{\phi_{m,n}^{(i+k_2)}}\,=\,\frac{\widetilde{\phi}_{m,n}^{(i+k_2)}}{\widetilde{\phi}_{m,n}^{(i+\ell_2)}} \end{array}  \right\}\,,
$$
where the auxiliary functions $\phi$ and $\widetilde{\phi}$ are related to $\chi$ and $\widetilde{\chi}$ by the B{\"a}cklund transformations ${\mathbb{B}}_{RA}(\phi,\chi;k_1,\ell_1,\alpha; k_2,\ell_2,\beta)$ and ${\mathbb{B}}_{RA}\left(\widetilde{\phi},\widetilde{\chi};\ell_1,k_1,\alpha^{-1};\ell_2,k_2,\beta^{-1}\right)$, respectively.
\end{Pro}

\begin{Rem} \label{rem:diagram-A-R}
In section \ref{sect:class-prob} we proposed a classification scheme of Lax pairs and corresponding integrable difference equations, in terms of their level structures.  Propositions \ref{prop:rat-pot-trans}, \ref{prop:bt-rat-add} and \ref{prop:bt-add-add} place these systems within families related with each other by B{\"a}cklund transformations. Schematically, it is encoded in the following diagram:
$$
\begin{array}{ccc} {\cal{R}}(\phi^{(i)};k_1,\ell_1,\alpha;k_2,\ell_2,\beta) & \rightleftarrows \quad {\cal{I}}
                        \quad \rightleftarrows &
                        {\cal{R}}\left(\widetilde{\phi}^{(i)};\ell_1,k_1,\frac{1}{\alpha};\ell_2,k_2,\frac{1}{\beta}\right) \\
                        \uparrow \downarrow & & \uparrow\downarrow  \\
                        {\mathbb{B}}_{RA}  & & {\mathbb{B}}_{RA}\\
                        \uparrow \downarrow & & \uparrow\downarrow  \\
                        {\cal{A}}(\chi^{(i)};k_1,\ell_1;k_2,\ell_2) & \leftrightarrows \quad {\mathbb{B}}_{AA} \quad
                        \leftrightarrows& {\cal{A}}(\widetilde{\chi}^{(i)};\ell_1,k_1;\ell_2,k_2)
                        \end{array}
$$
\end{Rem}

\subsection{The Degenerate Case}\label{degenerate}

The final class of coprime system discussed in our classification scheme of Section \ref{sect:class-prob} is the degenerate case, for which $b = 0$. From conditions (\ref{eq:dLP-coprime-case}) we set
\begin{equation}
u^{(N-1)}_{m,n}\,=\, a\, \prod_{i=0}^{N-2} \frac{1}{u^{(i)}_{m,n}}\,,\quad v^{(N-1)}_{m,n} = 0\,.
\end{equation}
If $v^{(N-1)}_{m,n}$ is the only zero component, then equation (\ref{eq:dLP-ex-cc-1}) for $i = N-1$ implies that $k_1 = 0$. Hence $\ell_1\neq 0$ and $\ell_2 \equiv k_2+\ell_1 \; (\bmod{N})$, which follows from the consistency condition (\ref{eq:dLP-nec-rel}).

Using (\ref{eq:dLP-ex-cc-1}) to eliminate $v^{(i)}_{m+1,n}$, equation (\ref{eq:dLP-ex-cc-2}) implies
\be\label{vi+l1}
v^{(i+\ell_1)}_{m,n} \,=\, u^{(i+\ell_2)}_{m,n}-u^{(i)}_{m,n+1}+ \frac{u^{(i)}_{m,n+1}}{u^{(i+\ell_2-\ell_1)}_{m,n}}\, v^{(i)}_{m,n}
\ee
With $i = N-1$, we have
$$
v^{(\ell_1-1)}_{m,n} \,=\, u^{(\ell_2-1)}_{m,n}\,-\,u^{(N-1)}_{m,n+1}.
$$
Using (\ref{vi+l1}) for the inductive step, we can show that
\be\label{Pq}
v^{(q \ell_1-1)}_{m,n}\,=\, u^{(\ell_2 + (q-1) \ell_1 -1)}_{m,n}\,-\, u^{(N-1)}_{m,n+1}\, P_q,
\ee
where
$$
P_{q+1}= \frac{u^{(q \ell_1-1)}_{m,n+1}}{u^{(\ell_2+(q-1)\ell_1-1)}_{m,n}} \, P_q, \quad\mbox{with}\quad P_1=1.
$$
This leads to the general formula given in (\ref{vql-1}) below.

\begin{Pro}\label{prop:idc}
Let $\left(0, \ell_1;k_2, \ell_2 \right) \in {\cal{Q}}_N$ with $\big(N,\ell_1\big) =  \big(N,k_2-\ell_2\big) =1$.
Consider the system (\ref{eq:dLP-gen}), with $U_{m,n}={\boldsymbol{u}}_{m,n},\; V_{m,n}={\boldsymbol{v}}_{m,n} \Omega^{k_2}$
where
$$
{\boldsymbol{u}}_{m,n} \,:=\, {\rm{diag}}\left(u^{(0)}_{m,n}\,,\cdots,u^{(N-1)}_{m,n}\right) \quad
                  {\mbox{with}}\quad \det ({\boldsymbol{u}}_{m,n}) \,=\,\prod_{i=0}^{N-1} u^{(i)}_{m,n} \,=\,a,
$$
and ${\boldsymbol{v}}_{m,n} \,:=\, {\rm{diag}}\left(v^{(0)}_{m,n}\,,\cdots,v^{(N-2)}_{m,n}\,,0\right)$.
Then with
\bea
&& v^{(\ell_1-1)}_{m,n} \,=\, u^{(\ell_2-1)}_{m,n}\,-\,u^{(N-1)}_{m,n+1}, \nn \\[-1mm]
    \label{vql-1}  \\[-1mm]
&&  v^{(q \ell_1-1)}_{m,n}\,=\, u^{(\ell_2 + (q-1) \ell_1 -1)}_{m,n}\,-\, u^{(N-1)}_{m,n+1}\, \prod_{r=1}^{q-1} u^{(r \ell_1-1)}_{m,n+1} \,\prod_{s=0}^{q-2}\frac{1}{u^{(\ell_2+s \ell_1-1)}_{m,n}}\,,\quad q = 2,\dots, N-1,  \nn
\eea
the system (\ref{eq:dLP-ex-cc-1}) leads to a system of equations for the components $u^{(i)}$.
\end{Pro}

We now present inequivalent integrable systems for $N=2$ and $N=3$, and give the system which corresponds to the level structure $(0,1;0,1)$ for any dimension $N$.  Whilst our general discussion has concentrated on the case $v^{(N-1)}=0$, with other components nonzero, we also present some cases, for $N=3$, for which $v^{(1)}_{m,n}=v^{(2)}_{m,n}=0$.

\subsubsection{Integrable Systems in Two Dimensions}

In two dimensions, there exists only one nontrivial system: \\

\noindent {\emph{Level Structure $\left[(0,1;0,1)\right]$}}\\

We set
$$
u_{m,n}^{(0)}\,=\,u_{m,n},\quad u_{m,n}^{(1)}\,=\,\frac{a}{u_{m,n}}\,,\quad v^{(0)}_{m,n}\,=\,u_{m,n}\,-\,\frac{a}{u_{m,n+1}}\,,
$$
and the resulting equation is Hirota's KdV equation,
\begin{equation} \label{eq:2D-Hirota}
\frac{a}{u_{m+1,n+1}}\,+\,u_{m,n+1}\,=\,u_{m+1,n}\,+\,\frac{a}{u_{m,n}}.
\end{equation}

\br
The system which follows from the Lax pair with structure $(0,1;1,0)$ can be easily shown to be reducible to an ordinary difference equation.
\er

\subsubsection{Integrable Systems in Three Dimensions}

In three dimensions, only two systems arise.\\

\noindent {\emph{Equivalence class $[(0,1;0,1)]$}}\\

In this case the system may be considered as the two-component analogue of Hirota's KdV equation (\ref{eq:2D-Hirota}), for the components $u^{(0)}_{m,n},\, u^{(1)}_{m,n}$.  We have
$$
u^{(2)}_{m,n} = \frac{a}{u^{(0)}_{m,n} u^{(1)}_{m,n}},\quad v^{(0)}_{m,n}=u^{(0)}_{m,n}-\frac{a}{u^{(0)}_{m,n+1} u^{(1)}_{m,n+1}},\quad
     v^{(1)}_{m,n} = u^{(1)}_{m,n} -\frac{a}{u^{(0)}_{m,n} u^{(1)}_{m,n+1}}\, ,
$$
together with the system
\begin{subequations}\label{eq:3D-Hirota}
\begin{eqnarray}
&& \frac{a}{u^{(0)}_{m+1,n+1} u^{(1)}_{m+1,n+1}}\,+\,u^{(0)}_{m,n+1}  \,=\,
                    u^{(0)}_{m+1,n}\,+\,\frac{a}{u^{(0)}_{m,n} u^{(1)}_{m,n+1}}\,,\label{eq:3D-Hirota-a}\\
&& \frac{a}{u^{(0)}_{m+1,n} u^{(1)}_{m+1,n+1}}\,+\,u^{(1)}_{m,n+1}  \,=\,
                      u^{(1)}_{m+1,n}\,+\,\frac{a}{u^{(0)}_{m,n} u^{(1)}_{m,n}}\,.\label{eq:3D-Hirota-b}
\end{eqnarray}
\end{subequations}

This example admits a reduction, with $v^{(1)}_{m,n}=0$, corresponding to $u^{(0)}_{m,n} = \frac{a}{u^{(1)}_{m,n} u^{(1)}_{m,n+1}}$.  In this case, (\ref{eq:3D-Hirota-b}) holds identically, whilst (\ref{eq:3D-Hirota-a}) takes the form of a six point equation
\be\label{eq:BMXa}
u^{(1)}_{m,n} \,+\,\frac{a}{u^{(1)}_{m+1,n}u^{(1)}_{m+1,n+1}}\,=\, u^{(1)}_{m+1,n+2}\,+\,\frac{a}{u^{(1)}_{m,n+1} u^{(1)}_{m,n+2}}.
\ee

\br
Another reduction, with $v^{(0)}_{m,n}=0$, corresponds to $u^{(1)}_{m,n} = \frac{a}{u^{(0)}_{m,n} u^{(0)}_{m,n-1}}$, in which case system (\ref{eq:3D-Hirota}) reduces again to equation (\ref{eq:BMXa}), but for $u^{(0)}$.
System (\ref{eq:3D-Hirota}) and its reductions (\ref{eq:BMXa}) were derived first in \cite{BMX} in a different context.
\er

\noindent {\emph{Equivalence class $[(0,1;1,2)]$}}\\

In this case we introduce variables $u_{m,n}$ and $v_{m,n}$ by
\begin{eqnarray*}
&& u^{(0)}_{m,n} = u_{m,n} v_{m,n}, \quad u^{(1)}_{m,n}= \frac{1}{u_{m,n}},\quad u^{(2)}_{m,n} = \frac{a}{v_{m,n}}\,,\\
    &&   \label{degen0112-vars}  \\
&& v^{(0)}_{m,n}= \frac{1}{u_{m,n}}\,-\,\frac{a}{v_{m,n+1}}\,,\quad v^{(1)}_{m,n} = a\, \left(\frac{1}{v_{m,n}} \,-\,u_{m,n} u_{m,n+1}\right),
\end{eqnarray*}
to derive the system
\begin{subequations}\label{degen0112}
\begin{eqnarray}
u_{m,n} v_{m,n} \,+\,\frac{a}{v_{m+1,n}} &= & \frac{1}{u_{m,n+1}}\,+\,a u_{m+1,n}\, u_{m+1,n+1}\,, \label{degen0112-a}\\
u_{m,n+1} v_{m,n+1} \,+\,\frac{a}{v_{m+1,n+1}} &= & \frac{1}{u_{m+1,n}}\,+\, a u_{m,n}\, u_{m,n+1}\,. \label{degen0112-b}
\end{eqnarray}
\end{subequations}
Noting that the left hand sides of these equations are related by a shift in the $n$ direction, we can derive an equation for the single component $u$:
\begin{equation}\label{eq:BMX}
a u_{m+1,n+1} u_{m+1,n+2}\,+\,\frac{1}{u_{m,n+2}}\,=\,\frac{1}{u_{m+1,n}}\,+\,a u_{m,n} u_{m,n+1}\,.
\end{equation}
Equation (\ref{degen0112-a}) is then a first order, ``driven'' difference equation for $v_{m,n}$.

\br
The further reduction $v^{(1)}_{m,n}=0$ corresponds to $v_{m,n}=\frac{1}{u_{m,n}\, u_{m,n+1}}$, after which (\ref{degen0112-a}) is identically satisfied, whilst (\ref{degen0112-b}) takes the form of (\ref{eq:BMX}). Similarly, the choice $v^{(0)}_{m,n}=0$ corresponds to $v_{m,n} = a u_{m,n-1}$ and reduces system (\ref{degen0112}) to equation (\ref{eq:BMX}) again. Equation (\ref{eq:BMX}) is related to
$$
u_{m+1,n} u_{m,n+1} \left(u_{m,n}+u_{m+1,n+1} \right)\,+\,\frac{1}{a} \,=\,0,
$$
found in \cite{MX}, which, up to inversion of one the lattice directions, is equation (\ref{eq:MX2}). Finally, equation (\ref{eq:BMX}) is also related to equation (\ref{eq:BMXa}) by the point transformation $(u_{i,j},a) \rightarrow (1/u_{-i,j},1/a)$.\\
\er

\noindent {\emph{Equivalence class $[(0,1;2,0)]$}}\\

As in the two-dimensional case, the resulting system can be reduced to a scalar ordinary difference equation, so is not considered here.
\\

\subsubsection{Integrable Systems in $N$ Dimensions}

The equivalence class $[(0,1;0,1)]$ is defined for any dimension $N$. From Proposition \ref{prop:idc} with $k_2 = 0$ and $\ell_1=\ell_2=1$, we find that
\begin{equation}\label{eq:N-Hirota-def-v}
v^{(i)}_{m,n}\,=\,u^{(i)}_{m,n}\,-\,a \, \prod_{r=0}^{i-1}\frac{1}{u^{(r)}_{m,n}}\,\prod_{s=i}^{N-2}\frac{1}{u^{(s)}_{m,n+1}}\,,
\end{equation}
in view of which we derive the system
\begin{equation}\label{eq:N-Hirota}
u^{(i)}_{m,n} u^{(i)}_{m+1,n}\,-\,\frac{a u^{(i)}_{m,n}}{\prod_{r=0}^{i-1} u^{(r)}_{m+1,n} \prod_{s=i}^{N-2} u^{(s)}_{m+1,n+1}}=u^{(i)}_{m,n} u^{(i)}_{m,n+1}\,-\,\frac{a u^{(i)}_{m,n+1}}{\prod_{r=0}^{i-1} u^{(r)}_{m,n} \prod_{s=i}^{N-2} u^{(s)}_{m,n+1}},
\end{equation}
where $i \in {\mathbb{Z}}_{N-1}$.

This system can be reduced further by setting $v^{(i)}=0$ for $i\ne 0$, after which all other variables can be written in terms of $u^{(1)}_{m,n}$ and its shifts.  Setting $u_{m,n}=\frac{1}{u^{(1)}_{m,n}}$ and employing $a\mapsto \frac{1}{a}$, we obtain
\be\label{eq:gen-dis-eq}
\frac{a}{u_{m+1,n+N-1}} +  \prod_{i=1}^{N-1} u_{m,n+i} = \prod_{i=0}^{N-2} u_{m+1,n+i}\,+\,\frac{a}{u_{m,n}},
\ee
which obviously involves $2 N$ points.  The above equation coincides with Hirota's KdV equation (\ref{eq:2D-Hirota}) for $N=2$, and, up to the inversion of $u$ and $a$, becomes equation (\ref{eq:BMXa}) when $N=3$.

\section{The Non-Coprime Case}\label{sect:noncoprime}
\setcounter{equation}{0}

Our classification of Section \ref{sect:class-prob} finished with the {\em non-coprime case}, which may be considered as representing a coupling between coprime systems.  This follows by the block structure described in Proposition \ref{prop:prop-block}.  In this short section we give some examples to illustrate this structure.

It should be emphasised that the permutation matrix $P$ depends only upon the greatest common divisor, $p=(N,\ell_i-k_i)$, so all matrices with the same $p$ can {\em simultaneously} be put in block form.  We can see from formula (\ref{plp-1}) that when $k_1$ (respectively $k_2$) is a multiple of $p$, then $L$ (respectively $M$) takes block diagonal form. If {\em both} $L$ {\em and} $M$ take block diagonal form, then the system decouples into $p$ copies of the same $q-$component system.  Otherwise, the system is organised as a coupling between the $q-$vectors ${\bf u}_i = (u^{(i)},u^{(i+p)},\dots ,u^{(i+p(q-1))}),\; i=0, \dots ,p-1$.

\bex[$N=4,\, \ell_i-k_i\equiv 2$]\label{ncp-exN=4}  {\em
Using the notation of Section \ref{gcd}, we have $p=2, q=2, r=1$.  Inequivalent choices of $(k_1,\ell_1)$ are $(0,2), (1,3)$ or $(2,0)$, in which case we have respectively
$$
L=\left(
    \begin{array}{cc}
      L^{(0,1)}_{02} & 0 \\
      0 & L^{(0,1)}_{13}
    \end{array}
  \right),\quad       L=\left(
                         \begin{array}{cc}
                         0 & L^{(0,1)}_{02} \\
                         L^{(1,0)}_{13} & 0
                         \end{array}  \right),\quad    L=\left(
                                                          \begin{array}{cc}
                                                          L^{(1,0)}_{02} & 0 \\
                                                          0 & L^{(1,0)}_{13}
                                                          \end{array} \right),
$$
where $L^{(k,\ell)}_{ij}$ is the $2\times 2$ matrix of level structure $(k,\ell)$ and depending on variables $(u^{(i)}_{m,n},u^{(j)}_{m,n})$.
We must also choose $M$ to have one of these structures.

If we choose {\em both} $L$ {\em and} $M$ to be block-diagonal, then components $(0,2)$ and $(1,3)$ {\em decouple}.  For example, with level structure $(0,2;0,2)$, we can use (\ref{eq:dLP-gen-ch-1}) to define $(\phi_{m,n}^{(0)},\alpha_0,\beta_0)$, with $\phi_{m,n}^{(2)}=(\phi_{m,n}^{(0)})^{-1}$ and $(\phi_{m,n}^{(1)},\alpha_1,\beta_1)$, with $\phi_{m,n}^{(3)}=(\phi_{m,n}^{(1)})^{-1}$ to obtain two copies of the modified KdV equation (\ref{eq:2D-0101}).

On the other hand, if we choose level structure $(1,3;1,3)$, then equation (\ref{eq:dLP-gen-cc}) yields a coupled system
\bea
L^{(0,1)}_{02}({\bf u}_{m,n+1})M^{(1,0)}_{13}({\bf v}_{m,n}) &=& M^{(0,1)}_{02}({\bf v}_{m+1,n})L^{(1,0)}_{13}({\bf u}_{m,n}),\nn\\
   L^{(1,0)}_{13}({\bf u}_{m,n+1})M^{(0,1)}_{02}({\bf v}_{m,n}) &=& M^{(1,0)}_{13}({\bf v}_{m+1,n})L^{(0,1)}_{02}({\bf u}_{m,n}). \nn
\eea
If we use (\ref{eq:dLP-gen-ch-1}) to rewrite the equations in quotient potential form, and set $\phi^{(0)}_{m,n}=1/\phi^{(2)}_{m,n}=\phi_{m,n}$ and  $1/\phi^{(1)}_{m,n}=\phi^{(3)}_{m,n}=\psi_{m,n}$, we obtain the following coupled two-component system.
$$
\phi_{m,n} \phi_{m+1,n+1} \,=\,\frac{\alpha \psi_{m+1,n} - \beta \psi_{m,n+1}}{\alpha \psi_{m,n+1} - \beta \psi_{m+1,n}}\,,\quad
      \psi_{m,n} \psi_{m+1,n+1} \,=\,\frac{\alpha \phi_{m,n+1} - \beta \phi_{m+1,n}}{\alpha \phi_{m+1,n} - \beta \phi_{m,n+1}}\,.
$$
This system can be decoupled for either of the functions involved in it to the following five-point equation.
\bea
&& \left(\frac{\alpha \phi_{m+1,n+1}\,-\,\beta \phi_{m+2,n}}{\beta \phi_{m+1,n+1}\,-\,\alpha \phi_{m+2,n}}\right)
   \left(\frac{\alpha \phi_{m+1,n+1}\,-\,\beta \phi_{m,n+2}}{\beta \phi_{m+1,n+1}\,-\,\alpha \phi_{m,n+2}}\right) =  \nn\\
&& \hspace{4cm} \left(\frac{\alpha  \phi_{m,n} \phi_{m+1,n+1}\,+\,\beta}{\beta \phi_{m,n} \phi_{m+1,n+1}\,+\,\alpha}\right)
       \left(\frac{\alpha  \phi_{m+1,n+1} \phi_{m+2,n+2}\,+\,\beta}{\beta \phi_{m+1,n+1} \phi_{m+2,n+2}\,+\,\alpha }\right),  \nn
\eea
which could be interpreted as a discrete version of the modified Hirota-Satsuma equation, which has a $4\times 4$ matrix Lax pair (see section $3$ of \cite{BEF}).
}\eex

\bex[$N=6,\, \ell_i-k_i\equiv 2$]\label{p2q3}  {\em
Using the notation of Section \ref{gcd}, we have $p=2, q=3, r=1$.  Inequivalent choices of $(k_1,\ell_1)$ are $(0,2), (1,3), (2,4), (3,5), (4,0)$ or $(5,1)$.  The choices $(0,2),(2,4)$ and $(4,0)$ give block diagonal forms for $L$.  The choices $(1,3)$ and $(3,5)$, respectively, give the following forms for $L$:
$$
L=\left(
    \begin{array}{cc}
      0 & L^{(0,1)}_{024} \\
      L^{(1,2)}_{135} & 0
    \end{array}
  \right),\qquad       L=\left(
                         \begin{array}{cc}
                         0 & L^{(1,2)}_{024} \\
                         L^{(2,0)}_{135} & 0
                         \end{array}  \right),
$$
where $L^{(k,\ell)}_{abc}$ is the $3\times 3$ Lax matrix of level structure $(k,\ell)$ and depending on variables $u^{(a)}_{m,n}, u^{(b)}_{m,n}$ and $u^{(c)}_{m,n}$.  To obtain a coupled system, we could choose $L$ to be one of these three structures, with $M$ being any of the six.

For example, the choice $(1,3;1,3)$ leads to the system
\bea
L^{(0,1)}_{024}({\bf u}_{m,n+1})M^{(1,2)}_{135}({\bf v}_{m,n}) &=& M^{(0,1)}_{024}({\bf v}_{m+1,n})L^{(1,2)}_{135}({\bf u}_{m,n}),\nn\\
   L^{(1,2)}_{135}({\bf u}_{m,n+1})M^{(0,1)}_{024}({\bf v}_{m,n}) &=& M^{(1,2)}_{135}({\bf v}_{m+1,n})L^{(0,1)}_{024}({\bf u}_{m,n}), \nn
\eea
whilst the choice $(1,3;3,5)$ leads to the system
\bea
L^{(0,1)}_{024}({\bf u}_{m,n+1})M^{(2,0)}_{135}({\bf v}_{m,n}) &=& M^{(1,2)}_{024}({\bf v}_{m+1,n})L^{(1,2)}_{135}({\bf u}_{m,n}),\nn\\
   L^{(1,2)}_{135}({\bf u}_{m,n+1})M^{(1,2)}_{024}({\bf v}_{m,n}) &=& M^{(2,0)}_{135}({\bf v}_{m+1,n})L^{(0,1)}_{024}({\bf u}_{m,n}). \nn
\eea
The explicit form of the latter, in quotient potential form, with
$\phi^{(0)}_{m,n}=1/\varphi^{(0)}_{m,n},\, \phi^{(2)}_{m,n}=\psi^{(0)}_{m,n},\, \phi^{(4)}_{m,n}=\varphi^{(0)}_{m,n}/\psi^{(0)}_{m,n}$
and
$\phi^{(1)}_{m,n}=1/\psi^{(1)}_{m,n},\, \phi^{(3)}_{m,n}=\varphi^{(1)}_{m,n},\, \phi^{(5)}_{m,n}=\psi^{(1)}_{m,n}/\varphi^{(1)}_{m,n}$, is
\bea
 \varphi^{(0)}_{m,n}\varphi^{(0)}_{m+1,n+1} =
    \frac{\alpha \varphi^{(1)}_{m+1,n}\psi^{(1)}_{m,n+1}-\beta}
                                {\alpha \psi^{(1)}_{m+1,n}-\beta \varphi^{(1)}_{m+1,n}\varphi^{(1)}_{m,n+1}}, &&
   \psi^{(0)}_{m,n} \psi^{(0)}_{m+1,n+1} =
    \frac{\alpha \varphi^{(1)}_{m,n+1}-\beta \psi^{(1)}_{m+1,n}\psi^{(1)}_{m,n+1}}
                    {\alpha \psi^{(1)}_{m+1,n}-\beta \varphi^{(1)}_{m+1,n}\varphi^{(1)}_{m,n+1}},  \nn\\
 \varphi^{(1)}_{m,n}\varphi^{(1)}_{m+1,n+1} =
    \frac{\alpha \varphi^{(0)}_{m,n+1}\psi^{(0)}_{m+1,n}-\beta}
                                {\alpha \psi^{(0)}_{m,n+1}-\beta \varphi^{(0)}_{m+1,n}\varphi^{(0)}_{m,n+1}}, &&
   \psi^{(1)}_{m,n} \psi^{(1)}_{m+1,n+1} =
    \frac{\alpha \varphi^{(0)}_{m+1,n}-\beta \psi^{(0)}_{m+1,n}\psi^{(0)}_{m,n+1}}
                    {\alpha \psi^{(0)}_{m,n+1}-\beta \varphi^{(0)}_{m+1,n}\varphi^{(0)}_{m,n+1}},  \nn
\eea
which should be compared with system (\ref{eq:3D-0112}).
}\eex

\bex[$N=6,\, \ell_i-k_i\equiv 3$]\label{p3q2}  {\em
Using the notation of Section \ref{gcd}, we have $p=3, q=2, r=1$.  Inequivalent choices of $(k_1,\ell_1)$ are $(0,3), (1,4), (2,5), (3,0), (4,1)$ or $(5,2)$.  The choices $(0,3)$ and $(3,0)$ give block diagonal forms for $L$.  The choices $(1,4)$ and $(2,5)$, for example, give the following forms for $L$:
$$
L=\left(
    \begin{array}{ccc}
      0 & L^{(0,1)}_{03} & 0 \\
      0 & 0 & L^{(0,1)}_{14}  \\
      L^{(1,0)}_{25} & 0 & 0
    \end{array}   \right),\quad
                        L=\left(
                          \begin{array}{ccc}
                         0 & 0 & L^{(0,1)}_{03} \\
                            L^{(1,0)}_{14}  & 0 & 0 \\
                             0 & L^{(1,0)}_{25} & 0
                              \end{array}   \right),
$$
where $L^{(k,\ell)}_{ab}$ is the $2\times 2$ Lax matrix of level structure $(k,\ell)$ and depending on variables $u^{(a)}_{m,n}$ and $u^{(b)}_{m,n}$.
The choice $(2,5;2,5)$ leads to the system
\begin{subequations}\label{mkdv}
\bea
L^{(0,1)}_{03}({\bf u}_{m,n+1})M^{(1,0)}_{25}({\bf v}_{m,n}) &=&  M^{(0,1)}_{03}({\bf v}_{m+1,n})L^{(1,0)}_{25}({\bf u}_{m,n}), \label{mkdv1}  \\
   L^{(1,0)}_{14}({\bf u}_{m,n+1})M^{(0,1)}_{03}({\bf v}_{m,n}) &=& M^{(1,0)}_{14}({\bf v}_{m+1,n})L^{(0,1)}_{03}({\bf u}_{m,n}), \label{mkdv2}  \\
L^{(1,0)}_{25}({\bf u}_{m,n+1})M^{(1,0)}_{14}({\bf v}_{m,n}) &=&  M^{(1,0)}_{25}({\bf v}_{m+1,n})L^{(1,0)}_{14}({\bf u}_{m,n}),  \label{mkdv3}
\eea
\end{subequations}
Writing the equations in potential form (\ref{eq:dLP-gen-ch-1}), with $\phi^{(0)}_{m,n}=1/\phi^{(3)}_{m,n}=\psi^{(0)}_{m,n}, \phi^{(4)}_{m,n}=1/\phi^{(1)}_{m,n}=\psi^{(1)}_{m,n}, \phi^{(2)}_{m,n}=1/\phi^{(5)}_{m,n}=\psi^{(2)}_{m,n}$, we obtain the system
\be\label{cmkdv}
\psi^{(i)}_{m+1,n+1} = \left(
   \frac{\alpha \psi^{(i+2)}_{m,n+1}-\beta \psi^{(i+2)}_{m+1,n}}{\alpha \psi^{(i+2)}_{m+1,n}-\beta \psi^{(i+2)}_{m,n+1}}
         \right)\, \psi^{(i+1)}_{m,n},\quad i\in {\mathbb{Z}}_3,
\ee
which is a coupled discrete MKdV system, reducing to (\ref{eq:2D-0101}) when all components are equal.

On the other hand, the choice $(1,4;2,5)$ leads to the system
\begin{subequations}\label{sg}
\bea
L^{(0,1)}_{03}({\bf u}_{m,n+1})M^{(1,0)}_{14}({\bf v}_{m,n}) &=& M^{(0,1)}_{03}({\bf v}_{m+1,n})L^{(1,0)}_{25}({\bf u}_{m,n}), \label{sg1}  \\
   L^{(0,1)}_{14}({\bf u}_{m,n+1})M^{(1,0)}_{25}({\bf v}_{m,n}) &=& M^{(1,0)}_{14}({\bf v}_{m+1,n})L^{(0,1)}_{03}({\bf u}_{m,n}), \label{sg2}  \\
L^{(1,0)}_{25}({\bf u}_{m,n+1})M^{(0,1)}_{03}({\bf v}_{m,n}) &=& M^{(1,0)}_{25}({\bf v}_{m+1,n})L^{(0,1)}_{14}({\bf u}_{m,n}). \label{sg3}
\eea
\end{subequations}
Writing the equations in potential form (\ref{eq:dLP-gen-ch-1}), with $\phi^{(0)}_{m,n}=1/\phi^{(3)}_{m,n}=\psi^{(0)}_{m,n}, \phi^{(4)}_{m,n}=1/\phi^{(1)}_{m,n}=\psi^{(1)}_{m,n}, \phi^{(2)}_{m,n}=1/\phi^{(5)}_{m,n}=\psi^{(2)}_{m,n}$, we now obtain the system
\be\label{csg}
\psi^{(i)}_{m,n}\psi^{(i)}_{m+1,n+1} =
   \frac{\alpha -\beta \psi^{(i+1)}_{m,n+1}\psi^{(i+2)}_{m+1,n}}{\alpha \psi^{(i+1)}_{m,n+1}\psi^{(i+2)}_{m+1,n}-\beta},\quad i\in {\mathbb{Z}}_3,
\ee
which is a coupled system of Hirota's discrete sine-Gordon equations, reducing to (\ref{eq:2D-0110}) when all components are equal.
}\eex

The above generalisations of the discrete MKdV and of Hirota's discrete sine-Gordon equations, with this $2\times 2$ block structure, are easily extended to an arbitrary number of components.

\section{Building the Lattice}\label{sect:lattice}
\setcounter{equation}{0}

We now consider some interesting lattices which can be built out of our systems.  Of course, we may take a copy of the {\em same} $L-M$ pair around {\em each} quadrilateral of the planar lattice.  However, in this section we wish to describe other, more exotic, lattices, formed by taking {\em different} systems around adjacent quadrilaterals.  Such systems have received much attention recently \cite{HV2,Bo}.

These are most conveniently described in terms of quotient potentials.

\subsection{The General Case}\label{gen-lattice}

\begin{figure}[ht]
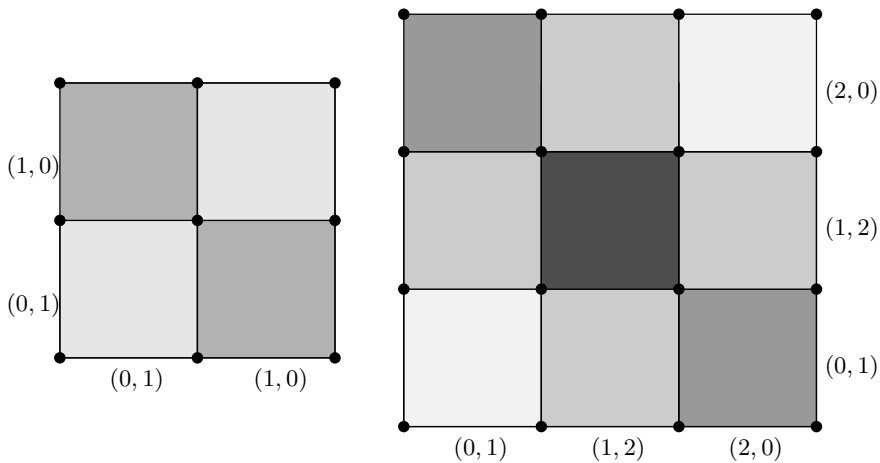

\begin{center}
\begin{texdraw} \setunitscale 0.36
\linewd 0.02 \arrowheadtype t:F
\htext(0 0.5) {\phantom{T}}
\move (-2 0) \lvec (0 0) \lvec (0 2) \lvec (-2 2) \lvec (-2 0) \lfill f:0.9
\move (0 0) \lvec (2 0) \lvec (2 2) \lvec (0 2) \lvec (0 0) \lfill f:0.7
\move (-2 2) \lvec (0 2) \lvec (0 4) \lvec (-2 4) \lvec (-2 4) \lfill f:0.7
\move (0 2) \lvec (2 2) \lvec (2 4) \lvec (0 4) \lvec (0 2) \lfill f:0.9
\move (-2 0) \fcir f:0.0 r:0.08 \move (0 0) \fcir f:0.0 r:0.08 \move (2 0) \fcir f:0.0 r:0.08
\move (-2 2) \fcir f:0.0 r:0.08 \move (0 2) \fcir f:0.0 r:0.08 \move (2 2) \fcir f:0.0 r:0.08
\move (-2 4) \fcir f:0.0 r:0.08 \move (0 4) \fcir f:0.0 r:0.08 \move (2 4) \fcir f:0.0 r:0.08
\move (-2 0) \lvec (2 0) \lvec(2 4) \lvec (-2 4) \lvec (-2 0)
\move (0 0) \lvec (0 4)
\htext (-1.3 -.5) {\footnotesize$(0,1)$} \htext (.8 -.5) {\footnotesize$(1,0)$}
\htext (-2.8 .6) {\footnotesize$(0,1)$} \htext (-2.8 2.6) {\footnotesize$(1,0)$}
\move (3 -1) \lvec (5 -1) \lvec (5 1) \lvec (3 1) \lvec (3 -1) \lfill f:0.95
\move (5 -1) \lvec (7 -1) \lvec (7 1) \lvec (5 1) \lvec (5 -1) \lfill f:0.8
\move (7 -1) \lvec (9 -1) \lvec (9 1) \lvec (7 1) \lvec (7 -1) \lfill f:0.6
\move (3 1) \lvec (5 1) \lvec (5 3) \lvec (3 3) \lvec (3 1) \lfill f:0.8
\move (5 1) \lvec (7 1) \lvec (7 3) \lvec (5 3) \lvec (5 1) \lfill f:0.3
\move (7 1) \lvec (9 1) \lvec (9 3) \lvec (7 3) \lvec (7 1) \lfill f:0.8
\move (3 3) \lvec (5 3) \lvec (5 5) \lvec (3 5) \lvec (3 3) \lfill f:0.6
\move (5 3) \lvec (7 3) \lvec (7 5) \lvec (5 5) \lvec (5 3) \lfill f:0.8
\move (7 3) \lvec (9 3) \lvec (9 5) \lvec (7 5) \lvec (7 3) \lfill f:0.95
\move (3 -1) \fcir f:0.0 r:0.08 \move (5 -1) \fcir f:0.0 r:0.08 \move (7 -1) \fcir f:0.0 r:0.08  \move (9 -1) \fcir f:0.0 r:0.08
\move (3 1) \fcir f:0.0 r:0.08 \move (5 1) \fcir f:0.0 r:0.08 \move (7 1) \fcir f:0.0 r:0.08  \move (9 1) \fcir f:0.0 r:0.08
\move (3 3) \fcir f:0.0 r:0.08 \move (5 3) \fcir f:0.0 r:0.08 \move (7 3) \fcir f:0.0 r:0.08  \move (9 3) \fcir f:0.0 r:0.08
\move (3 5) \fcir f:0.0 r:0.08 \move (5 5) \fcir f:0.0 r:0.08 \move (7 5) \fcir f:0.0 r:0.08  \move (9 5) \fcir f:0.0 r:0.08
\move (3 -1) \lvec (9 -1) \lvec(9 3) \lvec (3 3) \lvec (3 -1)
\move (5 -1) \lvec (5 4) \move (7 -1) \lvec (7 4)
\htext (3.7 -1.5) {\footnotesize$(0,1)$} \htext (5.7 -1.5) {\footnotesize$(1,2)$} \htext (7.7 -1.5) {\footnotesize$(2,0)$}
\htext (9.1 -.3) {\footnotesize$(0,1)$} \htext (9.1 1.7) {\footnotesize$(1,2)$} \htext (9.1 3.7) {\footnotesize$(2,0)$}
\end{texdraw}
\end{center}
\caption{{\emph{\small{Opposite edges carry matrices with exactly the same structure. Here it is demonstrated the simplest configuration for the two-dimensional case (left), which corresponds to a black-white lattice \cite{XP}, and the three-dimensional case (right). Quadrilaterals with the same colour carry equivalent integrable systems.}}}} \label{fig:bianchi}
\end{figure}

Here we exploit the condition (\ref{eq:dLP-nec-rel}) and build lattices for which the number $\ell_i-k_i$, calculated in each quadrilateral, is {\em fixed} $(\bmod{N})$.  This can be done by assigning to every edge of the lattice a Lax matrix with a certain level structure and by taking into account the following two rules:
\begin{enumerate}
\item Lax matrices on opposite edges have the same level structure and are related by shifts.
\item On every elementary quadrilateral, level structures assigned to the edges belong in ${\cal{Q}}_N$.
\end{enumerate}
Each quadrilateral will then have a specific choice of $(k_1,\ell_1;k_2,\ell_2)$, subject only to $\ell_i-k_i$ being fixed  $(\bmod{N})$.  To determine which combinations give {\em inequivalent} systems, we must take into account the equivalence relations of Definition \ref{def:equiv-LP} and Proposition \ref{prop:rat-pot-trans}.  As a result only {\em two} inequivalent systems arise in the $2-$dimensional case of Figure \ref{fig:bianchi} and only {\em four} in the $3-$dimensional case.
In this way, every quadrilateral carries a different discrete system deriving from the compatibility condition of the corresponding Lax pair.

We can then introduce a third direction.  Above a square with $(k_1,\ell_1;k_2,\ell_2)$, with $\ell_2-k_2\equiv\ell_1-k_1 \, (\bmod{N})$ we can place a third direction, with level structure $(k_3,\ell_3)$, such that $\ell_3-k_3\equiv\ell_1-k_1 \, (\bmod{N})$.  Above the lattices of Figure \ref{fig:bianchi} we therefore build respectively four or nine cubes.  For each cube, opposite faces have the same level structure (and therefore the same {\em planar} system of equations). For the examples of Figure \ref{fig:bianchi} all cubes either have {\em four} equivalent faces (with the remaining two being different) or {\em six} equivalent faces.  Whatever the choice of $(k_3,\ell_3)$, the cubes can be consistently placed above the planar lattice (with common faces having the same level structure). Around each cube the array of systems are consistent.  Since opposite faces have the same level structure, the ``ground floor'' planar lattice is reproduced on the ``first floor'', so the procedure can be repeated.  In this way, we build a $3D$ consistent lattice.

This $3D$ consistency is another manifestation of the integrability of these systems.  In fact, the two equations corresponding to the ``vertical faces'' of the cube can be interpreted as a B{\"a}cklund transformation for system on the ``horizontal face''.  In this way, two copies of system (\ref{eq:3D-0112}) serve as a B{\"a}cklund transformation for system (\ref{eq:3D-1212}).

\subsection{The Non-Coprime Case}\label{ncp-lattice}

We saw in Section \ref{sect:noncoprime} that when $(N,\ell_i-k_i)=p\ne 1$, then the discrete system (\ref{eq:dLP-ex-cc}) takes the form of a coupled system, involving $q\times q$ matrices, where $N=pq$.  We can, of course, follow the general procedure described above.

For example, with $N=6$ and $\ell_i-k_i\equiv 3$, we may consider a lattice that {\em looks like} the two-dimensional case of Figure \ref{fig:bianchi}, but with $3$ components at each vertex.  We just replace $(0,1)$ and $(1,0)$ by $(1,4)$ and $(2,5)$ to obtain the coupled discrete modified KdV (dMKdV) system (\ref{cmkdv}) in the light coloured squares and the coupled discrete sine-Gordon (dSG) system (\ref{csg}) in the dark squares.

However, we wish to introduce a different lattice, more closely reflecting the coupled system in the non-coprime case.  We describe the procedure within the context of Example \ref{p3q2}.

\bex[Coupled dMKdV and dSG Systems for $N=6$]\label{p3q2lat}  {\em
We note that for the dMKdV case, compatibility conditions (\ref{mkdv}) define three ``elementary quadrilaterals'', with edges labelled by the appropriate $(k,\ell,a,b)$, shown in Figure \ref{mkdv-quads;fig}.  The first and second of these share a common edge, indicated by $M^{(0,1)}_{03}$ and similarly for the second and third.  This is also true for the right edge of the third quadrilateral and the left edge of the first.  Similar identifications can be made between top and bottom edges, so it is possible to construct consistent arrays of elementary quadrilaterals to form a lattice.
We can draw similar elementary quadrilaterals for any of the coupled systems for which $\ell_i-k_i\equiv 3$.

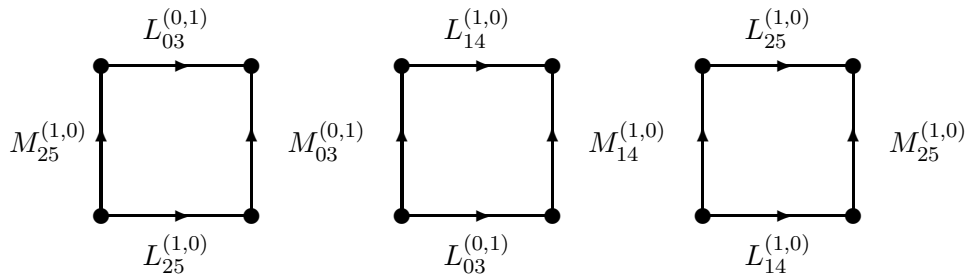
\begin{figure}[hbt]
\begin{center}
\unitlength=1mm
\begin{picture}(100,30)
%
%===========quad1=================
%
\put(0,0){\circle*{2}}
\put(20,0){\circle*{2}}
\put(0,20){\circle*{2}}
\put(20,20){\circle*{2}}
\thicklines
\put(0,0){\vector(1,0){12}}
\put(10,0){\line(1,0){10}}
\put(0,0){\vector(0,1){12}}
\put(0,10){\line(0,1){10}}
\put(0,20){\vector(1,0){12}}
\put(10,20){\line(1,0){10}}
\put(20,0){\vector(0,1){12}}
\put(20,10){\line(0,1){10}}
\put(10,-5){\makebox(0,0){$L^{(1,0)}_{25}$}}
\put(-7,10){\makebox(0,0){$M^{(1,0)}_{25}$}}
\put(10,25){\makebox(0,0){$L^{(0,1)}_{03}$}}
\put(30,10){\makebox(0,0){$M^{(0,1)}_{03}$}}
%
%===========quad2=================
%
\put(40,0){\circle*{2}}
\put(60,0){\circle*{2}}
\put(40,20){\circle*{2}}
\put(60,20){\circle*{2}}
\thicklines
\put(40,0){\vector(1,0){12}}
\put(50,0){\line(1,0){10}}
\put(40,0){\vector(0,1){12}}
\put(40,10){\line(0,1){10}}
\put(40,20){\vector(1,0){12}}
\put(50,20){\line(1,0){10}}
\put(60,0){\vector(0,1){12}}
\put(60,10){\line(0,1){10}}
\put(50,-5){\makebox(0,0){$L^{(0,1)}_{03}$}}
\put(50,25){\makebox(0,0){$L^{(1,0)}_{14}$}}
\put(70,10){\makebox(0,0){$M^{(1,0)}_{14}$}}
%
%===========quad3=================
%
\put(80,0){\circle*{2}}
\put(100,0){\circle*{2}}
\put(80,20){\circle*{2}}
\put(100,20){\circle*{2}}
\thicklines
\put(80,0){\vector(1,0){12}}
\put(90,0){\line(1,0){10}}
\put(80,0){\vector(0,1){12}}
\put(80,10){\line(0,1){10}}
\put(80,20){\vector(1,0){12}}
\put(90,20){\line(1,0){10}}
\put(100,0){\vector(0,1){12}}
\put(100,10){\line(0,1){10}}
\put(90,-5){\makebox(0,0){$L^{(1,0)}_{14}$}}
\put(90,25){\makebox(0,0){$L^{(1,0)}_{25}$}}
\put(110,10){\makebox(0,0){$M^{(1,0)}_{25}$}}
%
%===========end=================
%
\end{picture}
\end{center}
\caption{Elementary quadrilaterals for the coupled MKdV System}
\label{mkdv-quads;fig}
\end{figure}

When written in terms of the quotient potentials, the equations depicted by the three elementary quadrilaterals of Figure \ref{mkdv-quads;fig} are precisely equations (\ref{cmkdv}) for $i=0,1,2$ respectively.  Notice that, from the structure of these equations, each vertex of the quadrilateral is associated with a {\em specific variable} $\psi^{(i)},\, i \in \{0,1,2\}$.  Similarly, each of the three elementary quadrilaterals corresponding to equations (\ref{sg}) gives just one component of equations (\ref{csg}).

The elementary quadrilaterals can then be drawn in a very simple way, just indicating the specific variables which correspond to the particular quadrilateral.  These are depicted in Figure \ref{fig:lat-pat} (for both the dMKdV and dSG cases), where nine such elementary quadrilaterals are shown in the unique consistent configuration (for each choice of $i \in \{0,1,2\}$).  This configuration is then extended periodically to the whole plane.

\begin{figure}[ht]
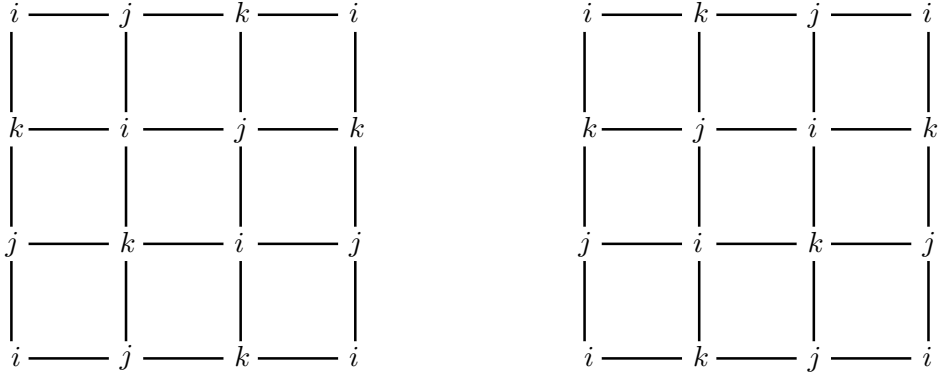

\begin{center}
\begin{texdraw} \setunitscale 0.6
\linewd 0.02 \arrowheadtype t:F
\htext(0 0.5) {\phantom{T}}
\move (-2 0) \lvec (1 0) \lvec (1 3) \lvec (-2 3) \lvec (-2 0)
\move (-1 0) \lvec (-1 3) \move (0 0) \lvec (0 3)
\move (-2 1) \lvec (1 1) \move (-2 2) \lvec (1 2)
\move (-2 0) \fcir f:1.0 r:0.15 \move (-1 0) \fcir f:1.0 r:0.15 \move (0 0) \fcir f:1.0 r:0.15 \move (1 0) \fcir f:1.0 r:0.15
\move (-2 1) \fcir f:1.0 r:0.15 \move (-1 1) \fcir f:1.0 r:0.15 \move (0 1) \fcir f:1.0 r:0.15 \move (1 1) \fcir f:1.0 r:0.15
\move (-2 2) \fcir f:1.0 r:0.15 \move (-1 2) \fcir f:1.0 r:0.15 \move (0 2) \fcir f:1.0 r:0.15 \move (1 2) \fcir f:1.0 r:0.15
\move (-2 3) \fcir f:1.0 r:0.15 \move (-1 3) \fcir f:1.0 r:0.15 \move (0 3) \fcir f:1.0 r:0.15 \move (1 3) \fcir f:1.0 r:0.15

\htext (-2 -.08){$i$} \htext (-1.05 -.11){$j$}\htext (-0.05 -.08){$k$}\htext (0.95 -.08){$i$}
\htext (-2.05 .88){$j$} \htext (-1.05 .92){$k$}\htext (-0.05 .92){$i$}\htext (0.95 .88){$j$}
\htext (-2.02 1.92){$k$} \htext (-1.05 1.92){$i$}\htext (-0.05 1.87){$j$}\htext (0.95 1.92){$k$}
\htext (-2.01 2.92){$i$} \htext (-1.05 2.88){$j$}\htext (-0.05 2.92){$k$}\htext (0.95 2.92){$i$}

\move (3 0) \lvec (6 0) \lvec (6 3) \lvec (3 3) \lvec (3 0)
\move (4 0) \lvec (4 3) \move (5 0) \lvec (5 3)
\move (3 1) \lvec (6 1) \move (3 2) \lvec (6 2)
\move (3 0) \fcir f:1.0 r:0.15 \move (4 0) \fcir f:1.0 r:0.15 \move (5 0) \fcir f:1.0 r:0.15 \move (6 0) \fcir f:1.0 r:0.15
\move (3 1) \fcir f:1.0 r:0.15 \move (4 1) \fcir f:1.0 r:0.15 \move (5 1) \fcir f:1.0 r:0.15 \move (6 1) \fcir f:1.0 r:0.15
\move (3 2) \fcir f:1.0 r:0.15 \move (4 2) \fcir f:1.0 r:0.15 \move (5 2) \fcir f:1.0 r:0.15 \move (6 2) \fcir f:1.0 r:0.15
\move (3 3) \fcir f:1.0 r:0.15 \move (4 3) \fcir f:1.0 r:0.15 \move (5 3) \fcir f:1.0 r:0.15 \move (6 3) \fcir f:1.0 r:0.15

\htext (3 -.08){$i$} \htext (3.95 -.08){$k$}\htext (4.95 -.12){$j$}\htext (5.95 -.08){$i$}
\htext (2.95 .88){$j$} \htext (3.95 .92){$i$}\htext (4.95 .92){$k$}\htext (5.95 .88){$j$}
\htext (2.98 1.92){$k$} \htext (3.95 1.88){$j$}\htext (4.95 1.92){$i$}\htext (5.95 1.92){$k$}
\htext (2.99 2.92){$i$} \htext (3.95 2.92){$k$}\htext (4.95 2.88){$j$}\htext (5.95 2.92){$i$}

\end{texdraw}
\end{center}
\caption{{\emph{\small{Patterns on the lattice with $(i,j,k) \in \left\{(0,1,2),(1,2,0),(2,0,1) \right\}$ :  The first involves a copy of modified KdV on every quadrilateral, whereas, in the second pattern, every quadrilateral carries a sine-Gordon equation.}}}} \label{fig:lat-pat}
\end{figure}

\br[Interlacing Columns]
We can also interlace columns of the dMKdV lattice with columns of the dSG lattice.  The only constraint is that right vertices of the left column must correspond to the left vertices of the right column.
\er

Corresponding to each choice of $i \in \{0,1,2\}$ there is also a unique $3D$ consistent cube, shown (for both the dMKdV and dSG cases) in Figure \ref{fig:3Dbianchi}. Each face of the dMKdV cube corresponds to an dMKdV equation.  It is not possible to build a consistent cube from just dSG faces. Two of the faces must be of dMKdV form (we already saw this in the 1 component case).

For the dMKdV case, opposite faces are related by the shift $i\mapsto i+1\, (\bmod{3})$.  Starting with a planar dMKdV lattice in the ``horizontal plane'', there is a unique configuration of dMKdV cubes with such ``bottom faces''.  On the ``first floor'' we now have the shifted planar lattice (with $i\mapsto i+1\, (\bmod{3})$), whilst on the ``second floor'' we have the twice shifted planar lattice (with $i\mapsto i+2\, (\bmod{3})$).  Since the whole process is periodic, the next shift gives back the original ``ground floor'' lattice.  In this way, we build a unique $3D$ consistent lattice.

\begin{figure}[ht]
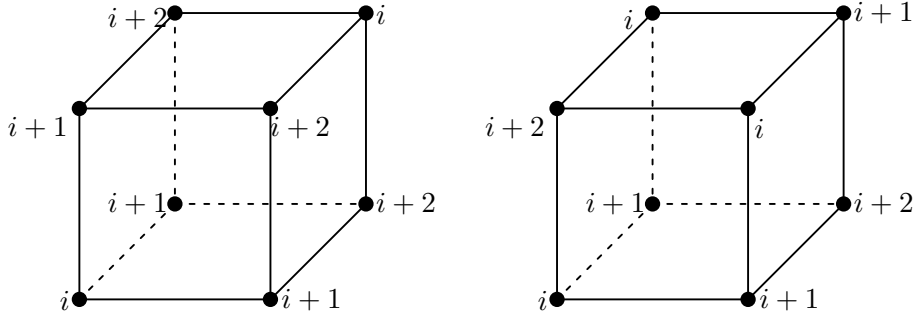

\begin{center}
\begin{texdraw} \setunitscale 0.5
\linewd 0.02 \arrowheadtype t:F
\htext(0 0.5) {\phantom{T}}
\move (-2 0) \lvec (0 0) \lvec (1 1) \lvec (1 3) \lvec (-1 3) \lvec (-2 2)  \lvec(-2 0)
\move (0 0) \lvec (0 2) \lvec (1 3) \move (-2 2) \lvec (0 2)
\lpatt (.05 .1) \move (-1 3) \lvec (-1 1) \lvec(-2 0) \move (-1 1) \lvec (1 1)
\lpatt ()
\move (-2 0) \fcir f:0.0 r:0.08 \move (0 0) \fcir f:0.0 r:0.08 \move (1 1) \fcir f:0.0 r:0.08
\move (-2 2) \fcir f:0.0 r:0.08 \move (0 2) \fcir f:0.0 r:0.08 \move (-1 1) \fcir f:0.0 r:0.08
\move (-1 3) \fcir f:0.0 r:0.08 \move (1 3) \fcir f:0.0 r:0.08 %\move (2 4) \fcir f:0.0 r:0.08

\htext (-2.2 -.1){$i$} \htext (.12 -.1) {$i+1$} \htext (.0 1.7) {$i+2$}
\htext (1.11 .9) {$i+2$} \htext (1.11 2.9) {$i$} \htext (-2.75 1.7){$i+1$}
\htext (-1.7 .9) {$i+1$} \htext (-1.7 2.85){$i+2$}

\move (3 0) \lvec (5 0) \lvec (6 1) \lvec (6 3) \lvec (4 3) \lvec (3 2)  \lvec(3 0)
\move (5 0) \lvec (5 2) \lvec (6 3) \move (3 2) \lvec (5 2)
\lpatt (.05 .1) \move (4 3) \lvec (4 1) \lvec(3 0) \move (4 1) \lvec (6 1)
\lpatt ()
\move (3 0) \fcir f:0.0 r:0.08 \move (5 0) \fcir f:0.0 r:0.08 \move (6 1) \fcir f:0.0 r:0.08
\move (3 2) \fcir f:0.0 r:0.08 \move (5 2) \fcir f:0.0 r:0.08 \move (4 1) \fcir f:0.0 r:0.08
\move (4 3) \fcir f:0.0 r:0.08 \move (6 3) \fcir f:0.0 r:0.08 %\move (2 4) \fcir f:0.0 r:0.08

\htext (2.8 -.1){$i$} \htext (5.12 -.1) {$i+1$} \htext (5.07 1.7) {$i$}
\htext (6.11 .9) {$i+2$} \htext (6.11 2.9) {$i+1$} \htext (2.25 1.7){$i+2$}
\htext (3.3 .9) {$i+1$} \htext (3.7 2.85){$i$}

\end{texdraw}
\end{center}
\caption{{\emph{\small{The first cube involves only copies of the modified KdV equation, whereas the second cube carries two copies of mKdV (bottom and top faces) and four copies of  the sine-Gordon equation.}}}} \label{fig:3Dbianchi}
\end{figure}

For the dSG case we can choose the top and bottom faces to be of dMKdV type.  We are then obliged to use a slightly different form of the dSG equation, corresponding to interchanging the positions of $\psi^{(i+1)}$ and $\psi^{(i+2)}$ in the $i^{th}$ equation.  This corresponds to the involution $\tilde f_{i,j}=f_{1-i,1-j}$, for any function $f$ located at vertex $(i,j)$ in the planar lattice.

Starting with an dMKdV planar lattice in the ``horizontal plane'', there is a unique configuration of dSG cubes with such ``bottom faces''.  On the ``first floor'' we now have the shifted planar lattice (with $i\mapsto i+2\, (\bmod{3})$), whilst on the ``second floor'' we have the twice shifted planar lattice (with $i\mapsto i+4 \equiv i+1\, (\bmod{3})$).  Since the whole process is periodic, the next shift gives back the original ``ground floor'' lattice.  In this way, we build a unique $3D$ consistent lattice.  The $3\times 3\times 3$ cube, with bottom and top faces in the configuration of the dMKdV lattice shown in Figure \ref{fig:lat-pat} has side faces with the dSG pattern of Figure \ref{fig:lat-pat}, but oriented so that vertices match at the base.

We can then place this side face on top of the planar dSG lattice of Figure \ref{fig:lat-pat}, and the $3\times 3\times 3$ cube above this planar dSG lattice is the unique consistent cube (exactly the one we could have built directly).

In this way we obtain two $3D$ lattices which consistently support a mixture of dMKdV and dSG equations on quadrilateral faces.
}\eex

\subsection{The Initial Value Problem}

The initial value problem for the standard planar lattice with a {\em single system} over the entire plane is just a multi-component version of the scalar case.  Since the equations can be solved for evolution in any direction, we can set initial conditions on (for example) a staircase.  This statement does not change if we introduce the ``multicoloured'' lattices of Section \ref{gen-lattice}.  The extension to the $3D$ lattice is standard.

However, for the lattices constructed in Section \ref{ncp-lattice}, the situation is different.  Since the equations can still be solved for evolution in any direction, we can again set initial conditions on (for example) an arbitrary staircase.  However, since each vertex only involves a subset of the systems components, we {\em cannot} determine the values of the other components at this vertex.

\begin{figure}[ht]
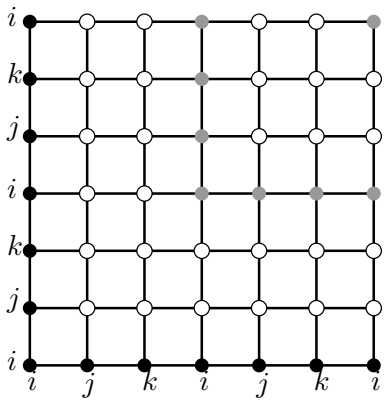

\begin{center}
\begin{texdraw} \setunitscale 0.6
\linewd 0.02 \arrowheadtype t:F
\htext(0 0.5) {\phantom{T}}
\move (-2 0) \lvec (1 0) \lvec (1 3) \lvec (-2 3) \lvec (-2 0)
\move (-1 0) \lvec (-1 3) \move (0 0) \lvec (0 3)
\move (-2 1) \lvec (1 1) \move (-2 2) \lvec (1 2)
\move (-1.5 0) \lvec (-1.5 3) \move (.5 0) \lvec (.5 3) \move (-.5 0) \lvec (-.5 3)
\move (-2 0.5) \lvec (1 .5) \move (-2 1.5) \lvec (1 1.5) \move (-2 2.5) \lvec (1 2.5)
\move (-2 0) \fcir f:0.0 r:0.06 \move (-1 0) \fcir f:0.0 r:0.06 \move (0 0) \fcir f:0.0 r:0.06 \move (1 0) \fcir f:0.0 r:0.06
\move (-1.5 0) \fcir f:0.0 r:0.06 \move (-.5 0) \fcir f:0.0 r:0.06 \move (0.5 0) \fcir f:0.0 r:0.06
\move (-2 0.5) \fcir f:0.0 r:0.06 \move (-2 1) \fcir f:0.0 r:0.06 \move (-2 1.5) \fcir f:0.0 r:0.06
\move (-2 2) \fcir f:0.0 r:0.06 \move (-2 2.5) \fcir f:0.0 r:0.06 \move (-2 3) \fcir f:0.0 r:0.06

\move (-1.5 .5) \lcir r:0.06 \move (-1 .5) \lcir r:0.06 \move (-.5 .5) \lcir r:0.06 \move (0 .5) \lcir r:0.06
\move (.5 .5) \lcir r:0.06 \move (1 .5) \lcir r:0.06
\move (-1.5 1) \lcir r:0.06 \move (-1 1) \lcir r:0.06 \move (-.5 1) \lcir r:0.06 \move (0 1) \lcir r:0.06
\move (.5 1) \lcir r:0.06 \move (1 1) \lcir r:0.06
\move (-1.5 1.5) \lcir r:0.06 \move (-1.5 2) \lcir r:0.06 \move (-1.5 2.5) \lcir r:0.06 \move (-1.5 3) \lcir r:0.06
\move (-1. 1.5) \lcir r:0.06 \move (-1. 2) \lcir r:0.06 \move (-1. 2.5) \lcir r:0.06 \move (-1. 3) \lcir r:0.06

\move (-1.5 .5) \fcir f:1.0 r:0.06 \move (-1 .5) \fcir f:1.0  r:0.06 \move (-.5 .5) \fcir f:1.0  r:0.06 \move (0 .5) \fcir f:1.0  r:0.06
\move (.5 .5) \fcir f:1.0  r:0.06 \move (1 .5) \fcir f:1.0  r:0.06
\move (-1.5 1) \fcir f:1.0  r:0.06 \move (-1 1) \fcir f:1.0  r:0.06 \move (-.5 1) \fcir f:1.0  r:0.06 \move (0 1) \fcir f:1.0  r:0.06
\move (.5 1) \fcir f:1.0  r:0.06 \move (1 1) \fcir f:1.0  r:0.06
\move (-1.5 1.5) \fcir f:1.0  r:0.06 \move (-1.5 2) \fcir f:1.0  r:0.06
\move (-1.5 2.5) \fcir f:1.0  r:0.06 \move (-1.5 3) \fcir f:1.0  r:0.06
\move (-1. 1.5) \fcir f:1.0  r:0.06 \move (-1. 2) \fcir f:1.0  r:0.06 \move (-1. 2.5) \fcir f:1.0  r:0.06 \move (-1. 3) \fcir f:1.0  r:0.06

\move (0 2) \lcir r:0.06 \move (0 2.5) \lcir r:0.06 \move (0 3) \lcir r:0.06
\move (.5 2) \lcir r:0.06 \move (.5 2.5) \lcir r:0.06 \move (.5 3) \lcir r:0.06
\move (1 2) \lcir r:0.06 \move (1 2.5) \lcir r:0.06

\move (0 2) \fcir f:1.0 r:0.06 \move (0 2.5)  \fcir f:1.0 r:0.06 \move (0 3)  \fcir f:1.0 r:0.06
\move (.5 2)  \fcir f:1.0 r:0.06 \move (.5 2.5)  \fcir f:1.0 r:0.06 \move (.5 3)  \fcir f:1.0 r:0.06
\move (1 2)  \fcir f:1.0 r:0.06 \move (1 2.5)  \fcir f:1.0 r:0.06

\move(-.5 1.5) \fcir f:0.6 r:0.06 \move(0 1.5) \fcir f:0.6 r:0.06 \move(.5 1.5) \fcir f:0.6 r:0.06
\move(1 1.5) \fcir f:0.6 r:0.06 \move (-.5 2) \fcir f:0.6 r:0.06   \move (-.5 2.5) \fcir f:0.6 r:0.06
\move (-.5 3) \fcir f:0.6 r:0.06   \move (1 3) \fcir f:.6 r:0.06

\htext (-2.02 -0.23) {$i$} \htext (-1.54 -0.29) {$j$} \htext (-1.02 -0.23) {$k$} \htext (-.52 -0.23) {$i$} \htext (-.02 -0.29) {$j$} \htext (.48 -0.23) {$k$}  \htext (.98 -0.23) {$i$}
\htext (-2.2 -.05) {$i$} \htext (-2.2 .45) {$j$} \htext (-2.2 .95) {$k$} \htext (-2.2 1.45) {$i$} \htext (-2.2 1.95) {$j$} \htext (-2.2 2.45) {$k$}  \htext (-2.2 2.95) {$i$}

\end{texdraw}
\end{center}
\caption{{\emph{\small{Patterns on the lattice and initial value problems with $(i,j,k) \in \left\{(0,1,2),(1,2,0),(2,0,1) \right\}$ : Every black vertex carries the initial value of the corresponding variable, e.g. the left bottom vertex carries the initial value of $\psi^{(i)}$. This initial pattern repeats after three diagonal steps leading to the updated gray vertices. }}}} \label{fig:ivp-pat}
\end{figure}

To be specific, suppose we consider the dMKdV system (\ref{cmkdv}) and the corresponding planar lattice shown in Figure \ref{fig:lat-pat}.  For simplicity, we consider the equation in the first quadrant $(m,n)$, with $m\ge 0,\, n\ge 0$ and set {\em initial conditions} on the axes.  It can be seen from Figure \ref{fig:lat-pat} that we have three possible configurations of initial conditions, depicted in Figure \ref{fig:ivp-pat} by the black vertices, where $(i,j,k) \in \left\{(0,1,2),(1,2,0),(2,0,1) \right\}$.  We can then use the equations to calculate specific values on the next (white) ``L'' shape (a similar configuration, but with index $i$ shifted by $2$ in the diagonal direction).  Each iteration involves this shift by $2$, so after $3$ iterations we return to the same configuration, as depicted by the grey vertices.  In this way we build a unique array of the letters $(i,j,k)$, in the pattern given in Figure \ref{fig:lat-pat}, and thus build {\em three} different lattices, depending upon the choice of values for these letters.

\br
The standard lattice, with the whole coupled MKdV system on each quadrilateral, is obtained by superposition of these three lattices.
\er

\br
Since we can calculate the value of $\psi^{(0)}$ (say) on a specific sub-array of vertices, we conjecture that a higher order equation exists for $\psi^{(0)}$ alone.  In fact, it follows from the symmetry of the system that each component $\psi^{(i)}$ would satisfy the same higher order equation.
\er

\section{Differential-Difference Equations}\label{continuous-defs}
\setcounter{equation}{0}

In this section we consider semi-discrete Lax pairs; the resulting isospectral flows are differential-difference equations. We first discuss continuous isospectral deformations of the matrix $L_{m,n}$, satisfying (\ref{eq:dLP-gen-L}).  The simplest solutions give rise to the lowest order autonomous flow $\pa_{t^1}u^{(i)}_{m,n}$ and a {\em master symmetry}, which generates an infinite hierarchy of autonomous flows.  When these isospectral deformations are made for {\em both} $L_{m,n}$ and $M_{m,n}$, compatibility is guaranteed on solutions of the fully discrete system  (\ref{eq:dLP-ex-cc}), in which case they are interpreted as symmetries of these discrete systems.

There is, in fact, a symmetry between the $L_{m,n}$ and $M_{m,n}$ calculations for the {\em non-degenerate case}.  This is broken in the degenerate case, so each system has to be analysed in a case by case fashion.  We present symmetries of the examples discussed in Section \ref{degenerate}.

Our general formulation concentrates on the coprime case, but we finish the section by briefly discussing the non-coprime example with $N=4,\, (k_1,\ell_1)=(1,3)$.

\subsection{Differential-Difference Equations as Isospectral Flows}\label{iso-m}

In this section we consider differential-difference systems which only involve shifts in the discrete variable $m$.  However, since we later wish to interpret these as symmetries of the fully-discrete system (\ref{eq:dLP-ex-cc}), we continue to write our functions as depending on {\em both} indices $m$ and $n$.  We have also deliberately not labelled $(k,\ell)$ as $(k_1,\ell_1)$, since when we later replace $L_{m,n}$ by $M_{m,n}$, we will need $(k,\ell)=(k_2,\ell_2)$.

Let us start with the $N \times N$ Lax pair
\be\label{eq:dLP-cont-m}  %
\Psi_{m+1,n}\,=\,L_{m,n} \Psi_{m,n} \,,\quad \partial_{t} \Psi_{m,n}\,=\,S_{m,n}\,\Psi_{m,n}\,,
\ee  %
with $L_{m,n}$ defined by (\ref{eq:dLP-gen-L}) and $S_{m,n}$ a $\lambda-$dependent matrix, to be determined.

The compatibility condition of the above system is
\begin{equation} \label{eq:dd-cc}
\partial_t L_{m,n} =  S_{m+1,n} L_{m,n} - L_{m,n}S_{m,n}.
\end{equation}
If we write this as
$$
L_{m,n}^{-1}\pa_t L_{m,n} = L_{m,n}^{-1} S_{m+1,n} L_{m,n} - S_{m,n},
$$
and note that
$$
\pa_t \log \left( \det L_{m,n} \right) = {\rm{Tr}}\left(L^{-1}_{m,n} \pa_t L_{m,n} \right),
$$
we obtain the conservation law
\be\label{dtlogdet}  %
\pa_t \log \left( \det L_{m,n} \right) = \Delta_m\, \left({\rm{Tr}} \left(S_{m,n}\right)\right).
\ee  %

Setting
\be\label{Q=LS}  %
S_{m,n} = L_{m,n}^{-1}Q_{m,n},
\ee  %
we can then write (\ref{eq:dd-cc}) as
\begin{equation} \label{eq:dd-cc-1}
\left(U_{m+1,n} + \lambda \Omega^{\ell}\right) \left(\pa_t U_{m,n} + Q_{m,n}\right) = Q_{m+1,n} \left(U_{m,n} + \lambda \Omega^{\ell}\right),
\end{equation}
which is used to determine $Q_{m,n}$ in terms of $U_{m,n}$, as well as the resulting differential-difference equations.

For the simplest solution, we assume that $Q_{m,n}$ is independent of the spectral parameter. We then collect the different powers of $\lambda$ in equation (\ref{eq:dd-cc-1}):
\begin{subequations}\label{eq:det-eq-Q}
\bea
&& Q_{m,n}U_{m-1,n}\,-\, U_{m,n} \Omega^{-\ell} Q_{m,n}\Omega^{\ell} = 0, \label{eq:det-eq-Q-1} \\[3mm]
&& \pa_{t} U_{m,n} = \Omega^{-\ell} Q_{m+1,n}\Omega^{\ell} \,-\,Q_{m,n}.  \label{eq:det-eq-Q-2}
\eea
\end{subequations}
Clearly, the second equation implies that $lev(Q_{m,n})  = lev(U_{m,n})$ and provides us with a set of $N$ differential-difference equations for the functions $u^{(i)}$, $i \in {\mathbb{Z}}_N$.

We write
\be\label{bold-u}  %
U_{m,n}={\rm{diag}}(u^{(0)}_{m,n},u^{(1)}_{m,n},\dots ,u^{(N-1)}_{m,n}) \Omega^k,\quad
   Q_{m,n}={\rm{diag}}(q^{(0)}_{m,n},q^{(1)}_{m,n},\dots ,q^{(N-1)}_{m,n}) \Omega^k.
\ee  %
Then, equations (\ref{eq:det-eq-Q}) give
\begin{subequations}
\bea  %
&&  q^{(i)}_{m,n}\, u^{(i+k)}_{m-1,n} = u^{(i)}_{m,n}\, q^{(i+k-\ell)}_{m,n},  \label{q-eqs}  \\
&&  \pa_t u^{(i)}_{m,n} = q^{(i-\ell)}_{m+1,n} - q^{(i)}_{m,n}.    \label{eq:gen-eq-sym}
\eea  %
\end{subequations}

\br  %
Notice that (\ref{q-eqs}) implies that
$$
\prod_{i=0}^{N-1} \frac{q^{(i)}_{m,n}}{q^{(i+k-\ell)}_{m,n}} = \prod_{i=0}^{N-1} \frac{u^{(i)}_{m,n}}{u^{(i+k)}_{m-1,n}}.
$$
The left hand side equals $1$, which tells us that $\prod_{i=0}^{N-1} u^{(i)}_{m,n}$ is a constant (in $m$) (compare with (\ref{eq:dLP-coprime-case}), which tells us that it is independent of $n$).
\er  %

To calculate $({\rm{Tr}} (L_{m,n}^{-1}\, Q_{m,n})$, we note that
$$
L_{m,n}^{-1} = \Omega^{-\ell} \left(\lambda I_N+ U \Omega^{-\ell}\right)^{-1}=
    \frac{1}{\lambda^{N}-(-1)^N a}\; \Omega^{-\ell} \left(\lambda^{N-1} I_N-\lambda^{N-2}{\cal D}+\cdots+(-{\cal D})^{N-1}\right),
$$
where ${\cal D} = U \Omega^{-\ell}$.  This follows from the property ${\cal D}^N= a\, I_N$ in the coprime case.
Thus
$$
{\rm{Tr}} \left(L_{m,n}^{-1}Q_{m,n}\right) = \frac{1}{\lambda^{N}-(-1)^N a}\;
   {\rm{Tr}} \left(\Omega^{-\ell}\left(\lambda^{N-1} I_N-\lambda^{N-2}{\cal D}+\cdots+(-{\cal D})^{N-1}\right)Q_{m,n}\right).
$$
Since $(N,k-\ell)=1$, the only diagonal term (level $N(k-\ell)$) is $\Omega^{-\ell}(-{\cal D})^{N-1} Q_{m,n}$, and ${\rm{Tr}}\left(\Omega^{-\ell}(-{\cal D})^{N-1} Q_{m,n}\right) = -\, {\rm{Tr}}\left((-{\cal D})^{N} U_{m,n}^{-1} Q_{m,n}\right)$, so we have
$$
{\rm{Tr}} \left(L_{m,n}^{-1}Q_{m,n}\right) = \frac{a}{a-(-\lambda)^N}\, \sum_{i=0}^{N-1} \frac{q^{(i)}_{m,n}}{u^{(i)}_{m,n}}.
$$
Noting that both $a$ and $\lambda$ are independent of $m$ and that $\det(L_{m,n})=a-(-\lambda)^N$, the conservation law (\ref{dtlogdet}) implies
\be\label{tracecon}
\pa_t (a-(-\lambda)^N) = a\, \Delta_m\, \left(\sum_{i=0}^{N-1} \frac{q^{(i)}_{m,n}}{u^{(i)}_{m,n}}\right).
\ee
Since the left hand side of this is independent of $m$, we have
\be\label{tracecon1}  %
 \sum_{i=0}^{N-1} \frac{q_{m,n}^{(i)}}{u_{m,n}^{(i)}} = \frac{c_0+c_1 m}{a} \quad\mbox{and}\quad \pa_t a = c_1,
\ee  %
where $c_0, c_1$ are constants.

Equations (\ref{q-eqs}) and (\ref{tracecon1}), fully determine the functions $q^{(i)}_{m,n}$ in terms of ${\bf u}_{m,n}$ and ${\bf u}_{m-1,n}$.

\br[Explicit formula for $q^{(i)}_{m,n}$]
It is, in fact, possible to derive a general formula for the functions $q^{(i)}_{m,n}$:
\be\label{q-genform}
q^{(i\delta)}_{m,n}= q^{(0)}_{m,n} \prod_{j=0}^{i-1} \frac{u^{(j \delta+k)}_{m-1,n}}{u^{(j \delta)}_{m,n}}\,,\quad\mbox{where}\;\;
     \delta  =  k - \ell\,,\;\; i =1,\ldots,N-1,
\ee
and where $q^{(0)}_{m,n}$ is given by
$$
\sum_{i=0}^{N-1} \frac{q^{(i)}_{m,n}}{u^{(i)}_{m,n}}\,=\,\frac{1}{a} \quad \Rightarrow \quad
  q^{(0)}_{m,n} = \frac{u^{(0)}_{m,n}}{a} \left(1 + \sum_{i=1}^{N-1} \frac{u^{(0)}_{m,n}}{u^{(i \delta)}_{m,n}}
    \prod_{j=0}^{i-1} \frac{u^{(j \delta+k)}_{m-1,n}}{u^{(j \delta)}_{m,n}} \right)^{-1}.
$$
\er

Thus, we have proven
\begin{Pro} \label{prop:sym-LP}
The system
$$
\Psi_{m+1,n}\,=\,(U_{m,n}+\lambda \Omega^\ell) \Psi_{m,n} \,,\quad \pa_{t} \Psi_{m,n}\,=\,(U_{m,n}+\lambda \Omega^\ell)^{-1}\,Q_{m,n}\,\Psi_{m,n}\,,
$$
where
$$
U_{m,n}={\rm{diag}}(u^{(0)}_{m,n},u^{(1)}_{m,n},\dots ,u^{(N-1)}_{m,n}) \Omega^k,\quad
   Q_{m,n}={\rm{diag}}(q^{(0)}_{m,n},q^{(1)}_{m,n},\dots ,q^{(N-1)}_{m,n}) \Omega^k,
$$
with
$$
q^{(i)}_{m,n}\, u^{(i+k)}_{m-1,n} = u^{(i)}_{m,n}\, q^{(i+k-\ell)}_{m,n} \quad\mbox{and}\quad
   \sum_{i=0}^{N-1} \frac{q^{(i)}_{m,n}}{u^{(i)}_{m,n}}\,=\,\frac{1}{a},
$$
is compatible if and only if $u^{(i)}_{m,n}$ satisfies
\be
\pa_t u^{(i)}_{m,n} = q^{(i-\ell)}_{m+1,n} - q^{(i)}_{m,n}.   \label{eq:gen-eq-sym1}
\ee
This leads to the (differential-difference) {\em local conservation law}
\be
\pa_t \left(\sum_{i=0}^{N-1}u^{(i)}_{m,n}\right) =  \Delta_m\, \left(\sum_{i=0}^{N-1}q^{(i)}_{m,n}\right).  \label{eq:sym-gen-cl}
\ee
\end{Pro}
An equivalent statement can be made for the case $c_0=0, c_1=1$, corresponding to the flow
\be\label{eq:gen-eq-msym1}
\pa_\tau u^{(i)}_{m,n} = (m+1)q^{(i-\ell)}_{m+1,n} - mq^{(i)}_{m,n},\quad\mbox{with}\quad \pa_\tau a = 1,
\ee
by making the replacement $q^{(i)}_{m,n}\rightarrow m q^{(i)}_{m,n}$ in the formulae.
We will see in Section \ref{master} that this defines a {\em master symmetry} for a hierarchy of autonomous flows, the first of which is (\ref{eq:gen-eq-sym1}).

\subsubsection{All Inequivalent Cases for $N=2$ and $N=3$}\label{SymmetriesN=2-3}

Up to equivalence defined by
$$
{\cal{T}} : (a,b) \mapsto (N-a,N-b),
$$
we list all semi-discrete Lax pairs in two and three dimensions. In the following lists, we present only the entries of matrices $Q$, for the autonomous case, and the corresponding differential-difference equations.  To obtain the non-autonomous solution, with $c_0=0, c_1=1$, we just multiply the formula for $q^{(i)}_{m,n}$ by $m$, obtaining the $\tau-$flow (\ref{eq:gen-eq-msym1}). Lax pairs can easily be constructed using the formulae of Section \ref{iso-m} with the given choices for $N$, $k$ and $\ell$.

\bex[$N=2$ : Level structure $(k,\ell) = (0,1)$]  {\em

Entries of matrix $Q_{m,n}$ are
$$
q^{(0)}_{m,n}\,=\, \frac{1}{u^{(0)}_{m-1,n} + u^{(1)}_{m,n}}\,,\quad
q^{(1)}_{m,n} \,=\,\frac{u^{(0)}_{m-1,n}}{u^{(0)}_{m,n} \left( u^{(0)}_{m-1,n} + u^{(1)}_{m,n}\right)}
$$
The corresponding differential-difference equations are
\be\label{eq:2D-sym-01}
\pa_t u^{(0)}_{m,n}\,=\, q^{(1)}_{m+1,n}-q^{(0)}_{m,n}\,,\quad \pa_t u^{(1)}_{m,n}\,=\, q^{(0)}_{m+1,n}-q^{(1)}_{m,n}\,.
\ee
}\eex

\bex[$N=2$ : Level structure $(k,\ell) = (1,0)$]  {\em  %

Entries of matrix $Q_{m,n}$ are
\begin{subequations}\label{eq:2D-sym-10}
\be
q^{(0)}_{m,n}\,=\, \frac{1}{u^{(1)}_{m-1,n} + u^{(1)}_{m,n}}\,,\quad
q^{(1)}_{m,n} \,=\,\frac{u^{(1)}_{m-1,n}}{u^{(0)}_{m,n} \left( u^{(1)}_{m-1,n} + u^{(1)}_{m,n}\right)}
\ee
The corresponding differential-difference equations are
\be
\pa_t u^{(0)}_{m,n}\,=\, q^{(0)}_{m+1,n}-q^{(0)}_{m,n}\,,\quad \pa_t u^{(1)}_{m,n}\,=\, q^{(1)}_{m+1,n}-q^{(1)}_{m,n}\,.
\ee
\end{subequations}
}\eex  %

\bex[$N=3$ : Level structure $(k,\ell) = (0,1)$]  {\em  %

Entries of matrix $Q_{m,n}$ are
\begin{subequations}\label{eq:3D-sym-01}
\be
q^{(0)}_{m,n}=\frac{u^{(1)}_{m-1,n}}{u^{(1)}_{m,n}}\, q^{(1)}_{m,n},\qquad  q^{(1)}_{m,n}=\frac{1}{\Gamma},\qquad
    q^{(2)}_{m,n}=\frac{u^{(0)}_{m-1,n}u^{(1)}_{m-1,n}}{u^{(0)}_{m,n}u^{(1)}_{m,n}}\, q^{(1)}_{m,n} ,
\ee
where $\Gamma = u^{(0)}_{m-1,n} u^{(1)}_{m-1,n} +  u^{(0)}_{m,n} u^{(2)}_{m,n} + u^{(1)}_{m-1,n} u^{(2)}_{m,n}$.
The corresponding differential-difference equations are
\be
\pa_t u^{(0)}_{m,n}\,=\, q^{(2)}_{m+1,n}-q^{(0)}_{m,n}\,,\quad \pa_t u^{(1)}_{m,n}\,=\, q^{(0)}_{m+1,n}-q^{(1)}_{m,n}\,,\quad
            \pa_t u^{(2)}_{m,n}\,=\, q^{(1)}_{m+1,n}-q^{(2)}_{m,n} \,.
\ee
\end{subequations}
}\eex  %

\bex[$N=3$ : Level structure $(k,\ell) = (1,0)$]  {\em  %

Entries of matrix $Q_{m,n}$ are
\begin{subequations}\label{eq:3D-sym-10}
\be%\label{eq:3D-sym-10}
q^{(0)}_{m,n}=\frac{1}{\Gamma},\qquad  q^{(1)}_{m,n}=\frac{u^{(1)}_{m-1,n}}{u^{(0)}_{m,n}}\, q^{(0)}_{m,n},\qquad
    q^{(2)}_{m,n}=\frac{u^{(1)}_{m-1,n}u^{(2)}_{m-1,n}}{u^{(0)}_{m,n}u^{(1)}_{m,n}}\, q^{(0)}_{m,n} ,
\ee
where $\Gamma = u^{(1)}_{m,n} u^{(2)}_{m,n} +  u^{(1)}_{m-1,n} u^{(2)}_{m-1,n} + u^{(1)}_{m-1,n} u^{(2)}_{m,n}$.
The corresponding differential-difference equations are
\be
\pa_t u^{(0)}_{m,n}\,=\, q^{(0)}_{m+1,n}-q^{(0)}_{m,n}\,,\quad \pa_t u^{(1)}_{m,n}\,=\, q^{(1)}_{m+1,n}-q^{(1)}_{m,n}\,,\quad
                     \pa_t u^{(2)}_{m,n}\,=\, q^{(2)}_{m+1,n}-q^{(2)}_{m,n} \,.
\ee
\end{subequations}
}\eex  %

\bex[$N=3$ : Level structure $(k,\ell) =  (1,2)$]  {\em  %

Entries of matrix $Q_{m,n}$ are
\begin{subequations}\label{eq:3D-sym-12}
\be
q^{(0)}_{m,n}=\frac{u^{(2)}_{m-1,n}}{u^{(1)}_{m,n}}\, q^{(1)}_{m,n},\qquad  q^{(1)}_{m,n}=\frac{1}{\Gamma},\qquad
    q^{(2)}_{m,n}=\frac{u^{(1)}_{m-1,n}u^{(2)}_{m-1,n}}{u^{(0)}_{m,n}u^{(1)}_{m,n}}\, q^{(1)}_{m,n} ,
\ee
where $\Gamma = u^{(1)}_{m-1,n} u^{(2)}_{m-1,n} +  u^{(0)}_{m,n} u^{(2)}_{m,n} + u^{(2)}_{m-1,n} u^{(2)}_{m,n}$.
The corresponding differential-difference equations are
\be
\pa_t u^{(0)}_{m,n}\,=\, q^{(1)}_{m+1,n}-q^{(0)}_{m,n}\,,\quad \pa_t u^{(1)}_{m,n}\,=\, q^{(2)}_{m+1,n}-q^{(1)}_{m,n}\,,\quad
                \pa_t u^{(2)}_{m,n}\,=\, q^{(0)}_{m+1,n}-q^{(2)}_{m,n} \,.
\ee
\end{subequations}
}\eex  %

\subsection{Hierarchies of Commuting Flows}\label{master}

The flow (\ref{eq:gen-eq-sym1}) is just the first of an infinite hierarchy of (autonomous) isospectral flows.  As is often the case, these can be constructed with the use of a {\em master symmetry}.  The results of this section are summarised in Proposition \ref{master-prop}.

We denote this first autonomous vector field by $X^1$:
$$
X^1 = \sum_{i=-\infty}^{\infty}\sum_{j=0}^{N-1} X^{1j}_{m+i,n} \pa_{u^{(j)}_{m+i,n}},
$$
where this infinite sum is the {\em formal prolongation} of the components $X^{1j}_{m,n}=q^{(j-\ell)}_{m+1}-q^{(j)}_m$ to the shifted variables.  Acting on functions of a {\em finite} number of shifts, this sum is finite, so well defined.  We denote the corresponding $t-$parameter as $t^1$.

\bd[Master Symmetry]
A vector field $X^M$, given by
$$
X^M = \sum_{i=-\infty}^{\infty}\sum_{j=0}^{N-1} X^{Mj}_{m+i,n} \pa_{u^{(j)}_{m+i,n}},
$$
for some functions $X^{Mj}_{m,n}$ is said to be a {\em master symmetry} of $X^1$ if
$$
[[X^M,X^1],X^1]=0, \quad\mbox{whilst}\;\; [X^M,X^1]\neq 0.
$$
We then define $X^k$ recursively by $X^{k+1}=[X^M,X^k]$.
\ed
\begin{Pro}
Given the sequence of vector fields $X^k$, defined above, we suppose that, for some $\ell\ge 2$, $\{X^1,\dots ,X^\ell\}$ pairwise commute.  Then $[X^i,X^{\ell+1}]=0$, for $1\leq i\leq \ell-1$.
\end{Pro}
This follows from an application of the Jacobi identity.
\br
We \underline{cannot} deduce that $[[X^M,X^\ell],X^\ell]=0$ by using the Jacobi identity.  Since we are \underline{given} this equality for $\ell=2$, we \underline{can} deduce that $[X^1,X^3]=0$ (see the discussion around Theorem 19 of \cite{Y}).  Nevertheless it \underline{is} possible to check this by hand for low values of $\ell$, for all the examples given in this paper.
\er

Since some of the ABS equations fall into our general class, we can generalise known results \cite{X} on master symmetries.  We thus consider two vector fields defined by:
\begin{enumerate}
\item  $S^1_{m,n}$, corresponding to the vector field $X^1$, defined by (\ref{eq:gen-eq-sym1}).
\item  $S^M_{m,n}$, corresponding to the vector field $X^M$, defined by (\ref{eq:gen-eq-msym1}).
\end{enumerate}
Note that $S^M_{m,n}=mS^1_{m,n}$.  Then
\be\label{SM}
\pa_{\tau} L_{m,n} =  S^M_{m+1,n} L_{m,n} - L_{m,n}S^M_{m,n} \quad\Rightarrow\quad
   \pa_{\tau} u^{(i)}_{m,n} = (m+1) q^{(i-\ell)}_{m+1,n} - m q^{(i)}_{m,n},
\ee
with $q^{(i)}_{m,n}$ defined as in the autonomous case.  This defines the components of the vector field $X^M$.  It can be checked that $[X^M,X^1]=X^2$ is nonzero and that $[X^1,X^2]=0$, so $X^M$ defines a master symmetry for $X^1$.

Consider the compatibility of the two equations
\be\label{X1XM}
X^1 \psi_{m,n} = S^1_{m,n}\psi_{m,n}, \quad\mbox{and}\quad      X^M \psi_{m,n} = S^M_{m,n}\psi_{m,n}.
\ee
Since $X^1$ and $X^M$ do not commute, we cannot consider $\psi_{m,n}$ as being {\em simultaneously} dependent on both $t^1$ and $\tau$.  The compatibility condition for these equations is
$$
X^2\psi_{m,n} = [X^M,X^1]\psi_{m,n} = (X^MS^1_{m,n} - X^1S^M_{m,n} + [S^1_{m,n},S^M_{m,n}])\psi_{m,n}= S^2_{m,n}\psi_{m,n},
$$
defining the next flow.
\bd[The $k^{th}$ flow]
We can recursively define the $k^{th}$ flow by
\be\label{Sk}
X^k\psi_{m,n} = S^k_{m,n}\psi_{m,n}, \quad\mbox{where}\quad
      S^k_{m,n} = X^MS^{k-1}_{m,n} - X^{k-1}S^M_{m,n} + [S^{k-1}_{m,n},S^M_{m,n}],\;\; k\geq 2.
\ee
\ed
\br[Assuming commutativity]
We know that $[X^1,X^2]=[X^1,X^3]=0$, but for the next calculation \underline{we assume} that $[X^i,X^j]=0$, for all $i,j\geq 1$.  This means that we have the {\em zero curvature} conditions
\be\label{zerocurv}
X^jS^i_{m,n}-X^iS^j_{m,n}+[S^i_{m,n},S^j_{m,n}]=0.
\ee
\er
Recalling that $S^M_{m,n}=mS^1_{m,n}$, the formula for $S^k_{m,n}$, using (\ref{zerocurv}), then gives
$$
S^k_{m,n} = X^MS^{k-1}_{m,n}-m (X^{k-1}S^1_{m,n} - [S^{k-1}_{m,n},S^1_{m,n}]) = X^M S^{k-1}_{m,n} - m X^1 S^{k-1}_{m,n} = {\cal{R}} S^{k-1}_{m,n},
$$
where (using (\ref{eq:gen-eq-sym}) and (\ref{SM}))
\be\label{R}
{\cal{R}} =  \sum_{i=-\infty}^{\infty} \sum_{j=0}^{N-1} \left((i+1) q^{(j-\ell)}_{m+i+1,n} - i q^{(j)}_{m+i,n}\right) \pa_{u^{(j)}_{m+i,n}}.
\ee
We now define a sequence of matrices $Q^k_{m,n}$, with $S_{m,n}^k=L_{m,n}^{-1}Q_{m,n}^k$, where $Q_{m,n}^1=Q_{m,n}$ (of Proposition \ref{prop:sym-LP}). We then have
$$
{\cal{R}} S^k_{m,n} = L_{m,n}^{-1} \left({\cal{R}} Q^k_{m,n} - ({\cal{R}} L_{m,n})L_{m,n}^{-1}Q_{m,n}^k\right),
$$
thus giving the recursion
\be\label{RQk}
Q^{k+1}_{m,n} = {\cal{R}} Q^k_{m,n} - ({\cal{R}} L_{m,n})L_{m,n}^{-1}Q_{m,n}^k =
                   {\cal{R}} Q^k_{m,n} - \Omega^{-\ell}Q_{m+1,n}^1 \Omega^{\ell} L_{m,n}^{-1}Q_{m,n}^k.
\ee
Starting with $Q^1_{m,n}$, which is independent of $\lambda$, we find
$$
Q^{2}_{m,n} = {\cal{R}} Q^1_{m,n} - \Omega^{-\ell}Q_{m+1,n}^1 \Omega^{\ell} L_{m,n}^{-1}Q_{m,n}^1
        = {\cal{R}} Q^1_{m,n} + \mbox{$\lambda$ dependent terms}.
$$
which is the sum of two parts, the first of which is independent of $\lambda$, whilst the second is rational in $\lambda$, with $\det L_{m,n}$ in the denominator.  Using the recursion, we find
\be\label{Qk}
Q^{k}_{m,n} = {\cal{R}}^{k-1} Q^1_{m,n} + \mbox{$\lambda$ dependent terms}.
\ee
Again, the first part is independent of $\lambda$ and the second a rational function (with $(\det L_{m,n})^{k-1}$ in the denominator).

The calculation of the evolution $\pa_{t^k} U_{m,n}$ leads to an analogous formula to (\ref{eq:dd-cc-1}):
\be \label{eq:dd-cc-k}
\left(U_{m+1,n} + \lambda \Omega^{\ell}\right) \left(\pa_{t^k} U_{m,n} + Q^k_{m,n}\right) = Q^k_{m+1,n} \left(U_{m,n} + \lambda \Omega^{\ell}\right).
\ee
It follows from the structure of $Q^k_{m,n}$ that
$$
\pa_{t^k} U_{m,n} =  \Omega^{-\ell} {\cal{S}}_m \left({\cal{R}}^{k-1} Q^1_{m,n}\right)\Omega^{\ell} - {\cal{R}}^{k-1} Q^1_{m,n}.
$$
We summarise these results in:

\begin{Pro}\label{master-prop}
The master symmetry
$$
\pa_\tau u^{(i)}_{m,n} = (m+1)q^{(i-\ell)}_{m+1,n} - mq^{(i)}_{m,n}
$$
generates the hierarchy of symmetries
\be \label{utk}
\pa_{t^k} u^{(i)}_{m,n}\,=\,{\cal{S}}_m \left( {\cal{R}}^{k-1}\left(q^{(i-\ell)}_{m,n}\right)\right)\,-\,{\cal{R}}^{k-1}\left(q^{(i)}_{m,n}\right),
\ee
where
$$
{\cal{R}} =  \sum_{i=-\infty}^{\infty} \sum_{j=0}^{N-1} \left((i+1) q^{(j-\ell)}_{m+i+1,n} - i q^{(j)}_{m+i,n}\right) \pa_{u^{(j)}_{m+i,n}},
$$
and $\pa_{t^1}u^{(i)}_{m,n}$ is given by (\ref{eq:gen-eq-sym1}).
Equation (\ref{utk}) is the compatibility condition of
$$
\Psi_{m+1,n}\,=\,L_{m,n} \Psi_{m,n} \,,\quad \pa_{t^k} \Psi_{m,n}\,=\,S_{m,n}^k\,\Psi_{m,n},
$$
where $S_{m,n}^k$ is defined recursively by $S_{m,n}^k={\cal{R}} S^{k-1}_{m,n}$.
\end{Pro}

\subsection{Symmetries of the Difference Equations}\label{symmetries}

In Sections \ref{iso-m} and \ref{master} we considered the compatibility of a discrete shift in the $m-$direction and a continuous evolution in $t$, given by (\ref{eq:dLP-cont-m}).  The simplest case is described in Proposition \ref{prop:sym-LP}.  The master symmetry of Section \ref{master} generates an infinite family of commuting symmetries.

The compatibility of discrete shifts in both $m-$ and $n-$directions (equations (\ref{eq:dLP-gen})) leads to our fully discrete system
(\ref{eq:dLP-gen-scc}) (written explicitly as (\ref{eq:dLP-ex-cc}) or (\ref{eq:dLP-ex-cc-s})).  We now consider the flows (\ref{utk}) as symmetries of this discrete system.

\subsubsection*{The Evolution $\pa_t\, v^{(i)}_{m,n}$}

For any time evolution (\ref{eq:dLP-cont-m}), the compatibility of
\be\label{eq:dMP-cont}  %
\Psi_{m,n+1} = M_{m,n} \Psi_{m,n},\quad \pa_{t} \Psi_{m,n} = S_{m,n}\,\Psi_{m,n},
\ee  %
with $M_{m,n}$ defined by (\ref{eq:dLP-gen-M}), gives the equation
\be\label{dtM}
\pa_t\, M_{m,n} = S_{m,n+1}\, M_{m,n} -M_{m,n}\, S_{m,n}.
\ee
Here $S_{m,n}=L_{m,n}^{-1}Q_{m,n}$, where $Q_{m,n}$ and $U_{m,n}$ must satisfy (\ref{eq:dd-cc-1}).  Using (\ref{eq:dd-cc}) and (\ref{dtM}), we then have
\bea
\pa_t (L_{m,n+1} M_{m,n} - M_{m+1,n} L_{m,n}) &=& S_{m+1,n+1} (L_{m,n+1} M_{m,n} - M_{m+1,n} L_{m,n}) \nn\\
    && \hspace{2cm} - (L_{m,n+1} M_{m,n} - M_{m+1,n} L_{m,n}) S_{m,n},  \nn
\eea
showing compatibility on solutions of the fully discrete system (\ref{eq:dLP-gen-scc}).

For the flow given in Proposition \ref{prop:sym-LP}, we find
$$
L_{m,n+1} \pa_t M_{m,n} = Q_{m,n+1}\,M_{m,n}-L_{m,n+1}\,M_{m,n}\,L_{m,n}^{-1}\,Q_{m,n} = Q_{m,n+1}\,M_{m,n}-M_{m+1,n}\,Q_{m,n},
$$
where we used the difference equation (\ref{eq:dLP-gen-cc}).

The explicit forms of $L$ and $M$ then lead to
\bea  %
  \pa_t V_{m,n} &=& \Omega^{-\ell_1} Q_{m,n+1}\Omega^{\ell_2}-\Omega^{\ell_2-\ell_1}Q_{m,n} , \nn\\
  U_{m,n+1}\, \pa_t V_{m,n} &=& Q_{m,n+1}\,V_{m,n}-V_{m+1,n}\,Q_{m,n}.  \nn
\eea  %
The first of these is just the $t-$evolution of $V_{m,n}$, which, in components, is just
\be\label{dtvi}
\pa_t v^{(i)}_{m,n}\,=\,q^{(i-\ell_1)}_{m,n+1}\,-\,q^{(i+\ell_2-\ell_1)}_{m,n},
\ee
whilst the second, using (\ref{eq:dLP-gen-scc-2}) to eliminate $V_{m+1,n}$, leads to
\be\label{vt-constraint}
\Omega^{\ell_1} V_{m,n} \Omega^{-\ell_1} Q_{m,n} - Q_{m,n+1}V_{m,n} = \Omega^{\ell_2}U_{m,n}\Omega^{-\ell_1}Q_{m,n}-U_{m,n+1}\Omega^{-\ell_1}Q_{m,n+1}\Omega^{\ell_2}.
\ee
\br
Each expression in this equation has level $k_1+k_2$ (requiring the condition (\ref{eq:dLP-nec-rel})).  Since $U,\, V$ and $Q$ are {\em known quantities}, this looks like an additional constraint, but it can be shown that this holds identically as a consequence of previous equations.
\er
Similar results can be calculated for the $\tau-$flow (\ref{eq:gen-eq-msym1}) (see Proposition \ref{prop:difdif-sym-dif}).
We can extend this discussion to the flows introduced in Section \ref{master}.  The analogous formula to (\ref{utk}) is
\be\label{vitk}
\pa_{t^k} v^{(i)}_{m,n}\,=\, {\cal{S}}_n \left(
                  {\cal{R}}^{k-1}\left(q^{(i-\ell_1)}_{m,n}\right)\right)\,-\,{\cal{R}}^{k-1}\left(q^{(i+\ell_2-\ell_1)}_{m,n}\right).
\ee
The remaining parts of (\ref{dtM}) would give several conditions analogous to (\ref{vt-constraint}).  We \underline{conjecture} that these hold identically, but have no general proof.  For all specific examples calculated, this is the case.

\subsubsection*{Symmetries in the $n-$Direction}

This whole structure can be repeated for continuous flows in the $n-$direction:
\be\label{dsPsi}
\pa_s \Psi_{m,n} = (V_{m,n}+\lambda \Omega^{\ell_2})^{-1} R_{m,n} \Psi_{m,n},
\ee
with the simplest choice being that $R_{m,n}$ is $\lambda-$independent.  The analogous formulae to (\ref{eq:det-eq-Q}) are
$$
R_{m,n}V_{m,n-1}-V_{m,n} \Omega^{-\ell_2} R_{m,n}\Omega^{\ell_2}=0 \quad\mbox{and}\quad
       \pa_s V_{m,n}= \Omega^{-\ell_2} R_{m,n+1}\Omega^{\ell_2}-R_{m,n+1}.
$$
The results following from the mutual compatibility of the four linear equations are summarised in the proposition below.

\begin{Pro}\label{prop:difdif-sym-dif}
Let $\left(k_1,\ell_1 \, ;\, k_2,\ell_2 \right) \in {\cal{Q}}_N$ with $(N,k_1-\ell_1) = (N,k_2-\ell_2) =1$, and consider the system of equations
\begin{subequations}\label{eq:all-LP}
\begin{eqnarray}
\Psi_{m+1,n} &=& \left(U_{m,n}+\lambda \Omega^{\ell_1}\right)\Psi_{m,n}\,,\label{eq:all-LP-m}\\
\Psi_{m,n+1} &=& \left(V_{m,n}+\lambda \Omega^{\ell_2}\right)\Psi_{m,n}\,,\label{eq:all-LP-n}\\
\partial_t \Psi_{m,n}&=& \left(U_{m,n}+\lambda \Omega^{\ell_1}\right)^{-1} Q_{m,n} \Psi_{m,n}\,,\label{eq:all-LP-t}\\
\partial_s \Psi_{m,n} &=& \left(V_{m,n}+\lambda \Omega^{\ell_2}\right)^{-1} R_{m,n} \Psi_{m,n}\,,\label{eq:all-LP-s}
\end{eqnarray}
\end{subequations}
where
\begin{eqnarray}
&& U_{m,n} = {\rm{diag}}\left(u^{(0)}_{m,n},\cdots,u^{(N-1)}_{m,n}\right) \Omega^{k_1}, \quad V_{m,n} = {\rm{diag}}\left(v^{(0)}_{m,n},\cdots,v^{(N-1)}_{m,n}\right) \Omega^{k_2}\,,\\
&& Q_{m,n}\,=\, {\rm{diag}}\left(q^{(0)}_{m,n},\cdots,q^{(N-1)}_{m,n}\right) \Omega^{k_1},\quad R_{m,n}\,=\, {\rm{diag}}\left(r^{(0)}_{m,n},\cdots,r^{(N-1)}_{m,n}\right) \Omega^{k_2}.
\end{eqnarray}

\noindent
Then, the system of difference equations (\ref{eq:dLP-ex-cc}) follows from the compatibility condition of equations (\ref{eq:all-LP-m}) and (\ref{eq:all-LP-n}).

\noindent
The differential-difference equations
\begin{subequations}\label{eq:LP-sym-gen-m-ma}
\bea
 \pa_t u^{(i)}_{m,n} = q^{(i-\ell_1)}_{m+1,n}\,-\,q^{(i)}_{m,n}\,,&&
\pa_t v^{(i)}_{m,n} = q^{(i-\ell_1)}_{m,n+1}\,-\,q^{(i+\ell_2-\ell_1)}_{m,n}\,,   \label{eq:LP-sym-gen-m}  \\
       \pa_{\tau} u^{(i)}_{m,n} = (m+1)q^{(i-\ell_1)}_{m+1,n} - mq^{(i)}_{m,n}\,,&&
\pa_{\tau} v^{(i)}_{m,n} =  m(q^{(i-\ell_1)}_{m,n+1}-q^{(i+\ell_2-\ell_1)}_{m,n})\,,   \label{eq:LP-sym-gen-mma}
\eea
\end{subequations}
where $i \in {\mathbb{Z}}_N$ and the functions $q^{(i)}_{m,n}$ are solutions of
$$
q^{(i)}_{m,n}\, u^{(i+k_1)}_{m-1,n} = u^{(i)}_{m,n}\, q^{(i+k_1-\ell_1)}_{m,n} \quad\mbox{and}\quad
   \sum_{i=0}^{N-1} \frac{q_{m,n}^{(i)}}{u_{m,n}^{(i)}} = \frac{1}{a},
$$
follow from  the compatibility condition of equation (\ref{eq:all-LP-t}) with (\ref{eq:all-LP-m}) and (\ref{eq:all-LP-n}), respectively, and define symmetries of system (\ref{eq:dLP-ex-cc}) in the $m$ direction.  The $\tau-$flow satisfies $\pa_\tau a = 1, \,  \pa_\tau b = 0$.

\noindent
The differential-difference equations
\begin{subequations}\label{eq:LP-sym-gen-n-ma}
\bea
  \pa_s u^{(i)}_{m,n}\,=\,r^{(i-\ell_2)}_{m+1,n}\,-\,r^{(i+\ell_1-\ell_2)}_{m,n}\,,&&
\pa_s v^{(i)}_{m,n}\,=\, r^{(i-\ell_2)}_{m,n+1}\,-\,r^{(i)}_{m,n} \,,  \label{eq:LP-sym-gen-n}  \\
   \pa_{\sigma} u^{(i)}_{m,n} = n(r^{(i-\ell_2)}_{m+1,n} - r^{(i+\ell_1-\ell_2)}_{m,n})\,,&&
\pa_{\sigma} v^{(i)}_{m,n} = (n+1)r^{(i-\ell_2)}_{m,n+1} - n r^{(i)}_{m,n} \,,  \label{eq:LP-sym-gen-nma}
\eea
\end{subequations}
where $i \in {\mathbb{Z}}_N$ and the functions $r^{(i)}_{m,n}$ are solutions of
$$
r^{(i)}_{m,n}\, v^{(i+k_2)}_{m,n-1} = v^{(i)}_{m,n}\, r^{(i+k_2-\ell_2)}_{m,n} \quad\mbox{and}\quad
   \sum_{i=0}^{N-1} \frac{r_{m,n}^{(i)}}{v_{m,n}^{(i)}} = \frac{1}{b},
$$
follow from  the compatibility condition of equation (\ref{eq:all-LP-s}) with (\ref{eq:all-LP-m}) and (\ref{eq:all-LP-n}), respectively, and define symmetries of system (\ref{eq:dLP-ex-cc}) in the $n$ direction.  The $\sigma-$flow satisfies $\pa_\sigma a = 0, \,  \pa_\sigma b = 1$.
\end{Pro}

\br[Master Symmetries]
The non-autonomous flows (\ref{eq:LP-sym-gen-mma}) and (\ref{eq:LP-sym-gen-nma}) are master symmetries for their respective hierarchies.
\er

\subsection{Symmetries in Potentials Variables}

We can explicitly write these symmetries in terms of the potentials introduced in Section \ref{sect:coprime}.

\subsubsection{Quotient Potentials}

Substituting $u^{(i)}_{m,n}=\alpha \frac{\phi^{(i)}_{m+1,n}}{\phi_{m,n}^{(i+k_1)}}$ (see section \ref{sect:coprime-quotient}) into the first of (\ref{eq:LP-sym-gen-m}), we obtain
$$
\alpha\, \frac{\phi^{(i)}_{m+1,n}}{\phi_{m,n}^{(i+k_1)}}\,\left(\frac{\pa_t \phi^{(i)}_{m+1,n}}{\phi^{(i)}_{m+1,n}} -
    \frac{\pa_t \phi^{(i+k_1)}_{m,n}}{\phi^{(i+k_1)}_{m,n}}\right)= q^{(i-\ell_1)}_{m+1,n} - q^{(i)}_{m,n}.
$$
Using the potential form of (\ref{q-eqs}),
$$
\frac{\phi_{m,n}^{(i+k_1)}}{\phi^{(i)}_{m+1,n}}\, q^{(i)}_{m,n} =
                   \frac{\phi^{(i+2k_1)}_{m-1,n}}{\phi_{m,n}^{(i+k_1)}}\, q^{(i+k_1-\ell_1)}_{m,n},
$$
we can write this as
\be\label{uit-phi}
\alpha \, \frac{\pa_t \phi^{(i+k_1)}_{m,n}}{\phi^{(i+k_1)}_{m,n}} - \frac{\phi_{m-1,n}^{(i+2k_1)}}{\phi^{(i+k_1)}_{m,n}}\, q^{(i+k_1-\ell_1)}_{m,n} =
    \alpha \, \frac{\pa_t \phi^{(i)}_{m+1,n}}{\phi^{(i)}_{m+1,n}} - \frac{\phi_{m,n}^{(i+k_1)}}{\phi^{(i)}_{m+1,n}}\, q^{(i-\ell_1)}_{m+1,n}.
\ee
Since the right side of (\ref{uit-phi}) is just the left side with $(m,i)\mapsto (m+1,i-k_1)$, we have
\be\label{phit-c}
\frac{\pa_t \phi^{(i)}_{m,n}}{\phi^{(i)}_{m,n}}= \frac{\phi^{(i+k_1)}_{m-1,n}}{\alpha\phi^{(i)}_{m,n}}\, q^{(i-\ell_1)}_{m,n} +c,
\ee
where $c$ is a constant, to be determined.  Since
$$
\pa_t \left(\log \prod_{i=0}^{N-1}\phi^{(i)}_{m,n}\right) =  \sum_{i=0}^{N-1}\left(
     \frac{\phi^{(i+k_1)}_{m-1,n}}{\alpha\phi^{(i)}_{m,n}}\, q^{(i-\ell_1)}_{m,n}\right) + c N = 0,
$$
then
$$
c N = - \sum_{i=0}^{N-1} \frac{q^{(i-\ell_1)}_{m,n}}{u_{m-1,n}^{(i)}}= - \sum_{i=0}^{N-1} \frac{q^{(i-k_1)}_{m,n}}{u_{m,n}^{(i-k_1)}}
                = - \, \frac{1}{\alpha^{N}}.
$$
Carrying out a similar calculation of the $s-$symmetry (\ref{eq:LP-sym-gen-n}), we obtain the formulae
\begin{subequations}\label{eq:phi-sys-sym}
\bea
\pa_t\,\phi^{(i)}_{m,n} &=& \alpha^{-1} \,q^{(i-\ell_1)}_{m,n}\, \phi_{m-1,n}^{(i+k_1)}
                      -\frac{\phi^{(i)}_{m,n}}{N\alpha^{N}}\,,  \label{eq:phi-sys-sym-1}   \\
\pa_s\,\phi^{(i)}_{m,n} &=& \beta^{-1} \,q^{(i-\ell_2)}_{m,n}\, \phi_{m,n-1}^{(i+k_2)}
                                 -\frac{\phi^{(i)}_{m,n}}{N\beta^{N}}.  \label{eq:phi-sys-sym-2}
\eea
\end{subequations}

\br
These choices of constants have made the vector fields {\em tangent} to the level surfaces $\prod_{i=0}^{N-1} \phi^{(i)}_{m,n}=\mbox{constant}$, so these symmetries survive the reduction to $N-1$ components, which we always make in our examples.
\er

The {\em master symmetries} are calculated in the same way, but we must take into account that $\pa_\tau \alpha = 1/(N \alpha^{N-1})$ and $\pa_\sigma \beta = 1/(N \beta^{N-1})$, which implies that the additional constants must depend upon $m$ and $n$ respectively, giving
\begin{subequations}\label{eq:phi-sys-msym}
\bea
\pa_{\tau}\,\phi^{(i)}_{m,n} &=& m\,\alpha^{-1} \,q^{(i-\ell_1)}_{m,n}\, \phi_{m-1,n}^{(i+k_1)}
                                     -\frac{m\,\phi^{(i)}_{m,n}}{N\alpha^{N}}\,,  \label{dtauphim} \\
\pa_{\sigma}\,\phi^{(i)}_{m,n} &=& n\, \beta^{-1} \,q^{(i-\ell_2)}_{m,n}\, \phi_{m,n-1}^{(i+k_2)}
                                    -\frac{n\,\phi^{(i)}_{m,n}}{N\beta^{N}}.  \label{dsigmaphim}
\eea
\end{subequations}

\bex[$(k_1,\ell_1;k_2,\ell_2)=(1,2;1,2)$]  {\em  %

We use the variables of Remark \ref{rem:rat-pot-1}.  The system (\ref{eq:3D-1212}) admits two point symmetries generated by
$$
\pa_\epsilon \phi^{(0)}_{m,n} = \omega^{n+m} \phi^{(0)}_{m,n},\;\; \pa_\epsilon \phi^{(1)}_{m,n} = 0 \quad\mbox{and}\quad
       \pa_\eta \phi^{(0)}_{m,n} = 0, \;\; \pa_\eta \phi^{(1)}_{m,n} = \omega^{n+m} \phi^{(1)}_{m,n},
$$
where $\omega^2+\omega+1=0$.  Written in terms of these potentials, the symmetries (\ref{eq:3D-sym-12}) take the form
\be\label{eq:3D-1212-sym-1}
 \pa_{t^1} \phi^{(0)}_{m,n} = -\,\frac{\phi^{(0)}_{m,n}}{\alpha^3}\, \left(\frac{\gamma^{(1)}_{m,n}}{{\cal{F}}_{m,n}}-\frac{1}{3}\right),
        \quad
 \pa_{t^1} \phi^{(1)}_{m,n} =  \frac{\phi^{(1)}_{m,n}}{\alpha^3}\, \left(\frac{\gamma^{(0)}_{m,n}}{{\cal{F}}_{m,n}}-\frac{1}{3}\right),
\ee
where $\gamma^{(i)}_{m,n}=\phi^{(i)}_{m+1,n}\phi^{(i)}_{m,n} \phi^{(i)}_{m-1,n}$ and ${\cal{F}}_{m,n}=1+\gamma^{(0)}_{m,n}+\gamma^{(1)}_{m,n}$.

The local master symmetry
$$
\pa_\tau \phi^{(0)}_{m,n} \,=\, m\, \pa_{t^1} \phi^{(0)}_{m,n}\,,\quad \pa_\tau \phi^{(1)}_{m,n} \,=\, m\, \pa_{t^1} \phi^{(1)}_{m,n}\,,\quad
             \pa_\tau \alpha \,=\,\frac{1}{3\alpha^2},
$$
allows us to construct a hierarchy of symmetries of system (\ref{eq:3D-1212}) in the $m$-direction. The formula $[X^M,X^1]=X^2$ gives a linear combination of a genuinely new symmetry and $X^1$.  The ``new'' part of $X^2$ is given by
\bea
\pa_{t^2} \phi^{(0)}_{m,n} =  \frac{\phi^{(0)}_{m,n}}{\alpha^6}\,
   \frac{\gamma^{(1)}_{m,n}}{{\cal{F}}_{m,n}} \left(
      {\cal{S}}_m+1\right)\left(\frac{\Delta_m\,\left(\gamma^{(0)}_{m-1,n}\right)}{{\cal{F}}_{m,n} {\cal{F}}_{m-1,n}} \right), \nn\\
\pa_{t^2} \phi^{(1)}_{m,n} = \frac{\phi^{(1)}_{m,n}}{\alpha^6}\, \frac{\gamma^{(0)}_{m,n}}{{\cal{F}}_{m,n}} \left({\cal{S}}_m+1\right)\left(
  \frac{\Delta_m\,\left(\gamma^{(1)}_{m-1,n}\right)}{{\cal{F}}_{m,n} {\cal{F}}_{m-1,n}} \right).   \nn
\eea
Similar considerations hold for symmetries in the $n$-direction. They actually follow from the above ones by employing the invariance of system (\ref{eq:3D-1212}) under the interchange of lattice variables and parameters.
}\eex  %

\subsubsection{Additive Potentials}

We consider symmetries of the discrete system (\ref{eq:dLP-gen-sys-2}).  An obvious Lie point symmetry is $\pa_{\varepsilon}\chi^{(i)}_{m,n} = 1$.  In terms of the potentials $\chi^{(i)}$, defined by (\ref{eq:dLP-gen-ch-2}), the differential-difference systems (\ref{eq:LP-sym-gen-m}) and (\ref{eq:LP-sym-gen-n}) take the form
$$
\pa_t\,\chi^{(i)}_{m,n}\,=\,q^{(i-\ell_1)}_{m,n}\,,\quad
    \pa_s\,\chi^{(i)}_{m,n}\,=\,r^{(i-\ell_2)}_{m,n}\,, \quad i \in {\mathbb{Z}}_N \,.
$$

\bex[The Discrete Boussinesq System]  {\em
It is worth separating the two component system ($N=3$). This system can be decoupled for either of the variables $\chi^{(0)}$ and $\chi^{(1)}$ to the nine-point scalar equation known as the Boussinesq equation. The lowest order symmetries for the two component system are generated by
\bea
\pa_{t^1} \chi^{(0)}_{m,n} &=& \frac{\chi_{m,n}^{(0)} - \chi_{m-1,n}^{(1)}}{\alpha^3 + (\chi_{m+1,n}^{(0)}-\chi_{m,n}^{(1)}) (\chi_{m,n}^{(0)}-\chi_{m-1,n}^{(1)}) (\chi_{m+1,n}^{(1)}-\chi_{m-1,n}^{(0)})}\,,   \nn\\
\pa_{t^1} \chi^{(1)}_{m,n} &=& \frac{\chi_{m+1,n}^{(0)} - \chi_{m,n}^{(1)}}{\alpha^3 + (\chi_{m+1,n}^{(0)}-\chi_{m,n}^{(1)}) (\chi_{m,n}^{(0)}-\chi_{m-1,n}^{(1)}) (\chi_{m+1,n}^{(1)}-\chi_{m-1,n}^{(0)})}\,,  \nn
\eea
and a master symmetry is
$$
\partial_\tau \chi^{(i)}_{m,n} \,=\, m\,\partial_{t^1} \chi_{m,n}^{(i)}\,,\quad \partial_\tau \alpha \,=\, \frac{1}{3 \alpha ^2}\,,\quad i=0,1.
$$
}\eex  %

\subsection{Symmetries in the Degenerate Case}

We briefly consider the reduction of the symmetries of our fully discrete system to the degenerate case discussed in Section \ref{degenerate}.  Whereas the $t-$flow (\ref{eq:LP-sym-gen-m}) can always be reduced to this subcase, the $s-$flow (\ref{eq:LP-sym-gen-n}) presents difficulties.

Analogous to the formula (\ref{dtlogdet}), we have
\be\label{dslogdetM}  %
\pa_s \log \left(\det M_{m,n} \right) = \Delta_n\, \left({\rm{Tr}} \left(M_{m,n}^{-1}R_{m,n}\right)\right).
\ee  %
For the generic case ($b\ne 0$) we can calculate the formula which is analogous to (\ref{tracecon})
\be\label{traceconb}
\pa_s (b-(-\lambda)^N) = b\, \Delta_n\, \left(\sum_{i=0}^{N-1} \frac{r^{(i)}_{m,n}}{v^{(i)}_{m,n}}\right) =
    \Delta_n\, \left(\sum_{i=0}^{N-1} r^{(i)}_{m,n}\, \prod_{j\ne i}v^{(j)}_{m,n}\right)
\ee
If we now let $v_{m,n}^{(N-1)}\rightarrow 0$ (so $b\rightarrow 0$), then the left hand side vanishes, leading to
\be\label{rN-1}
 r^{(N-1)}_{m,n} \, \prod_{i=0}^{N-2} v^{(i)}_{m,n} = \mbox{constant}.
\ee
If the above product is \underline{nonzero}, then we can set this constant to be $1$.  Otherwise, this equation trivialises.

In order to define the symmetry (\ref{eq:LP-sym-gen-n}), we need to solve the equations
\be\label{rv-eqs}
r^{(i)}_{m,n}\, v^{(i+k_2)}_{m,n-1} = v^{(i)}_{m,n}\, r^{(i+k_2-\ell_2)}_{m,n},\quad i \in {\mathbb{Z}}_N,
\ee
with $v^{(N-1)}_{m,n}=0$, in conjunction with (\ref{rN-1}).

\br[Further degeneration and nonlocal symmetries]
If we have (say) $v^{(N-2)}_{m,n}=0$, then the system of equations for the components $r_{m,n}^{(i)}$ is under-determined, so, at this stage, we have at least one arbitrary function.

When $k_2=0$, each $v^{(i)}$ only occurs in \underline{one} of the equations in (\ref{rv-eqs}), which is then trivially satisfied if that $v^{(i)}$ is zero.  On the other hand, of $k_2\ne 0$, then successive equations imply that successive components $r_{m,n}^{(i)}$ must vanish (or that further $v^{(i)}$ components must vanish).  This can lead to either a \underline{trivial} symmetry (all components are zero) or to arbitrary functions.

In either case, the symmetry equations can impose conditions on the surviving components, leading to a nonlocal symmetry.  Some examples of this phenomenon are given below.
\er

\bex[Hirota's KdV Equation: $N=2$, equivalence class $(0,1;0,1)$]  {\em
The $t-$flows for Hirota's KdV equation (\ref{eq:2D-Hirota}) are direct reductions of those for the $2-$component system. The first symmetry (\ref{eq:2D-sym-01}) reduces to the single component
$$
\pa_{t^1} u_{m,n} = u_{m,n} \Delta_m\, \left(\frac{1}{u_{m,n} u_{m-1,n} +a}\right).
$$
Since the $t-$hierarchy is impervious to the degeneration of $V$, the corresponding master symmetry survives:
$$
\pa_{\tau} u_{m,n} = u_{m,n} \Delta_m\,\left(\frac{m}{u_{m,n} u_{m-1,n} +a}\right)\,,\quad \pa_\tau a =1.
$$
Whilst the $s^1-$flow
$$
\pa_{s^1} u_{m,n} = u_{m,n} \Delta_n\,\left(\frac{1}{u_{m,n} u_{m,n-1} -a}\right)
$$
survives the reduction, not all symmetries do, indicating that the $2-$component master symmetry does not reduce.  However, since equation (\ref{eq:2D-Hirota}) is invariant under a simple discrete symmetry $(m,n,a)\mapsto (n,m,-a)$ (under which the $t^1$ and $s^1$ flows interchange), we can write the master symmetry for the $s-$hierarchy:
$$
\pa_{\sigma} u_{m,n} = u_{m,n} \Delta_n\,\left(\frac{n}{u_{m,n} u_{m,n-1} -a}\right)\,,\quad \pa_\sigma a =-1.
$$
To derive this from the reduced spectral problem requires the consideration of \underline{non-isospectral} flows.
}\eex

\bex[$N=3$, equivalence class $(0,1;0,1)$]  {\em
A two-component analogue of Hirota's KdV equation is given by (\ref{eq:3D-Hirota}).  The $t^1$ flow is exactly (\ref{eq:3D-sym-01}), whilst the
$s^1$ flow is given by
\bea
\pa_{s^1} u^{(0)}_{m,n} &=&
   \frac{\left(u^{(0)}_{m,n}\right)^2 u^{(1)}_{m,n} u^{(1)}_{m,n+1}}{u^{(0)}_{m,n} u^{(1)}_{m,n} u^{(1)}_{m,n+1}-a}
         \Delta_n\,\left(\frac{1}{u^{(0)}_{m,n-1} u^{(0)}_{m,n} u^{(1)}_{m,n}-a}\right),  \nn \\
\pa_{s^1} u^{(1)}_{m,n} &=&
     \frac{ u^{(0)}_{m,n-1} u^{(0)}_{m,n} \left(u^{(1)}_{m,n}\right)^2}{u^{(0)}_{m,n-1} u^{(0)}_{m,n} u^{(1)}_{m,n}-a}
         \Delta_n\,\left(\frac{1}{u^{(0)}_{m,n-1} u^{(1)}_{m,n-1} u^{(1)}_{m,n}-a}\right).  \nn
\eea
The master symmetry for the $t-$hierarchy is given by
$$
\pa_{\tau} u^{(0)}_{m,n} = u^{(0)}_{m,n} \Delta_m\,\left(\frac{m u^{(1)}_{m-1,n}}{\Gamma}\right),\quad
    \pa_{\tau} u^{(1)}_{m,n} = u^{(1)}_{m,n} \Delta_m\,\left(\frac{m u^{(0)}_{m,n}}{\Gamma}\right),
$$
where $\Gamma = a (u^{(0)}_{m,n}+u^{(1)}_{m-1,n}) + u^{(0)}_{m,n} u^{(0)}_{m-1,n} u^{(1)}_{m,n} u^{(1)}_{m-1,n}$, and we have
$\pa_\tau a = 1$.  This time we have no simple symmetry and no master symmetry to generate a hierarchy of $s-$flows.

We showed that the $2-$component system (\ref{eq:3D-Hirota}) can be reduced further to the $1-$component equation (\ref{eq:BMXa}).  In this case the $t^1$ flow and the master symmetry reduce to
$$
\pa_{t^1} u_{m,n} = u_{m,n} \Delta_m\,\left(\frac{u_{m-1,n+1}}{\Gamma}\right),\quad
    \pa_{\tau} u_{m,n} = u_{m,n} \Delta_m\,\left(\frac{m u_{m-1,n+1}}{\Gamma}\right),
$$
where $\Gamma = a (u_{m,n}+u_{m-1,n+1}) + u_{m,n} u_{m-1,n} u_{m,n+1} u_{m-1,n+1}$, with $u_{m,n}=u^{(1)}_{m,n}$ and $\pa_\tau a = 1$.
}\eex

\bex[The equivalence class $(0,1;1,2)$, for $N=3$ and $v^{(2)}_{m,n}=0$]  {\em
This is the case we discussed in Section \ref{degenerate}, which lead to the $2-$component system (\ref{degen0112}).
We wrote all functions in terms of two functions $u_{m,n}, v_{m,n}$ and their shifts, given by (\ref{degen0112-vars}).
The symmetry in the $m-$direction is given by (\ref{eq:3D-sym-01}), but under this reduction:
$$
\pa_t u = -u_{m,n} \Delta_m\, \left(\frac{u_{m-1,n}u_{m,n}v_{m,n}}{\Gamma}\right),\quad
      \pa_t v = -v_{m,n} \Delta_m\, \left(\frac{u_{m-1,n}v_{m-1,n}v_{m,n}}{\Gamma}\right),
$$
where $\Gamma = u_{m-1,n}v_{m,n}(a u_{m,n}+v_{m-1,n})+a$.  On the other hand, since $k_2=1$, equations (\ref{rv-eqs}) imply that $r^{(0)}_{m,n}=r^{(2)}_{m,n}=0$, with no constraint on $r^{(1)}_{m,n}$, leading to
$$
\pa_s u_{m,n}=0 \quad\mbox{and}\quad   \pa_s v_{m,n} = v_{m,n}^2 r^{(1)}_{m,n} .
$$
Consistency with (\ref{degen0112-vars}) (or alternatively, the equation (\ref{degen0112})) implies that $r^{(1)}_{m,n}$ must satisfy the {\em difference equation}
$$
r^{(1)}_{m+1,n} = u_{m,n} v_{m,n}^2 r^{(1)}_{m,n},
$$
so the symmetry is {\em nonlocal}.

In Section \ref{degenerate}, we saw that the $2-$component system (\ref{degen0112}) could be reduced to the single $6-$point equation (\ref{eq:BMX}). In this case the $t-$symmetry reduces to
\be\label{utbmx}
 \pa_t u_{m,n} = -u_{m,n} \Delta_m\,
      \left(\frac{u_{m-1,n} u_{m,n} u_{m-1,n+1}}{1+u_{m,n} u_{m-1,n+1} (u_{m-1,n}+u_{m,n+1})}\right).
\ee
On the other hand, the $s-$symmetry trivialises, with $r^{(1)}_{m,n}=0$.
}\eex

\bex[The equivalence class $(0,1;0,1)$, for arbitrary $N$ and $v^{(i)}_{m,n}=0$ for $i\ne 0$]  {\em
The symmetries of the scalar equation (\ref{eq:gen-dis-eq}) are given by
\be \label{eq:sym-n-all}
\pa_t u_{m,n}= u_{m,n}  \Delta_m\,(\Upsilon),\quad  \pa_{\tau} u_{m,n}= u_{m,n}  \Delta_m\,(m\Upsilon),
\ee
where
$$
\Upsilon = \frac{u_{m-1,n} \prod_{j=0}^{N-2}u_{m,n+j}}{a+\sum_{j=1}^{N-1} \prod_{r=1}^{j} u_{m,n+N-r-1} \,\prod_{s=0}^{N-j-1} u_{m-1,n+s}}.
$$
}\eex

\subsection{Differential-Difference Equations in the Non-Coprime Case}\label{symms:noncoprime}

It is straightforward to extend our discussion of symmetries to the non-coprime case, but rather than develop many general formulae, we just present a simple example.  The only additional element is $Q_{m,n}={\bf q}_{m,n}\, \Omega^{k_1}$ (in the case of the $m-$direction), which, under the action of the permutation $P$, takes the form
$$
Q_{m,n} = {\rm{diag}}({\bf q}_0,\dots , {\bf q}_{p-1})\omega_p^{k_1},\quad\mbox{where}\quad
                          {\bf q}_i = (q^{(i)},q^{(i+p)},\dots ,q^{(i+p(q-1))}),\; i=0, \dots ,p-1.
$$
We now consider one case of Example \ref{ncp-exN=4}.

\bex[$N=4,\, (k_1,\ell_1)=(1,3)$]  {\em
Using the notation of Section \ref{gcd}, we have $p=2, q=2, r=1$.  With
$$
L=\left( \begin{array}{cc}
              0 & L^{(0,1)}_{02} \\
              L^{(1,0)}_{13} & 0
              \end{array}  \right),\quad    Q=\left(
                                                \begin{array}{cc}
                                                  0 & {\bf q}_0 \\
                                                  {\bf q}_1\Omega_2 & 0
                                                          \end{array} \right),
$$
where $L^{(k,\ell)}_{ij}$ is the $2\times 2$ Lax matrix of level structure $(k,\ell)$ and depending on variables $(u^{(i)}_{m,n},u^{(j)}_{m,n})$,
the matrix symmetry equation (\ref{eq:dd-cc-1}) leads to
\begin{subequations}\label{noncoprime-Q}
\bea
&&  u^{(0)}_{m,n} q^{(2)}_{m,n} = u^{(1)}_{m-1,n} q^{(0)}_{m,n},\qquad  u^{(2)}_{m,n} q^{(0)}_{m,n} = u^{(3)}_{m-1,n} q^{(2)}_{m,n},\\
&&    u^{(1)}_{m,n} q^{(3)}_{m,n} = u^{(2)}_{m-1,n} q^{(1)}_{m,n},\qquad  u^{(3)}_{m,n} q^{(1)}_{m,n} = u^{(0)}_{m-1,n} q^{(3)}_{m,n},\\
&&  \pa_t u^{(0)}_{m,n} = q^{(1)}_{m+1,n} - q^{(0)}_{m,n}, \qquad \pa_t u^{(1)}_{m,n} = q^{(2)}_{m+1,n} - q^{(1)}_{m,n}, \\
&&    \pa_t u^{(2)}_{m,n} = q^{(3)}_{m+1,n} - q^{(2)}_{m,n}, \qquad  \pa_t u^{(3)}_{m,n} = q^{(0)}_{m+1,n} - q^{(3)}_{m,n}.
\eea
\end{subequations}
Notice that the first two of these equations implies
$$
u^{(0)}_{m+1,n}u^{(2)}_{m+1,n}=u^{(1)}_{m,n}u^{(3)}_{m,n}\quad\mbox{and}\quad u^{(1)}_{m+1,n}u^{(3)}_{m+1,n}=u^{(0)}_{m,n}u^{(2)}_{m,n}.
$$
In order to consider the conservation law (\ref{dtlogdet}), we next calculate ${\rm{Tr}} \left(L_{m,n}^{-1}Q_{m,n}\right)$.  The block structure gives
$$
{\rm{Tr}} \left(L_{m,n}^{-1}Q_{m,n}\right) = {\rm{Tr}} \left(\left(L^{(0,1)}_{02}\right)^{-1}{\bf q}_0\right) +
     {\rm{Tr}} \left(\left(L^{(1,0)}_{13}\right)^{-1}{\bf q}_1\right),
$$
so the conservation law (\ref{dtlogdet}) takes the form
$$
\frac{\pa_t(u^{(0)}_{m,n}u^{(2)}_{m,n})}{u^{(0)}_{m,n}u^{(2)}_{m,n}-\lambda^2}+
      \frac{\pa_t(u^{(1)}_{m,n}u^{(3)}_{m,n})}{u^{(1)}_{m,n}u^{(3)}_{m,n}-\lambda^2}=
       \Delta_m\, \left(\frac{u^{(2)}_{m,n}q^{(0)}_{m,n}+u^{(0)}_{m,n}q^{(2)}_{m,n}}{u^{(0)}_{m,n}u^{(2)}_{m,n}-\lambda^2} +
            \frac{u^{(3)}_{m,n}q^{(1)}_{m,n}+u^{(1)}_{m,n}q^{(3)}_{m,n}}{u^{(1)}_{m,n}u^{(3)}_{m,n}-\lambda^2}\right).
$$
Since the left hand side is independent of $m$ (being {\em symmetric} in $u^{(0)}_{m,n}u^{(2)}_{m,n}$ and $u^{(1)}_{m,n}u^{(3)}_{m,n}$), this gives us two conditions analogous to (\ref{tracecon1}):
\be\label{tracecon-noncop}  %
 u^{(2)}_{m,n}q^{(0)}_{m,n}+u^{(0)}_{m,n}q^{(2)}_{m,n} = c_0+c_1 m \quad\mbox{and}\quad
               u^{(3)}_{m,n}q^{(1)}_{m,n}+u^{(1)}_{m,n}q^{(3)}_{m,n} = c_0+c_1 m,
\ee  %
with $\pa_t (u^{(0)}_{m,n}u^{(2)}_{m,n}) = \pa_t (u^{(1)}_{m,n}u^{(3)}_{m,n}) = c_1$, where $c_0, c_1$ are constants.

The first two of (\ref{noncoprime-Q}) are then taken with (\ref{tracecon-noncop}) to determine the form of $q^{(i)}_{m,n}$.
When $c_0=1, c_1=0$, we have
$$
q^{(0)}_{m,n}=\frac{1}{u^{(1)}_{m-1,n}+u^{(2)}_{m,n}}, \quad q^{(1)}_{m,n}=\frac{1}{u^{(2)}_{m-1,n}+u^{(3)}_{m,n}},\quad
q^{(2)}_{m,n}=\frac{u^{(1)}_{m-1,n}}{u^{(0)}_{m,n}}\, q^{(0)}_{m,n} , \quad
                q^{(3)}_{m,n}=\frac{u^{(2)}_{m-1,n}}{u^{(1)}_{m,n}}\, q^{(1)}_{m,n}.
$$
The second pair of (\ref{noncoprime-Q}) then give us the symmetry.
}\eex

\section{Nonlocal Symmetries and the Relation to $2D$ Toda Systems}\label{2dtoda}
\setcounter{equation}{0}

In the previous section, we derived (generically) {\em local} continuous symmetries for the discrete system (\ref{eq:dLP-ex-cc}), as described in Proposition \ref{prop:difdif-sym-dif}.  In that case, $S_{m,n}$ of the Lax pair (\ref{eq:dLP-cont-m}) had specific forms given by (\ref{eq:all-LP-t}) and (\ref{eq:all-LP-s}).

In the current section we present two different solutions for $S_{m,n}$, one linear in $\lambda$ and the other proportional to $\lambda^{-1}$, which give {\em nonlocal} symmetries of the entire class of fully discrete systems (\ref{eq:dLP-ex-cc}) (only the first of these is nontrivial for the {\em degenerate case}).

The Lax pair given in \cite{FG3} for the $2D$ Toda lattice is generalized to (7.1) yielding nonlocal symmetries for our fully discrete system.
In this way, we associate a $2D$ Toda lattice with each equation in the entire class discussed in this paper. In the generic (non-degenerate and non-reduced) case, the nonlocal symmetries are just the components of the corresponding B\"acklund transformation and the discrete system the corresponding nonlinear superposition formula.

In the case of reduced equations, such as (\ref{eq:MX2}), this correspondence is not so straight forward.  However, in this case, we use our nonlocal symmetries to derive Schief's B\"acklund transformation \cite{S} for the Tzitzeica equation.  It is still not clear what role is being played by the discrete equation (\ref{eq:MX2}).

\subsection{Nonlocal Symmetries}

We consider the compatibility of the discrete Lax equations (\ref{eq:dLP-gen-L}) and (\ref{eq:dLP-gen-M}) with the continuous evolutions
\begin{subequations}\label{eq:all-nl-LP1}
\bea
\pa_x \Psi_{m,n} &=& \left(A_{m,n}+\lambda \Omega^{\ell_1-k_1}\right)\Psi_{m,n}\,,\label{eq:all-nl-LP-x}\\
\pa_y \Psi_{m,n} &=& \frac{1}{\lambda} B_{m,n} \Psi_{m,n}\,.\label{eq:all-nl-LP-y}
\eea
\end{subequations}
For compatibility with either (\ref{eq:dLP-gen-L}) or (\ref{eq:dLP-gen-M}), $A$ and $B$ must have the following level structure
\bea
&& A_{m,n} = {\rm{diag}}(a^{(0)}_{m,n},a^{(1)}_{m,n},\dots ,a^{(N-1)}_{m,n}),  \nn\\
&&   B_{m,n} = {\rm{diag}}(b^{(0)}_{m,n},b^{(1)}_{m,n},\dots ,b^{(N-1)}_{m,n}) \Omega^{k_1-\ell_1} =
                      {\rm{diag}}(b^{(0)}_{m,n},b^{(1)}_{m,n},\dots ,b^{(N-1)}_{m,n})\Omega^{k_2-\ell_2}.\nn
\eea
The calculations for the two cases are identical, just requiring the change of labels, $(u,k_1,\ell_1)\mapsto (v,k_2,\ell_2)$.  The condition (\ref{eq:dLP-nec-rel}) allows them to be {\em simultaneously} compatible.  The results can be summarised as follows:

\begin{Pro}\label{prop:nonlocsym}

The compatibility of (\ref{eq:all-nl-LP-x}) with (\ref{eq:dLP-gen-L}) and (\ref{eq:dLP-gen-M}) leads to the following equations
\begin{subequations}\label{dxLM}
\bea  %
\pa_{x} u^{(i)}_{m,n} &=& u^{(i)}_{m,n}\,\left(a^{(i)}_{m+1,n}\,-\,a^{(i+k_1)}_{m,n} \right),  \label{dxL}  \\
    \pa_{x} v^{(i)}_{m,n} &=& v^{(i)}_{m,n}\,\left(a^{(i)}_{m,n+1}\,-\,a^{(i+k_2)}_{m,n} \right),  \label{dxM}
\eea  %
\end{subequations}
where
\begin{subequations}\label{a-diff}
\bea  %
    a^{(i)}_{m+1,n}-a^{(i+\ell_1)}_{m,n} &=&   u_{m,n}^{(i)}- u_{m,n}^{(i+\ell_1-k_1)},  \\
    a^{(i)}_{m,n+1}-a^{(i+\ell_2)}_{m,n} &=&   v_{m,n}^{(i)}- v_{m,n}^{(i+\ell_2-k_2)}
\eea  %
\end{subequations}
whilst the compatibility of (\ref{eq:all-nl-LP-y}) with (\ref{eq:dLP-gen-L}) and (\ref{eq:dLP-gen-M}) leads to
\begin{subequations}\label{dyLM}
\bea  %
\pa_{y} u^{(i)}_{m,n} &=& b^{(i)}_{m+1,n}\,-\,b^{(i+\ell_1)}_{m,n},  \label{dyL}  \\
    \pa_{y} v^{(i)}_{m,n} &=& b^{(i)}_{m,n+1}\,-\,b^{(i+\ell_2)}_{m,n},  \label{dyM}
\eea  %
\end{subequations}
where
\begin{subequations}\label{b-diff}
\bea %
    b^{(i)}_{m+1,n}\, u^{(i+k_1-\ell_1)}_{m,n} &=&   u_{m,n}^{(i)}\, b_{m,n}^{(i+k_1)},  \\
    b^{(i)}_{m,n+1}\, v^{(i+k_2-\ell_2)}_{m,n} &=&   v_{m,n}^{(i)}\, b_{m,n}^{(i+k_2)},
\eea  %
\end{subequations}
Equations (\ref{dxLM}) and (\ref{dyLM}) define nonlocal symmetries for our general discrete system (\ref{eq:dLP-ex-cc}).
\end{Pro}
The $y-$flow does not always exist for the degenerate cases of Section \ref{degenerate}.

\subsubsection{Quotient Potential Form}

In the quotient potential form (\ref{eq:dLP-gen-ch-1}), equations (\ref{dxLM}) immediately integrate to give $a^{(i)}_{m,n}=\pa_x(\log \phi^{(i)}_{m,n})$, whilst (\ref{a-diff}) lead to
 \begin{subequations}\label{dxphi}
\bea
\pa_x \log \left(\frac{\phi_{m+1,n}^{(i)}}{\phi^{(i+\ell_1)}_{m,n}}\right) &=& \alpha \,\left( \frac{\phi_{m+1,n}^{(i)}}{\phi^{(i+k_1)}_{m,n}}- \frac{\phi_{m+1,n}^{(i+\ell_1-k_1)}}{\phi^{(i+\ell_1)}_{m,n}} \right), \label{phi-nl-Lx}  \\
\pa_x \log \left(\frac{\phi_{m,n+1}^{(i)}}{\phi^{(i+\ell_2)}_{m,n}}\right) &=& \beta \,\left( \frac{\phi_{m,n+1}^{(i)}}{\phi^{(i+k_2)}_{m,n}}\,-\, \frac{\phi_{m,n+1}^{(i+\ell_1-k_1)}}{\phi^{(i+\ell_2)}_{m,n}} \right), \label{phi-nl-Mx}
\eea
\end{subequations}
On the other hand, (\ref{b-diff}) lead to $b^{(i)}_{m,n}=\frac{\phi_{m,n}^{(i)}}{\phi_{m,n}^{(i+k_1-\ell_1)}}$, with (\ref{dyLM}) giving
\begin{subequations}\label{dyphi}
\bea
\pa_y \log \left(\frac{\phi_{m+1,n}^{(i)}}{\phi^{(i+k_1)}_{m,n}}\right) &=& \frac{1}{\alpha} \,\left( \frac{\phi_{m,n}^{(i+k_1)}}{\phi^{(i+k_1-\ell_1)}_{m+1,n}}- \frac{\phi_{m,n}^{(i+\ell_1)}}{\phi^{(i)}_{m+1,n}} \right), \label{phi-nl-Ly} \\
\pa_y \log \left(\frac{\phi_{m,n+1}^{(i)}}{\phi^{(i+k_2)}_{m,n}}\right) &=& \frac{1}{\beta} \,\left( \frac{\phi_{m,n}^{(i+k_2)}}{\phi^{(i+k_2-\ell_2)}_{m,n+1}}- \frac{\phi_{m,n}^{(i+\ell_2)}}{\phi^{(i)}_{m,n+1}} \right). \label{phi-nl-My}
\eea
\end{subequations}
Equations (\ref{dxphi}) and (\ref{dyphi}) define nonlocal symmetries for the potential system (\ref{eq:dLP-gen-sys-1}).

\bex[Modified KdV : $N=2$, structure $(0,1;0,1)$, quotient potential]  {\em
The $x$ flow (see also \cite{ABS}) takes the form
\bea  %
&& \pa_x \log\left(\phi_{m,n}\phi_{m+1,n}\right) =
    \alpha \left(\frac{\phi_{m+1,n}}{\phi_{m,n}} \,-\,\frac{\phi_{m,n}}{\phi_{m+1,n}} \right),  \nn\\
&& \pa_x \log\left(\phi_{m,n}\phi_{m,n+1}\right) = \beta \left(\frac{\phi_{m,n+1}}{\phi_{m,n}} \,-\,\frac{\phi_{m,n}}{\phi_{m,n+1}} \right). \nn
\eea  %
The $y$ flow is
\bea  %
\alpha \pa_y \log\left(\frac{\phi_{m+1,n}}{\phi_{m,n}}\right) = \phi_{m,n}\phi_{m+1,n} \,-\,\frac{1}{\phi_{m,n}\phi_{m+1,n}} ,\nn\\
 \beta \pa_y \log\left(\frac{\phi_{m,n+1}}{\phi_{m,n}}\right) = \phi_{m,n}\phi_{m,n+1} \,-\,\frac{1}{\phi_{m,n}\phi_{m,n+1}}.  \nn
\eea  %
}\eex

\subsubsection{Additive Potential Form}

In the additive potential form (\ref{eq:dLP-gen-ch-2}), equations (\ref{dyLM}) immediately integrate to give $b^{(i)}_{m,n}=\pa_y\chi^{(i)}_{m,n})$, whilst (\ref{b-diff}) lead to
 \begin{subequations}\label{dychi}
\bea
\pa_y \chi^{(i)}_{m+1,n} (\chi^{(i+k_1-\ell_1)}_{m+1,n}-\chi^{(i+k_1)}_{m,n}) &=&
                   (\chi^{(i)}_{m+1,n}-\chi^{(i+\ell_1)}_{m,n}) \pa_y \chi^{(i+k_1)}_{m,n},   \\
\pa_y \chi^{(i)}_{m,n+1} (\chi^{(i+k_2-\ell_2)}_{m,n+1}-\chi^{(i+k_2)}_{m,n}) &=&
                   (\chi^{(i)}_{m,n+1}-\chi^{(i+\ell_2)}_{m,n}) \pa_y \chi^{(i+k_2)}_{m,n}.
\eea
\end{subequations}
On the other hand, (\ref{a-diff}) lead to $a^{(i)}_{m,n}=\chi^{(i)}_{m,n}-\chi^{(i+\ell_1-k_1)}_{m,n}$, with (\ref{dxLM}) giving
\begin{subequations}\label{dxchi}
\bea
\pa_x \left(\log \left(\chi^{(i)}_{m+1,n}-\chi^{(i+\ell_1)}_{m,n}\right)\right) &=& \chi^{(i)}_{m+1,n}-\chi^{(i+k_1)}_{m,n}-
     \left(\chi^{(i+\ell_1-k_1)}_{m+1,n}-\chi^{(i+\ell_1)}_{m,n}\right),  \\
\pa_x \left(\log \left(\chi^{(i)}_{m,n+1}-\chi^{(i+\ell_2)}_{m,n}\right)\right) &=& \chi^{(i)}_{m,n+1}-\chi^{(i+k_2)}_{m,n}-
     \left(\chi^{(i+\ell_2-k_2)}_{m,n+1}-\chi^{(i+\ell_2)}_{m,n}\right).
\eea
\end{subequations}
Equations (\ref{dxchi}) and (\ref{dychi}) define nonlocal symmetries for the potential system (\ref{eq:dLP-gen-sys-2}).

\bex[Potential KdV : $N=2$, structure $(0,1;0,1)$, additive potential]  {\em
The $x$ flows (see also \cite{ABS}), after some manipulation, take the form
\bea  %
&& \pa_x \left(\chi_{m+1,n}+\chi_{m,n}\right) = (\chi_{m+1,n}-\chi_{m,n})^2+\alpha^2 \,,\nn\\
 && \pa_x \left(\chi_{m,n+1}+\chi_{m,n}\right) = \left(\chi_{m,n+1}-\chi_{m,n}\right)^2+\beta^2 . \nn
\eea  %
The second function $\chi^{(1)}$ cannot be removed from the $y$ flow. It remains in the formulae as a pseudo-potential $\psi$.  Explicitly,
$$
 \pa_y \chi_{m+1,n} = \frac{ (\chi_{m+1,n}-\psi_{m,n})^2}{\alpha^2}\partial_y \chi_{m,n}\,,\quad
 \pa_y \chi_{m,n+1} = \frac{ (\chi_{m,n+1}-\psi_{m,n})^2}{\beta^2}\partial_y \chi_{m,n},
$$
where $\psi_{m,n}$ is pseudo-potential such that
$$
(\chi_{m+1,n}-\psi_{m,n})  (\chi_{m,n}-\psi_{m+1,n})= \alpha^2,\quad
      (\chi_{m,n+1}-\psi_{m,n}) (\chi_{m,n}-\psi_{m,n+1}) = \beta^2.
$$
In fact, $\psi$ is another solution of H1 related to $\chi$ via the above B{\"a}cklund transformation.
}\eex

\bex[Schwarzian KdV : $N=2$, structure $(1,0;1,0)$, additive potential]  {\em
The $x$ flows, after some manipulation, take the form
\bea  %
 \pa_x \left(\chi_{m+1,n}-\chi_{m,n}\right) &=&
   \left(\chi_{m+1,n}-\chi_{m,n}\right) \left(\chi_{m+1,n}+\chi_{m,n}-\psi_{m+1,n}-\psi_{m,n} \right) \,,\nn\\
 \pa_x \left(\chi_{m,n+1}-\chi_{m,n}\right) &=&
        \left(\chi_{m,n+1}-\chi_{m,n}\right) \left(\chi_{m,n+1}+\chi_{m,n}-\psi_{m,n+1}-\psi_{m,n} \right) , \nn
\eea  %
where the pseudo-potential $\psi$ is determined by the relations
$$
\psi_{m+1,n}-\psi_{m,n}\,=\,\frac{\alpha^2}{\chi_{m+1,n}-\chi_{m,n}}\,,\quad
     \psi_{m,n+1}-\psi_{m,n}\,=\,\frac{\beta^2}{\chi_{m,n+1}-\chi_{m,n}}.
$$
The $y$ flow (see also \cite{ABS}) takes the form
$$
\left( \pa_y \chi_{m+1,n}\right) \left( \pa_y \chi_{m,n}\right) = \frac{\left(\chi_{m+1,n}-\chi_{m,n}\right)^2}{\alpha^2}\,,\quad
    \left( \pa_y \chi_{m,n+1}\right) \left( \pa_y \chi_{m,n}\right) = \frac{\left(\chi_{m,n+1}-\chi_{m,n}\right)^2}{\beta^2}.
$$
}\eex

\subsection{Associated $2D$ Toda Lattices}\label{2dtoda-der}

The compatibility condition of (\ref{eq:all-nl-LP-x}) and (\ref{eq:all-nl-LP-y}) leads to the equations
\begin{subequations}\label{eq:psixy}
\bea
\pa_{y} a^{(i)}_{m,n} &=& b^{(i)}_{m,n}\,-\,b^{(i+\ell_1-k_1)}_{m,n}, \label{eq:psixy-ay}  \\
       \pa_{x} \log \left(b^{(i)}_{m,n}\right) &=& a^{(i)}_{m,n} - a^{(i+k_1-\ell_1)}_{m,n}.  \label{eq:psixy-bx}
\eea
\end{subequations}

\subsubsection{Quotient Potential Form}

Here we have
$$
a^{(i)}_{m,n}=\pa_x(\log \phi^{(i)}_{m,n})\quad\mbox{and}\quad  b^{(i)}_{m,n}=\frac{\phi_{m,n}^{(i)}}{\phi_{m,n}^{(i+k_1-\ell_1)}},
$$
so (\ref{eq:psixy-bx}) is identically satisfied, whilst (\ref{eq:psixy-ay}) leads to
\be \label{eq:Toda}
\pa_{x}\pa_y \theta^{(i)}_{m,n} = \exp(\theta^{(i)}_{m,n}-\theta^{(i+k_1-\ell_1)}_{m,n})-
                                     \exp(\theta^{(i+\ell_1-k_1)}_{m,n}-\theta^{(i)}_{m,n}),
\ee
where $\phi^{(i)}_{m,n}=\exp(\theta_{m,n}^{(i)})$.  These are the {\em Toda lattice equations} for $\theta^{(i)}_{m,n}$.

The nonlocal symmetries (\ref{phi-nl-Lx}) and (\ref{phi-nl-Ly}) then take the form of the {\em B\"acklund transformation}
\bea
  \pa_x \left( \theta_{m+1,n}^{(i)}-\theta_{m,n}^{(i+\ell_1)}\right) & = &
    \alpha \left(\exp(\theta_{m+1,n}^{(i)}-\theta_{m,n}^{(i+k_1)}) - \exp(\theta_{m+1,n}^{(i+\ell_1-k_1)}-\theta_{m,n}^{(i+\ell_1)})\right),\nn \\
    &&  \label{btm}   \\
 \pa_y \left( \theta_{m+1,n}^{(i)}-\theta_{m,n}^{(i+k_1)}\right) &=&
    \frac{1}{\alpha} \left(\exp(\theta_{m,n}^{(i+k_1)}-\theta_{m+1,n}^{(i+k_1-\ell_1)}) -
                                      \exp(\theta_{m,n}^{(i+\ell_1)}-\theta_{m+1,n}^{(i)})\right), \nn
\eea
whilst the nonlocal symmetries (\ref{phi-nl-Mx}) and (\ref{phi-nl-My}) take the form of the {\em B\"acklund transformation}
\bea
   \pa_x \left( \theta_{m,n+1}^{(i)}-\theta_{m,n}^{(i+\ell_2)}\right) &=&
   \beta \left( \exp(\theta_{m,n+1}^{(i)}-\theta_{m,n}^{(i+k_2)}) - \exp(\theta_{m,n+1}^{(i+\ell_2-k_2)}-\theta_{m,n}^{(i+\ell_2)})\right), \nn \\
      &&   \label{btn}   \\
 \pa_y \left( \theta_{m,n+1}^{(i)}-\theta_{m,n}^{(i+k_2)}\right) &=&
    \frac{1}{\beta}\left(\exp(\theta_{m,n}^{(i+k_2)}-\theta_{m,n+1}^{(i+k_2-\ell_2)}) -
                          \exp(\theta_{m,n}^{(i+\ell_2)}-\theta_{m,n+1}^{(i)})\right).   \nn
\eea

\br[$k_2-\ell_2 \equiv k_1-\ell_1 \; (\bmod N)$]
Notice that only the combination $k_1-\ell_1$ appears in the Toda lattice equations.  The relation (\ref{eq:dLP-nec-rel}) therefore guarantees that both of these B\"acklund transformations lead to the same Toda lattice equations.  Furthermore, \underline{different} choices of $k_i,\ell_i$, subject only to their \underline{difference} $k_i-\ell_i$ remaining constant, lead to \underline{different} forms of the B\"acklund transformation for the \underline{same} Toda Lattice equation.
\er

\br[Sum of the B\"acklund Equations]
Summing the equations of (\ref{btm}), we find
$$
\pa_x \left(\sum_{i=0}^{N-1} \theta_{m+1,n}^{(i)}\right) = \pa_x \left(\sum_{i=0}^{N-1} \theta_{m,n}^{(i+\ell_1)}\right) \quad\mbox{and}\quad
   \pa_y \left(\sum_{i=0}^{N-1} \theta_{m+1,n}^{(i)}\right) = \pa_y \left(\sum_{i=0}^{N-1} \theta_{m,n}^{(i+k_1)}\right),
$$
and similarly for (\ref{btn}).  In fact, as a consequence of (\ref{eq:LP-ir-g-rat-2}), we have $\sum_{i=0}^{N-1} \theta_{m,n}^{(i)}=0$.

As a consequence, only $N-1$ of the B\"acklund equations are independent, for each of the four sets of equations in (\ref{btm}) and (\ref{btn}).
\er

\bex[The Tzitzeica Reduction]  {\em
The choice $N=3$, with $k_i=1,\ell_i=2$ corresponds to the level structure $(1,2;1,2)$ and equation (\ref{eq:3D-1212}), with reduction (\ref{eq:MX2}). In the present context, it can be seen that equations (\ref{eq:Toda}) admit the reduction $\theta^0_{m,n} = -\theta^1_{m,n} = \theta_{m,n}, \,\theta^2_{m,n}=0$, giving rise to the single component equation
\be\label{tz-eqn}
\pa_x\pa_y\theta_{m,n} = e^{\theta_{m,n}} - e^{-2\theta_{m,n}},
\ee
known as the Tzitzeica equation.  However, the reduction of the B\"acklund equations to this case is not so simple. Setting $\theta^{(2)}_{m+1,n}= -\theta^{(0)}_{m+1,n}-\theta^{(1)}_{m+1,n}$, the four independent equations of (\ref{btm}) are
\begin{subequations}\label{tz-bt}
\bea
\pa_x \theta^{(0)}_{m+1,n} &=& \alpha \left(e^{\theta^{(0)}_{m+1,n}+\theta_{m,n}}-e^{\theta^{(1)}_{m+1,n}}\right), \label{tz-bt0x} \\
\pa_x (\theta^{(1)}_{m+1,n}-\theta_{m,n}) &=& \alpha \left(e^{\theta^{(1)}_{m+1,n}} - e^{-\theta^{(0)}_{m+1,n} - \theta^{(1)}_{m+1,n} - \theta_{m,n}}\right), \label{tz-bt1x} \\
\pa_y (\theta^{(0)}_{m+1,n}+\theta_{m,n}) &=& \frac{1}{\alpha} \left(e^{\theta^{(0)}_{m+1,n} + \theta^{(1)}_{m+1,n} - \theta_{m,n}} - e^{-\theta^{(0)}_{m+1,n}}\right), \label{tz-bt0y} \\
\pa_y \theta^{(1)}_{m+1,n} &=& \frac{1}{\alpha} \left(e^{-\theta^{(0)}_{m+1,n}} - e^{\theta_{m,n}-\theta^{(1)}_{m+1,n}}\right). \label{tz-bt1y}
\eea
\end{subequations}
Equations (\ref{tz-bt0x}) and (\ref{tz-bt1y}) imply
\be\label{tz-conlaw}
\pa_x \left(\frac{1}{\alpha} e^{-\theta^{(0)}_{m+1,n}}\right) = \pa_y \left(\alpha e^{\theta^{(1)}_{m+1,n}}\right).
\ee
One option is to set $\theta^{(1)}_{m+1,n}=-\theta^{(0)}_{m+1,n}$, but this immediately implies
$$
(\pa_x-\alpha^2\pa_y)\theta^{(0)}_{m+1,n}=0 \quad\mbox{and}\quad (\pa_x-\alpha^2\pa_y)\theta_{m,n}=0,
$$
with each function satisfying (\ref{tz-eqn}), so they represent {\em travelling wave solutions}, related by
$$
\pa_z \theta^{(0)}_{m+1,n} = \frac{1}{\alpha}\left(e^{\theta^{(0)}_{m+1,n}+\theta_{m,n}}-e^{-\theta^{(0)}_{m+1,n}}\right),\quad
\pa_z \theta_{m,n} = \frac{1}{\alpha}\left(e^{\theta^{(0)}_{m+1,n}+\theta_{m,n}}-e^{-\theta_{m,n}}\right),
$$
where each of the two functions depends only upon $z=\alpha^2 x+ y$.

To avoid this degeneration, we use (\ref{tz-conlaw}) to define the {\em potential function} $w(x,y)$, satisfying
$$
\frac{1}{\alpha} e^{-\theta^{(0)}_{m+1,n}} = -\, \pa_y \log(w), \quad \alpha e^{\theta^{(1)}_{m+1,n}} = -\, \pa_x \log(w).
$$
The remaining equations of (\ref{tz-bt}) then imply
$$
w_{xx}= \theta_x w_x - \alpha^3 e^{-\theta} w_y, \quad w_{xy} = e^\theta w,\quad w_{yy}= \theta_y w_y - \frac{e^{-\theta} w_x}{\alpha^3},
$$
where $\theta = \theta_{m,n}$ and $x$ and $y$ suffices denote partial derivatives.  These are the equations derived by Schief in \cite{Sc,S} for the B\"acklund transformation of the Tzitzeica equation.
}\eex

\subsubsection{Additive Potential Form}

Here we have
$$
a^{(i)}_{m,n}=\chi^{(i)}_{m,n}-\chi^{(i+\ell_1-k_1)}_{m,n} \quad\mbox{and}\quad
             b^{(i)}_{m,n}=\pa_y\chi^{(i)}_{m,n},
$$
so (\ref{eq:psixy-ay}) is identically satisfied, whilst (\ref{eq:psixy-bx}) leads to
$$
\pa_x \log \left( \pa_y\,\chi^{(i)}_{m,n}\right) =  2 \chi^{(i)}_{m,n} - \chi^{(i+k_1-\ell_1)}_{m,n} - \chi^{(i+\ell_1-k_1)}_{m,n}.
$$
Differentiating this with respect to $y$ and defining $\rho_{m,n}^{(i)}=\log (\pa_y\,\chi^{(i)}_{m,n})$, we obtain an alternative form of the Toda equations
\be \label{eq:Toda-rho}
\pa_x\pa_y \rho_{m,n}^{(i)} = 2 \exp(\rho^{(i)}_{m,n}) - \exp(\rho^{(i+k_1-\ell_1)}_{m,n}) - \exp(\rho^{(i+\ell_1-k_1)}_{m,n}).
\ee

\subsection{Degenerate Case and Nonlocal Symmetries}

We do not have general formulae in this case, so just present an explicit example.

\bex[Hirota's KdV Equation]  {\em  %

For Hirota's KdV equation (\ref{eq:2D-Hirota}), a nonlocal symmetry is generated by
\begin{subequations} \label{eq:2D-Hirota-nl-sym}
\begin{equation}
\pa_x \log \left(u_{m,n}\right)\,=\,-2 a^{(0)}_{m,n} +u_{m,n} \,-\,\frac{a}{u_{m,n}}\,,
\end{equation}
where
\begin{equation}
a^{(0)}_{m+1,n}=u_{m,n}\,-\,\frac{a}{u_{m,n}}\,-\,a^{(0)}_{m,n}\,,\quad a^{(0)}_{m,n+1}\,=\,u_{m,n}\,-\,\frac{a}{u_{m,n+1}}\,-\,a^{(0)}_{m,n}\,.
\end{equation}
\end{subequations}
}\eex  %

\section{Conclusions \& discussion}\label{conclusions}

\setcounter{equation}{0}

In this paper we considered the class of discrete Lax pairs (\ref{eq:dLP-gen}) and a classification problem for such systems. We were naturally led to considering {\em coprime} vs {\em non-coprime} cases and focussed mainly in the analysis of the coprime case.  We were also naturally led to considering {\em generic} vs {\em degenerate} cases.  All aspects of the generic case can be systematically analysed, but the degenerate systems requires a case by case analysis, so their classification is far from complete.

The {\em generic coprime} case has two natural descriptions in terms of potential functions.  These are related through a B\"acklund transformation, but some well known, low dimensional, examples fit naturally into each description.  We presented all the inequivalent systems in two and three dimensions. As well as the very well known systems, we found several new ones, including (\ref{eq:3D-0112}), (\ref{eq:3D-0120}), (\ref{eq:3D-1212}), and (\ref{eq:3D-0101-a}). The degenerate case lead to Hirota's KdV equation and its multi-component generalisations (\ref{eq:gen-dis-eq}). Reductions of the three dimensional systems resulted in new $2-$component integrable systems (\ref{eq:3D-Hirota}) and (\ref{degen0112}), with further reductions leading to integrable scalar equations (\ref{eq:BMXa}) and (\ref{eq:BMX}) which are defined on six-point stencils, which may be considered as the discrete analogue of higher order hyperbolic partial differential equations \cite{ASh}.

In Section \ref{sect:noncoprime} we presented some examples of {\em non-coprime} systems, which give a mechanism for coupling coprime systems.  In particular, we presented coupled systems of the discrete MKdV equation and of Hirota's discrete sine-Gordon equation.  In Section \ref{sect:lattice} we discussed the problem of building $2D$ lattices with a mixture of equations in adjacent quadrilaterals and also the extension to $3D$ consistent systems.

Also for the {\em generic coprime} case we gave a complete description of isospectral deformations of $L$ (or $M$), corresponding to hierarchies of {\em autonomous} differential-difference equations.  Furthermore, we presented a {\em non-autonomous flow} which played the role of a {\em master symmetry}, which was used to generate the autonomous hierarchy.  When we ask for compatibility between these continuous flows and {\em both} $L$ {\em and} $M$, then these differential-difference equations are interpreted as symmetries of the fully discrete system (\ref{eq:dLP-ex-cc}).  This is another manifestation of the integrability of the fully discrete system.  Once again, the degenerate case is less systematic.  Whilst the compatibility with $L$ is is still guaranteed under this reduction, the compatibility with $M$ depends upon the specific values of $k_2$ and $\ell_2$.  Whilst the $t-$flows continue to exist, the $s-$flows may trivialise or become {\em nonlocal}. As a result, it is evident that our general construction of the corresponding master symmetry ($\sigma$ flow) is no longer valid. Nevertheless it is sometimes possible to generalise this construction by assuming that the matrix $R_{m,n}$ (see (\ref{dsPsi})) depends on $\lambda$, and to allow non-isospectrality, i.e. $\pa_\sigma\lambda\ne 0$. This approach was used for the derivation of the master symmetry in the $n$ direction for the Hirota's KdV equation, for which we found that
$$
R_{m,n}\,=\,\left(\begin{array}{cc}
                  \frac{n-1}{v^{(0)}_{m,n-1}} & \frac{1}{2 \lambda} \\[3mm]
                   0 &  \frac{n}{v^{(0)}_{m,n}}
                   \end{array} \right)\,.
$$
The study of this class of Lax pairs will be considered elsewhere.

Finally, we looked at a specific class of {\em nonlocal symmetries} in Section \ref{2dtoda} and again were able to give a complete analysis in the {\em generic coprime} case.  We gave two such symmetries (labelled $x-$ and $y-$flows), which again had natural representations in both the {\em quotient} and {\em additive} potential forms, giving simple forms for nonlocal symmetries of the corresponding potential forms of the fully discrete system.  These two forms naturally gave rise to two forms of the Toda lattice equations.  The quotient form immediately gives the standard form of the Toda lattice equations, with the nonlocal symmetries taking the form of components of the corresponding B\"acklund transformation.  In fact, for {\em each} Toda lattice equation, we obtain a sequence of different B\"acklund transformations.  Thus, in the three component case with $\ell_i-k_i =1$, we have three such transformations. These can be reduced to the B{\"a}cklund transformation for the Tzitzeica equation \cite{Sc,S}, but only one case possesses a reduced discrete system (equation (\ref{eq:MX2})).

There are a number of open questions.  We currently have no {\em proof} that our master symmetries generate {\em commuting hierarchies}, but have not found a counter-example in our examples.  We have found several {\em reductions} to lower dimensional systems (such as (\ref{eq:MX2}) and (\ref{eq:gen-dis-eq})) but have no systematic way of analysing these.  This includes the extension to non-isospectrality, as mentioned above.  The connection with well known reduced PDEs, such as the Sawada-Kotera and Hirota-Satsuma equations is also not clear.  In this paper, we restricted our Lax pairs to be {\em linear} in $\lambda$. Similar Lax pairs, polynomial in $\lambda$ would be interesting to consider.

\subsection*{Acknowledgements}

PX acknowledges support from the EPSRC grant {\emph{Structure of partial difference equations with continuous symmetries and conservation laws}}, EP/I038675/1.  We thank Frank Nijhoff for bringing the paper \cite{NPbog} to our attention.


\begin{thebibliography}{99}
\bibitem{A} V.E. Adler (2012) On a discrete analog of the Tzitzeica equation {\emph{ arXiv:1103.5139}}

\bibitem{ABS} V. E. Adler, A. I. Bobenko, Yu. B. Suris (2003) Classification of integrable equations on quad-graphs. The consistency approach {\emph{Comm. Math. Phys.}} {\bf{233}} 513--543

\bibitem{ASh} V.E. Adler, A.B. Shabat (2012) Toward a theory of integrable hyperbolic equations of third order  {\emph{J. Phys. A: Math. Theor.}} {\bf{45}} 395207 (17pp)

\bibitem{ALN} J. Atkinson, S. B. Lobb, F. W. Nijhoff (2012) An integrable multicomponent quad-equation and its Lagrangian formulation, {\emph{Theor. Math. Phys.}} {\bf 173} 1644--1653

\bibitem{BEF} S. Baker, V.Z. Enolskii, A.P. Fordy (1995) Integrable quartic potentials and coupled KdV equations {\emph{Phys.Letts.A}} {\bf{201}} 167--74

\bibitem{BMX} G. Berkeley, A. Mikhailov, P. Xenitidis (2014) {\emph{in preparation}}

\bibitem{Bo} R. Boll (2011) Classification of 3D consistent quad-equations {\emph{J. Nonlin. Math. Phys.}} {\bf{18}} 337--365

\bibitem{DJM} E. Date, M. Jimbo, T. Miwa (1983) Method for Generating Discrete Soliton Equation III {\emph{J. Phys. Soc. Japan}} {\bf{52}} 388--393

\bibitem{FG1} A. P. Fordy, J. Gibbons (1980) Factorization of operators I. Miura transformations {\emph{J. Math. Phys.}} {\bf{21}} 2508--2510

\bibitem{FG2} A. P. Fordy, J. Gibbons (1981) Factorization of operators II {\emph{J. Math. Phys.}} {\bf{22}} 1170--1175

\bibitem{FG3} A. P. Fordy, J. Gibbons (1980) Integrable Nonlinear Klein-Gordon Equations and Toda Lattices {\emph{Commun. Math. Phys.}} {\bf{77}} 21--30

\bibitem{HV2} J. Hietarinta and C. Viallet (2012) Weak Lax pairs for lattice equations {\em{Nonlinearity}} {\bf{25}} 1955 doi:10.1088/0951-7715/25/7/1955

\bibitem{M1} A.V. Mikhailov (1979) Integrability of a two-dimensional generalization of the Toda chain {\emph{JETP Lett}} {\bf{30}} 443--448

\bibitem{MX1} A.V. Mikhailov (2009) From automorphic Lie Algebras to discrete integrable systems, https://www.newton.ac.uk/seminar/20090617140014501

\bibitem{MX} A.V. Mikhailov, P. Xenitidis (2013) Second order integrability conditions for difference equations. An integrable equation {\emph{Lett. Math. Phys.}} doi 10.1007/s111005-013-0668-8

\bibitem{NC} F. W. Nijhoff, H. W. Capel (1995) The Discrete Korteweg-De Vries Equation {\emph{Acta Applicandae Mathematica}} {\bf{39}} 133--158

\bibitem{NPCQ} F. W. Nijhoff, V. G. Papageorgiou, H. W. Capel, G. R. W. Quispel (1992) The lattice Gel'fand-Dikii hierarchy {\emph{Inverse Problems}} {\bf{8}} 597--621

\bibitem{NPbog} F. W. Nijhoff, V. G. Papageorgiou (1996) On some integrable discrete-time systems associated with the Bogoyavlensky lattices {\emph{Physica A}} {\bf{228}} 172-188

\bibitem{Sc} W. K. Schief  (1996) Self-dual Einstein spaces via a permutability theorem for the Tzitzeica equation, {\emph{Phys. Lett. A}} {\bf{223}}, 55--62

\bibitem{S}  W. K. Schief (1996) The Tzitzeica equation: A B{\"a}cklund transformation interpreted as truncated Painlev\'e expansion, {\emph{J. Phys. A: Math. Theor.}} {\bf{29}} 5153–-5155

\bibitem{SHL} C. Scimiterna, M Hay, D. Levi (2014) On the integrability of a new lattice equation found by multiple scale analysis, {\emph{J. Phys. A: Math. Theor.}} {\bf{47}} 265204

\bibitem{X1} P. Xenitidis  (2009) Integrability and symmetries of difference equations: the Adler--Bobenko--Suris case. In {\em Proceedings of the 4th Workshop ``Group Analysis of Differential Equations and Integrable Systems'', Cyprus, 2008}, arXiv: 0902.3954

\bibitem{X} P. Xenitidis (2011) Symmetries and conservation laws of the ABS equations and corresponding differential–difference equations of Volterra type {\emph{J. Phys. A: Math. Theor.}} {\bf{44}} 435201

\bibitem{XN} P. Xenitidis and F. W. Nijhoff (2012) Symmetries and conservation laws of lattice Boussinesq equations {\emph{Phys. Lett. A}} {\bf{376}} 2394--2401

\bibitem{XN1}  P.  Xenitidis, F. Nijhoff (2012)  Lattice Schwarzian Boussinesq equation and two-component systems {\emph{ arXiv:1202.5767}}

\bibitem{XP} P. Xenitidis and V. G. Papageorgiou  (2009) Symmetries and integrability of discrete equations defined on a black--white lattice {\emph{J. Phys. A: Math. Theor.}} {\bf{42}} 454025

\bibitem{Y} R. Yamilov (2006) Symmetries as integrability criteria for differential difference equations {\emph{J. Phys. A: Math. Theor.}} {\bf{39}} R541-R623

\end{thebibliography}
\end{document}